\documentclass[nofootinbib,preprint,preprintnumbers,a4paper,11pt]{revtex4}
\usepackage{epsfig}
\usepackage{footnote}
\usepackage{ulem}
\usepackage{color}
\usepackage{array}
\usepackage{amssymb}
\usepackage{amsmath}
\usepackage{graphicx}
\usepackage{subfigure}
\usepackage{longtable}
\usepackage{verbatim}
\usepackage{amsfonts}
\usepackage{hyperref}
\usepackage{multirow}
\usepackage{cancel}
\usepackage{booktabs}
\usepackage[margin=1.6cm]{geometry}

\usepackage{simplewick}

\begin{document}

\title{Topological Diagrams and Hadronic Rescattering: \\ A Framework for CP Violation in Charmed Baryon Decays}

\author{Ying-Xin Lai$^{1}$}
\author{Di Wang$^{1}$}\email{wangdi@hunnu.edu.cn}
\address{%
$^1$Department of Physics, Hunan Normal University, Changsha 410081, China
}

\begin{abstract}
Establishing CP violation in the charmed baryon sector remains a central challenge in particle physics.
In this work, we develop a framework for CP violation in charmed baryon decays by combining quark-level topological amplitudes with the hadron-level rescattering mechanism.
The two descriptions are connected through (1,1)-rank amplitudes, which are linear combinations of topological amplitudes and can be matched directly to the chiral Lagrangian.
We verify that the rescattering amplitudes derived from topological amplitudes are consistent with those obtained directly from the chiral Lagrangian and show explicitly that the isospin sum rules are preserved.
The $u$-, $t$-, and $s$-channel rescattering amplitudes contributing to each (1,1)-rank amplitude in the $SU(3)_F$ limit are derived.
It is found that the long-distance rescattering contributions to quark-loop amplitudes can be related to those of tree amplitudes without introducing additional nonperturbative parameters.
We perform global fits to available branching-fraction and decay-parameter data and predict the CP asymmetries in charmed baryon decays.
The $\Lambda^+_c\to p\eta^\prime$ mode, with a predicted CP asymmetry of order $10^{-3}$, is identified as a promising channel for CP violation searches in charmed baryon sector.
We also suggest measuring more observables in $\Xi^0_c$ decay modes to reduce the uncertainties in the CP violation predictions.
In addition, the K\"orner-Pati-Woo theorem is found to be in tension with  rescattering dynamics, casts doubt on the validity of its conventional proof.
A precise measurement of the branching fraction of the $\Lambda^+_c\to \Sigma^+K^0_S$ mode would provide a stringent test the K\"orner-Pati-Woo theorem.

\end{abstract}
\maketitle

{\tableofcontents}

\section{Introduction}

Charmed baryon decays provide an important platform for studying nonperturbative baryonic transitions.
Among them, decays of the charmed baryon anti-triplet
($B_{c\overline 3}$) into a baryon octet ($B_{8}$) and a pseudoscalar meson ($P$) are the most widely investigated.
Extensive data on the $B_{c\overline 3}\to B_{8} P$ decays have been collected by BESIII \cite{BESIII:2023wrw,BESIII:2026lqo,BESIII:2026qbp,BESIII:2024sfz,BESIII:2024cbr,BESIII:2025vvd,BESIII:2022wxj,BESIII:2023uvs,BESIII:2023vfi,BESIII:2023ooh,BESIII:2023rky,BESIII:2023iwu,
BESIII:2022udq,BESIII:2022izy,BESIII:2022vrr,BESIII:2022xne,BESIII:2022onh,
BESIII:2022aok,BESIII:2022bkj,BESIII:2021fqx,BESIII:2020kap,BESIII:2020cpu,
BESIII:2019odb,BESIII:2018qyg,
Ablikim:2018jfs,Ablikim:2018woi,Ablikim:2018bir,Ablikim:2015prg,
Ablikim:2015flg,Ablikim:2016tze,Ablikim:2016mcr,
Ablikim:2016vqd,Ablikim:2017ors,Ablikim:2017iqd,BESIII:2025zbz,Li:2025nzx},
Belle (II) \cite{Belle:2025qrq,Belle:2025skz,Belle:2024xcs,Belle:2021dgc,Belle:2022bsi,
Belle:2021btl,Pal:2017ypp,Berger:2018pli,Belle:2024ikp,Zupanc:2013iki,Yang:2015ytm}
and LHCb \cite{LHCb:2026wcy,Aaij:2017xva,Aaij:2017nsd} over recent decades.
Meanwhile, CP violation has been established in $K$, $B$, and $D$ mesons and in bottom baryon systems \cite{LHCb:2025ray,LHCb:2019hro,Belle:2001zzw,Christenson:1964fg,BaBar:2001ags}.
However, it has not yet been observed in charmed baryon decays, although many experimental efforts have been performed \cite{BESIII:2018ciw,LHCb:2017hwf,Belle:2021crz,Belle:2022uod,Belle-II:2025xvc}.
Establishing CP violation in the charmed baryon sector is one of the primary goals in particle physics at the present stage.

From a theoretical perspective, QCD-inspired approaches do not work well for charmed hadron decays due to the large expansion parameters $\alpha_s(m_c)$ and $\Lambda_{\rm QCD}/m_c$.
Phenomenological approaches, such as final-state interactions (FSIs), are therefore valuable for analyzing charmed baryon decays.
The rescattering mechanism played a key role in the theoretical analysis of doubly charmed baryon decays \cite{Yu:2017zst}, contributing to the discovery of the doubly charmed baryon $\Xi_{cc}^{++}$ \cite{LHCb:2017iph}.
Final-state interactions are also important for understanding CP violation in the charm sector.
In Ref.~\cite{Cheng:2012xb}, the assumption $P_L\simeq E_L$ was proposed to estimate the difference in CP asymmetries $\Delta A_{CP}=A_{CP}(D^0\to K^+K^-)-A_{CP}(D^0\to \pi^+\pi^-)$, where $P_L$ and $E_L$ denote the long-distance contributions to the quark-loop diagram and the $W$-exchange diagram, respectively.
The theoretical prediction is consistent with the LHCb measurement \cite{LHCb:2019hro}.
However, the dynamical basis for the relation $P_L\simeq E_L$ was not elaborated in Ref.~\cite{Cheng:2012xb}.
Our previous work \cite{Wang:2021rhd} found that the rescattering triangle diagrams contributing to the quark-loop diagram $P$ are identical to those contributing to the $W$-exchange diagram $E$ in the $SU(3)_F$ limit, thereby providing a dynamical basis for the hypothesis $P_L\simeq E_L$.
Recently, the rescattering mechanism has been applied to estimate CP asymmetries in charmed baryon decays \cite{Yang:2025orn,He:2024unv,Jia:2024pyb,Cheng:2025oyr}.

In the rescattering mechanism, a charmed baryon first decays into a baryon and a meson via a short-distance emission diagram $T$.
The $t$-, $s$-, and $u$-channel meson-baryon scattering provides the long-distance contributions.
There are two approaches for obtaining the FSI amplitudes contributing to a decay channel: deriving hadron-level Feynman diagrams directly using the chiral Lagrangian \cite{Li:2020qrh,Jia:2024pyb,Jiang:2018oak,Hu:2024uia,Han:2021gkl,Han:2021azw,Yu:2017zst}, or extracting them from topological diagrams \cite{Wang:2022wrb,Wang:2021rhd,Cheng:2004ru,Ablikim:2002ep}.
In the second method, the hadron-level rescattering amplitudes are obtained by inserting strong vertices and hadron propagators into topological amplitudes written in tensor form.
The coefficients of the rescattering amplitudes can be derived from topological amplitudes \cite{Wang:2022wrb,Wang:2021rhd}.
The theoretical framework for topological amplitudes in $B_{c\overline{3}}\to B_8 P$ decays was established in Ref.~\cite{Wang:2024ztg}.
However, it is not straightforward to extend the derivation of rescattering amplitudes from topological amplitudes to the $B_{c\overline{3}}\to B_8 P$ decays.
The chiral Lagrangian for meson-baryon coupling is constructed using (1,1)-rank octet tensors, while the topological amplitudes for $B_{c\overline{3}}\to B_8 P$ decays are constructed using third-rank octet tensors in which each flavor index represents a quark line.
To connect the chiral Lagrangian at the hadron level with topological diagrams at the quark level, a bridge between the two descriptions is required.
In Ref.~\cite{Wang:2024ztg}, we introduced the "(1,1)-rank amplitudes" to derive linear relations between topological amplitudes and the $SU(3)$ irreducible amplitudes.
These (1,1)-rank amplitudes also provide a convenient basis for studying the connection between topological amplitudes and rescattering dynamics.

In this work, we generalize the theoretical framework proposed in \cite{Wang:2021rhd,Wang:2022wrb} to charmed baryon decays.
The possible meson-meson or meson-baryon coupling configurations are constructed through tensor contractions.
The rescattering amplitudes derived from topological amplitudes are consistent with those obtained directly from the chiral Lagrangian.
The preservation of the complete set of isospin sum rules provides a nontrivial consistency check of the framework.
The K\"orner-Pati-Woo (KPW) theorem \cite{Korner:1970xq,Pati:1970fg}, which states that the two quarks produced at the weak vertex are antisymmetric in flavor if both enter the low-lying baryon, is found to be in tension with rescattering dynamics.
Furthermore, a loophole in the conventional proof of the K\"orner-Pati-Woo theorem is identified: it assumes that the color assignments of the quarks are preserved from the weak vertex until their confinement into the final baryon, whereas gluon exchanges can change these colors.
To test the K\"orner-Pati-Woo theorem, we suggest a more precise measurement of the branching fraction of the $\Lambda^+_c\to \Sigma^+K^0_S$ mode at Belle II.

For phenomenological applications, we perform global fits to the latest branching fraction and decay parameter data under exact $SU(3)_F$ symmetry  and with a subset of $SU(3)_F$ breaking effects.
Many branching fractions that were not well reproduced in previous analyses are accommodated by our fits, supporting our conclusion that the validity of the K\"orner-Pati-Woo theorem warrants re-examination.
To analyze CP asymmetries in charmed baryon decays, we derive the $u$-, $t$-, and $s$-channel long-distance contributions for each (1,1)-rank amplitude in the $SU(3)_F$ limit.
An important finding is that the rescattering amplitudes contributing to quark-loop diagrams can be expressed in terms of those contributing to tree diagrams without introducing additional nonperturbative parameters, thereby enabling predictions for CP asymmetries.
In this work, we predict the CP asymmetries of charmed baryon decays using the theoretical parameters extracted from global fits.
We suggest searching for CP violation in the $\Lambda^+_c\to p\eta^\prime$ decay as the CP asymmetry can reach the $10^{-3}$ level while retaining a comparatively small uncertainty.
We also encourage measurements of additional observables in $\Xi^0_c$ decays to reduce the theoretical uncertainties of CP-asymmetry predictions and to test the K\"orner-Pati-Woo theorem.

This paper is organized as follows.
In Sec.~\ref{ttor}, we construct a theoretical framework relating topological amplitudes and rescattering amplitudes, and derive the $u$-, $t$-, and $s$-channel long-distance contributions for each amplitude.
In Sec.~\ref{IK}, we test the isospin symmetry in the $\Xi_c\to \Xi\pi$ system in terms of the rescattering amplitudes and discuss the reliability of the K\"orner-Pati-Woo theorem.
In Sec.~\ref{pa}, we present the global fits to branching fractions and decay parameters and predictions for CP asymmetries.
Sec.~\ref{sum} is a brief summary.
The decay amplitudes for charmed baryon decays are presented in Appendix~\ref{amplitudes}.
The possible rescattering structures are presented in Appendix~\ref{scatter}.
The rescattering amplitudes for other isospin systems and the corresponding isospin-symmetry tests are presented in Appendix~\ref{exam}.
The fitting results for all two-body nonleptonic decays of charmed baryons are shown in Appendix~\ref{fit}.

\section{From topologies to rescattering}\label{ttor}

In this section, we construct a theoretical framework relating topological amplitudes and rescattering amplitudes, and derive the $u$-, $t$-, and $s$-channel long-distance contributions for each amplitude.

\subsection{Constructing rescattering amplitudes from topological diagrams}

The effective Hamiltonian for charm quark weak decays in the SM is \cite{Buchalla:1995vs}
 \begin{equation}\label{hsm}
 \mathcal H_{\rm eff}=\frac{G_F}{\sqrt 2}
 \left[\sum_{q_{1,2}=d,s}V_{cq_1}^*V_{uq_2}\left(\sum_{q=1}^2C_i
 (\mu)\mathcal{O}_i(\mu)\right)
 -V_{cb}^*V_{ub}\left(\sum_{i=3}^6C_i(\mu)\mathcal{O}_i(\mu)
 +C_{8g}(\mu)\mathcal{O}_{8g}(\mu)\right)\right].
 \end{equation}
The magnetic-penguin contributions can be absorbed into the Wilson coefficients for the penguin operators
\cite{Beneke:2003zv,Beneke:2000ry,Beneke:1999br}.
In the $SU(3)$ picture, the weak Hamiltonian for charm decay can be written as
 \begin{equation}\label{h}
 \mathcal H_{\rm eff}= \sum_{i,j,k=1}^3 H_{ij}^{k}\mathcal{O}_{ij}^{k},\qquad {\rm where}\qquad
\mathcal{O}_{ij}^{k} = \frac{G_F}{\sqrt{2}} \sum_{\rm color} \sum_{\rm current}C\,(\overline q_iq_k)(\overline q_jc).
\end{equation}
Analogously to the effective Hamiltonian, the initial and final states, including the pseudoscalar meson $P$, the octet baryon $B_8$, and the charmed anti-triplet baryon $B_{c\overline 3}$, can be written as
\begin{align}
  |P^\alpha\rangle = (P^\alpha)^{i}_{j}|P^{i}_{j} \rangle,\qquad |B^\alpha_8 \rangle = (B^\alpha_8)^{ijk}|B^{ijk}_8\rangle,\qquad  |B^\alpha_{c\overline 3}\rangle = |(B^\gamma_{c\overline 3})_{rs}|[B_{c\overline 3}]_{rs}\rangle,
\end{align}
where $|P^{i}_{j} \rangle$, $|B^{ijk}_8\rangle$ and $|[B_{c\overline 3}]_{rs}\rangle$ are the quark compositions of the states, and $(P^\alpha)$, $(B^\alpha_8)^{ijk}$ and $(B^\gamma_{c\overline 3})_{rs}$ are the coefficient matrices.
The decay amplitude for the $B^\gamma_{c\overline 3}\to B^\alpha_8 P^\beta$ mode is constructed as
\begin{align}\label{amp}
\mathcal{A}(B^\gamma_{c\overline 3}\to B_{8}^\alpha P^\beta)& = \langle B^\alpha_8 P^\beta |\mathcal{H}_{\rm eff}|B^\gamma_{c\overline 3}\rangle\nonumber\\&= \sum_{\rm Per.}\,(B^\alpha_8)^{ijk}\langle B^{ijk}_8|(P^\beta)^l_m\langle P^l_m||H_{np}^q\mathcal{O}_{np}^q||(B^\gamma_{c\overline 3})_{rs}|[B_{c\overline 3}]_{rs}\rangle\nonumber\\& = \sum_{\rm Per.}\,\langle B^{ijk}_8 P^l_m |\mathcal{O}_{np}^q|[B_{c\overline 3}]_{rs}\rangle \times (B^\alpha_8)^{ijk}(P^\beta)^l_m H_{np}^q(B^\gamma_{c\overline 3})_{rs}\nonumber\\& = \sum_\omega X_{\omega}(C_\omega)_{\alpha\beta\gamma}.
\end{align}
In the above formula, $\sum_{\rm Per.}$ denotes the sum over all possible complete contractions, $\omega$ is the reduced matrix element, and $(C_\omega)_{\alpha\beta\gamma}$ is the Clebsch-Gordan coefficient.
According to the Wigner-Eckhart theorem \cite{Eckart30,Wigner59}, $X_\omega$ is independent of the indices $\alpha$, $\beta$ and $\gamma$.
All information about the initial and final states is absorbed into the Clebsch-Gordan coefficient $(C_\omega)_{\alpha\beta\gamma}$.
There are two distinct light baryon octets, $B_8^S$ and $B_8^A$. The flavor wavefunctions of $B_8^S$ and $B_8^A$ are symmetric and antisymmetric under $q_1 \leftrightarrow q_2$, respectively.
Because of their different spin wavefunctions, the topological diagrams of charmed baryon decays into octet baryons fall into two different sets.
The total amplitude for the $B_{c\overline 3}\to B_8 P$ decay is the sum of the amplitudes for the $B_{c\overline 3}\to B_8^S P$ and $B_{c\overline 3}\to B_8^A P$ transitions,
\begin{align}
  \mathcal{A}(B_{c\overline 3}\to B_8 P) = \mathcal{A}^S(B_{c\overline 3}\to B_8^S P) + \mathcal{A}^A(B_{c\overline 3}\to B_8^A P).
\end{align}
The framework of topological amplitudes for $B_{c\overline 3}\to B_8 P$ decays was established in Ref.~\cite{Wang:2024ztg}, in which a one-to-one mapping between the topological diagram and the invariant tensor was built, and the completeness of topologies and the decomposition of $SU(3)$ irreducible amplitudes in terms of topologies were clarified.

The leading order chiral Lagrangians of the $VPP$, $BBP$, and $BBV$ interactions are given by \cite{Scherer:2002tk,Meissner:1987ge,Bernard:1995dp}
\begin{align}\label{x6}
\mathcal{L}_{\mathcal{V}\mathcal{P}\mathcal{P}} &= \frac{ig_{VPP}}{\sqrt{2}}
\,Tr\,(\mathcal{V}^{\mu}\,[\mathcal{P},\partial_\mu \mathcal{P}]),\nonumber\\
  \mathcal{L}_{\mathcal{B}\mathcal{B}\mathcal{P}}&= \sqrt{2}D \,Tr(\overline {\mathcal{B}}\,i\gamma_5\{ \mathcal{P},\mathcal{B}\}) + \sqrt{2}F \,Tr(\overline {\mathcal{B}}\,i\gamma_5[ \mathcal{P},\mathcal{B}]),\nonumber\\
  \mathcal{L}_{\mathcal{B}\mathcal{B}\mathcal{V}}&=  \sqrt{2}D_V Tr(\overline {\mathcal{B}}\gamma^\mu\{\mathcal{V}_\mu,\mathcal{B}\})+ \sqrt{2}F_V Tr(\overline {\mathcal{B}}\gamma^\mu[\mathcal{V}_\mu,\mathcal{B}])-\sqrt{2}(D_V-F_V) Tr(\overline {\mathcal{B}}\gamma^\mu\mathcal{B})Tr(\mathcal{V}_\mu).
\end{align}
The charmed baryon anti-triplet is
\begin{eqnarray}
 \mathcal{B}_{c\overline 3}=\left( \begin{array}{ccc}
     \Xi_c^0 \\
    -\Xi_c^+  \\
    \Lambda_c^+ \\
  \end{array}\right).
\end{eqnarray}
The light pseudoscalar meson nonet is
\begin{eqnarray}
 \mathcal{P}=  \left( \begin{array}{ccc}
   \frac{1}{\sqrt 2} \pi^0+  \frac{1}{\sqrt 6} \eta_8     & \pi^+  & K^+ \\
    \pi^- &   - \frac{1}{\sqrt 2} \pi^0+ \frac{1}{\sqrt 6} \eta_8  & K^0 \\
    K^- & \overline K^0 & -\sqrt{2/3}\eta_8 \\
  \end{array}\right).
\end{eqnarray}
The vector meson nonet is
\begin{eqnarray}\label{v}
 \mathcal{V}=  \left( \begin{array}{ccc}
   \frac{1}{\sqrt 2} \rho^0+  \frac{1}{\sqrt 2} \omega    & \rho^+  & K^{*+} \\
    \rho^- &   - \frac{1}{\sqrt 2} \rho^0+ \frac{1}{\sqrt 2} \omega   & K^{*0} \\
    K^{*-} & \overline K^{*0} & \phi \\
  \end{array}\right).
\end{eqnarray}
The baryon octet is given by
\begin{eqnarray}\label{B8}
 \mathcal{B}_8=  \left( \begin{array}{ccc}
   \frac{1}{\sqrt 2} \Sigma^0+  \frac{1}{\sqrt 6} \Lambda^0    & \Sigma^+  & p \\
    \Sigma^- &   - \frac{1}{\sqrt 2} \Sigma^0+ \frac{1}{\sqrt 6} \Lambda^0   & n \\
    \Xi^- & \Xi^0 & -\sqrt{2/3}\Lambda^0 \\
  \end{array}\right).
\end{eqnarray}

Note that $\mathcal{L}_{\mathcal{B}\mathcal{B}\mathcal{P}}$ and $\mathcal{L}_{\mathcal{B}\mathcal{B}\mathcal{V}}$ are constructed from (1,1)-rank octet tensors.
It is not straightforward to correlate Eq.~\eqref{x6} with topological amplitudes through Eq.~\eqref{amp}.
To study the relation between them, we must establish a bridge between third-rank and (1,1)-rank octet tensors.
Refs.~\cite{Wang:2025bdl,Wang:2024ztg} show that third-rank tensors $(B^S_8)^{ijk}$ and $(B^A_8)^{ijk}$ can be expressed in terms of (1,1)-rank tensors as
\begin{eqnarray}\label{sy1}
(B^S_8)^{ijk} = \frac{1}{\sqrt{6}}\left[\epsilon^{kil}(B_8)^j_l+\epsilon^{kjl}(B_8)^i_l\right],\qquad (B^A_8)^{ijk}= \frac{1}{\sqrt{2}}\epsilon^{ijl}(B_8)^k_l.
\end{eqnarray}
The (1,1)-rank amplitudes are constructed as follows:
\begin{align}\label{amp3}
 \mathcal{A}(B_{c\overline 3}\to B_8P) &=  A_{1}B_{c\overline 3}^{i} H^j_{kl}P^l_iB^k_j
 +A_{2}B_{c\overline 3}^{i} H^j_{lk}P^l_jB^k_i+ A_{3}B_{c\overline 3}^{i} H^j_{lk}P^l_iB^k_j+A_{4}B_{c\overline 3}^{i} H^j_{kl}P^l_jB^k_i\nonumber\\
&+ A_5B_{c\overline 3}^{i} H^j_{ik}P^l_jB^k_l+ A_6B_{c\overline 3}^{i} H^j_{il}P^l_kB^k_j+A_7B_{c\overline 3}^{i} H^j_{ki}P^l_jB^k_l  + A_8B_{c\overline 3}^{i} H^j_{li}P^l_kB^k_j\nonumber\\
&+ A_9B_{c\overline 3}^{i} H^j_{ik}P^l_lB^k_j
+A_{10}B_{c\overline 3}^{i} H^j_{ki}P^l_lB^k_j
+A_{11}B_{c\overline 3}^{i} H^j_{kj}P^l_iB^k_l
+ A_{12}B_{c\overline 3}^{i} H^j_{lj}P^l_kB^k_i\nonumber\\
& + A_{13}B_{c\overline 3}^{i} H^j_{kj}P^l_lB^k_i
+A_{14}B_{c\overline 3}^{i} H^j_{ij}P^l_kB^k_l+A_{15}B_{c\overline 3}^{i} H^j_{jk}P^l_iB^k_l
+ A_{16}B_{c\overline 3}^{i} H^j_{jl}P^l_kB^k_i\nonumber\\
& + A_{17}B_{c\overline 3}^{i} H^j_{jk}P^l_lB^k_i
+A_{18}B_{c\overline 3}^{i} H^j_{ji}P^l_kB^k_l.
\end{align}
The (1,1)-rank amplitudes can be decomposed in terms of topological amplitudes constructed from third-rank octet tensors.
See Ref.~\cite{Wang:2024ztg} for details.
In Eq.~\eqref{amp3}, both the tree and penguin operators can be inserted into the vertices of topological amplitudes.
Following Ref.~\cite{Wang:2024ztg}, we classify the amplitudes into tree- and penguin-induced amplitudes according to the type of four-quark operator inserted at the effective vertex.
We use the superscript $P$ to distinguish the penguin-induced amplitudes from the tree-induced amplitudes.
In the Standard Model, the tree-induced amplitudes $A_1 \sim A_{10}$ are found to receive contributions only from tree diagrams without quark-loop diagrams.
The tree-induced amplitudes $A_{11} \sim A_{14}$ include contributions from quark-loop diagrams, and tree operators in the SM cannot be inserted into $A_{15} \sim A_{18}$ terms of Eq.~\eqref{amp3}.
The decay amplitudes for charmed baryon decays presented in Ref.~\cite{Wang:2024ztg} are written in the flavor eigenstates, $\eta_8$ and $\eta_1$.
For convenience in the subsequent analysis, we rewrite the amplitude decompositions of charmed baryon decays in terms of mass eigenstates, which are listed in Tables~\ref{ta7cf}, \ref{ta8a}, and \ref{ta8b} in Appendix~\ref{amplitudes}.
The $\eta-\eta^\prime$ mixing is expressed as
\begin{align}
  \left(
    \begin{array}{c}
      \eta \\
      \eta^\prime \\
    \end{array}
  \right) =
  \left(
   \begin{array}{cc}
   \cos\phi & -\sin\phi \\
   \sin\phi & \cos\phi \\
   \end{array}
   \right)\left(
            \begin{array}{c}
              \eta_q \\
             \eta_s  \\
            \end{array}
          \right).
\end{align}
The amplitudes contributing to the modes involving $\eta$ and $\eta^\prime$ are presented explicitly in Appendix~\ref{amplitudes}.
The tree-induced amplitudes $A_{11} \sim A_{14}$ and all the penguin-induced amplitudes can be absorbed into four amplitudes in the $SU(3)_F$ limit \cite{Wang:2024ztg},
\begin{align}
 & A^{\prime}_{11}  = A_{11}+A^P_{3}+A^P_{5}+A^P_{11}+3A^P_{17},\qquad A^{\prime}_{12}  = A_{12}+A^P_{4}+A^P_{6}+A^P_{12}+3A^P_{15},\nonumber\\
& A^{\prime}_{13}  = A_{13}+A^P_{2}+A^P_{9}+A^P_{13}+3A^P_{16},\qquad A^{\prime}_{14}  = A_{14}+A^P_{7}+A^P_{8}+A^P_{14}+3A^P_{18}.
\end{align}
Eq.~\eqref{amp3} is constructed in the $SU(3)_F$ limit.
When $SU(3)_F$ breaking effects are included, the baryon singlet, which is expressed by the tensor $B^l_l$, does not vanish.
In this work, we aim to identify relations between topological amplitudes and rescattering dynamics.
When only isospin symmetry is imposed, we also consider decay amplitudes beyond the $SU(3)_F$ limit:
\begin{align}
 \mathcal{A}^{\rm add}(B_{c\overline 3}\to B_8P) &=  A_{19}B_{c\overline 3}^{i} H^j_{ki}P^k_jB^l_l
 +A_{20}B_{c\overline 3}^{i} H^j_{ik}P^k_jB^l_l+ A_{21}B_{c\overline 3}^{i} H^j_{kj}P^k_iB^l_l +A_{22}B_{c\overline 3}^{i} H^j_{ij}P^k_kB^l_l.
\end{align}

In the rescattering mechanism, a charmed baryon first decays into a baryon and a meson via a short-distance emission diagram $T$, and then non-factorizable QCD effects are modeled as $u$-, $t$-, and $s$-channel meson-baryon scattering between the two particles generated from $T$.
According to Ref.~\cite{Wang:2024ztg}, the emission diagram contributes only to the (1,1)-rank topological amplitude $A_2$ defined in Eq.~\eqref{amp3}.
If we neglect the short-distance contributions of other topological diagrams, as assumed in rescattering dynamics \cite{Li:2020qrh,Jia:2024pyb,Jiang:2018oak,Hu:2024uia,Han:2021gkl,Han:2021azw,Yu:2017zst}, the long-distance rescattering contributions describe transitions from $A_2$ to itself and to the other amplitudes.

In tensor language, the amplitudes for the $VPP$, $BBP$, and $BBV$ strong interactions can be written as
\begin{align}\label{x7}
  \mathcal{A}_{VPP}& = \alpha^+P^j_i V^i_k P^k_j +\alpha^-P^j_i V^k_jP^i_k,\nonumber\\  \mathcal{A}_{BBP} &= \beta^+B^j_i P^i_k B^k_j +\beta^-B^j_i P^k_j B^i_k,\nonumber\\  \mathcal{A}_{BBV} &= \gamma^+B^j_i V^i_k B^k_j +\gamma^-B^j_i V^k_j B^i_k+\gamma^0B^j_i V^k_k B^i_j.
\end{align}
By comparing Eq.~\eqref{x6} with Eq.~\eqref{x7}, we find that the amplitudes $\alpha^{\pm}$, $\beta^{\pm}$ and $\gamma^{\pm,0}$ match the chiral Lagrangian as
\begin{align}
  \alpha^+&=-\alpha^-=\frac{ig_{VPP}}{\sqrt{2}}V^\mu P \partial_\mu P,\nonumber\\
  \beta^+&=\sqrt{2}(D+F)\,\overline {B}\,i\gamma_5  PB,\qquad \beta^-=\sqrt{2}(D-F)\,\overline {B}\,i\gamma_5 B P,\nonumber\\
   \gamma^+&=\sqrt{2}(D_V+F_V)\, \overline {B}\gamma^\mu V_\mu B,\qquad \gamma^-=-\gamma^0=\sqrt{2}(D_V-F_V)\, \overline {B}\gamma^\mu BV_\mu.
\end{align}
For example, the tensor contraction for $p\pi^0p$ in Eq.~\eqref{x7} is $\mathcal{A}_{p\pi^0p} = \beta^+\,(p)^3_1 (\pi^0)^1_1 (p)^1_3 = \beta^+/\sqrt{2}$, which is in agreement with the chiral-Lagrangian
coupling $g_{{\pi^0}pp}=(D+F)$.

\begin{figure}[t!]
  \centering
  \includegraphics[width=10cm]{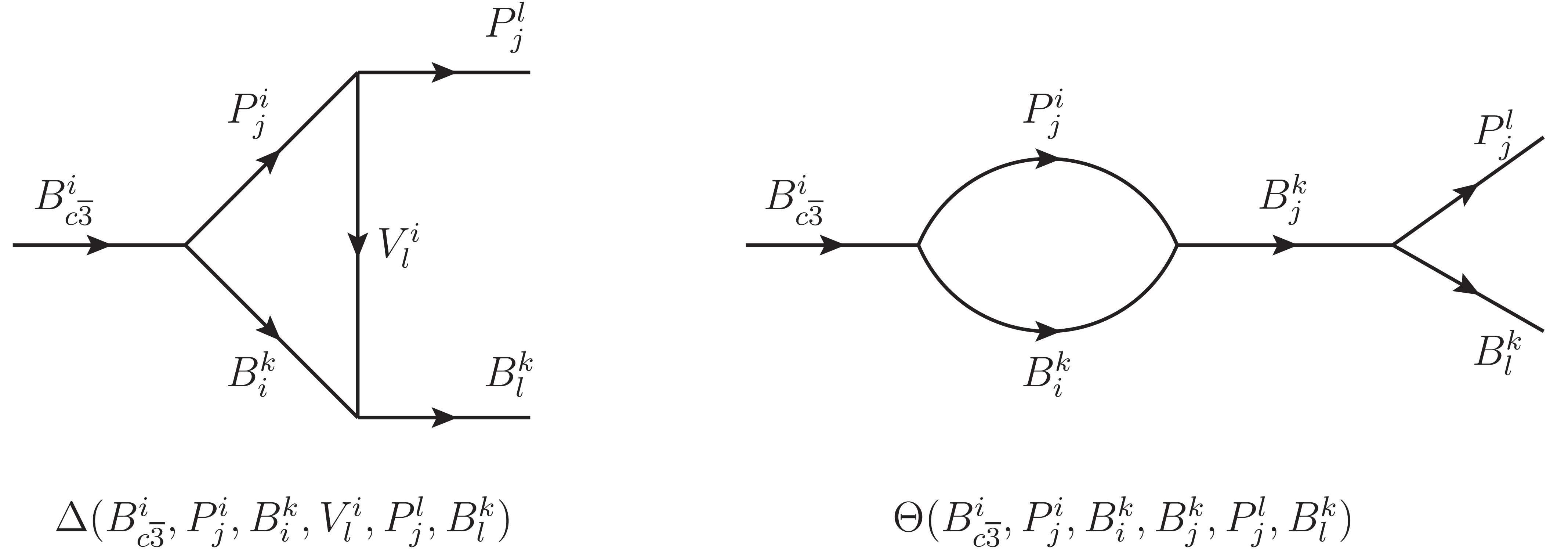}
  \caption{The $u$- and $s$-channel rescattering amplitudes contributing to the $A_5$ amplitude.}\label{f2}
\end{figure}
Following Ref.~\cite{Wang:2021rhd}, we express the transitions from $A_2$ to $A_i$ in tensor form.
Taking the $A_2\to A_5$ transition as an example, the $u$-channel rescattering amplitude is written as
\begin{align}\label{le}
U(A_5)[i,j,k,l]\,=\, B_{c\overline 3}^qH_{pm}^n\contraction[1mm]{}{P}{_n^pB_q^m\,\cdot\,}{P}P_n^p
\contraction[2mm]{}{B}{_q^m\,\cdot\,P_p^nV_l^q P_j^l\,\cdot\,}{B}B_q^m\,\cdot\,P_p^n
\contraction[1mm]{}{V}{_l^q P_j^l\,\cdot\,B_m^q}{V}V_l^q P_j^l\,\cdot\,B_m^qV_q^l B_l^k\,\cdot\,
\delta_{ip}\delta_{km}\delta_{jn}\,\cdot\,\delta_{iq}.
\end{align}
It can be understood as a triangle diagram in which $A_2$, shown on the left, serves as the weak vertex, and $PVP$ and $BVB$ represent the meson-meson and meson-baryon scattering vertices, respectively.
The index contractions $\contraction[1mm]{}{P}{^p_n}{P}P^p_nP^n_p $,  $\contraction[1mm]{}{V}{^q_l}{V}V^q_l V^l_q$ and $\contraction[1mm]{}{B}{^{m}_q}{B}B^{m}_qB_{m}^q$ represent propagators.
Kronecker delta symbols are used to set certain indices equal.
In addition to $u$-channel rescattering, the $A_2\to A_5$ transition also receives an $s$-channel rescattering contribution,
\begin{align}
S(A_5)[i,j,k,l]\,=\, B_{c\overline 3}^qH_{pm}^n\contraction[1mm]{}{P}{_n^pB_q^m\,\cdot\,}{P}P_n^p
\contraction[2mm]{}{B}{_q^m\,\cdot\,P_p^n}{B}B_q^m\,\cdot\,P_p^nB_m^q
\contraction[1mm]{}{B}{_n^m\,\cdot\,}{B}B_n^m\,\cdot\,B_k^j P_j^l B_l^k\,\cdot\,
\delta_{ip}\delta_{km}\delta_{jn}\,\cdot\,\delta_{iq}.
\end{align}
It can be understood as a bubble diagram.
To distinguish between the $u$- and $s$-channel rescattering amplitudes, we denote them by $\Delta$ and $\Theta$, respectively.
The triangle and bubble diagrams contributing to $A_5$ are shown in Fig.~\ref{f2}.
Analogously to the $A_2\to A_5$ transition, we list all possible rescattering structures involved in the transitions from $A_2$ to the $A_i$ amplitudes in  Appendix~\ref{scatter}.
These structures constitute a nonredundant set of possible meson-baryon rescattering substructures.
In what follows, we use these rescattering structures to derive the rescattering amplitudes.

\subsection{Flavor $SU(3)$ limit}

\begin{table*}[t!]
\caption{Rescattering amplitudes contributing to the $(1,1)$-rank decay amplitudes in the $SU(3)_F$ limit. }\label{tab}
\begin{tabular}{|c|c|c|c|}
\hline\hline
 & $u$-channel  & $t$-channel & $s$-channel  \\\hline
 $A_1$ &    &  \quad$-\frac{1}{3}\Delta_{(\beta^++\beta^-),(\beta^++\beta^-)}$~~~~  &    \\\hline
 $A_3$ & $\Delta_{\alpha^-,\gamma^-}$   & $\Delta_{\beta^-,\beta^-}$   &    \\\hline
 $A_4$ &  $\Delta_{\alpha^+,\gamma^+}$  &   $\Delta_{\beta^+,\beta^+}$  &    \\\hline
 $A_5$ &  $\Delta_{\alpha^+,\gamma^-}$  &    &  $\Theta_{\beta^-,\beta^-}$  \\\hline
 $A_6$ &    &   $\Delta_{\beta^-,\beta^+}$  & $\Theta_{\beta^-,\beta^+}$   \\\hline
 $A_7$ &  \quad$-\frac{1}{3}\Delta_{\alpha^+,(\gamma^++\gamma^-)}$~~~~  &  $-\frac{1}{3}\Delta_{\beta^+,(\beta^++\beta^-)}$  &  $-\frac{1}{3}\Theta_{(\beta^++\beta^-),\beta^-}$  \\\hline
 $A_8$ &  $-\frac{1}{3}\Delta_{\alpha^-,(\gamma^++\gamma^-)}$  & $-\frac{1}{3}\Delta_{\beta^-,(\beta^++\beta^-)}$   &  $-\frac{1}{3}\Theta_{(\beta^++\beta^-),\beta^+}$  \\\hline
 $A_{10}$ &    & $\frac{1}{9}\Delta_{(\beta^++\beta^-),(\beta^++\beta^-)}$   &    \\\hline
 $A_{11}$ &    & $\Delta_{\beta^+,\beta^-}$   &  $\Theta_{\beta^+,\beta^-}$  \\\hline
 $A_{12}$ & $\Delta_{\alpha^-,\gamma^+}$   &    &  $\Theta_{\beta^+,\beta^+}$  \\\hline
 $A_{14}$ &    &    &  \quad$-\frac{1}{3}\Theta_{(\beta^++\beta^-),(\beta^++\beta^-)}$~~~~ \\
\hline\hline
\end{tabular}
\end{table*}

In the $SU(3)_F$ limit, the rescattering amplitudes take a compact form and can be derived through tensor contractions.
For $u$-channel rescattering $B_{c\overline3}\to P_2B_2\to PB$ mediated by an intermediate vector meson, the tensor structure can be written as
\begin{align}
 \mathcal{R}^{u} &=\sum_{P_2,B_2,V}(B_{c\overline3})^iH^j_{lk}(P_2)^l_j(B_2)^k_i\times[\alpha^+(P_2)^p_mV^m_nP^n_p
  +\alpha^-(P_2)^m_pV^n_mP^p_n]\\&\qquad\qquad \times[\gamma^+(B_2)^c_aV^a_bB^b_c
  +\gamma^-(B_2)^a_cV^b_aB^c_b+\gamma^0(B_2)^a_bV^c_cB^b_a].
\end{align}
Similarly, the tensor structure of $s$-channel rescattering $B_{c\overline3}\to P_2B_2\to PB$ mediated by an intermediate baryon $B_3$ can be written as
\begin{align}
 \mathcal{R}^{s}&= \sum_{P_2,B_2,B_3}(B_{c\overline3})^iH^j_{lk}(P_2)^l_j(B_2)^k_i\times[\beta^+(P_2)^m_p(B_2)^n_m(B_3)^p_n
  +\beta^-(P_2)^p_m(B_2)^m_n(B_3)^n_p]\\&\qquad\qquad\times[\beta^+(B_3)^c_aP^a_bB^b_c
  +\beta^-(B_3)^a_cP^b_aB^c_b].
\end{align}
The tensor structure of $t$-channel rescattering $B_{c\overline3}\to P_2B_2\to PB$ mediated by a baryon $B_3$ can be written as
\begin{align}
  \mathcal{R}^{t}&=  \sum_{P_2,B_2,B_3}(B_{c\overline3})^iH^j_{lk}(P_2)^l_j(B_2)^k_i\times[\beta^+(P_2)^m_p(B_3)^n_mB^p_n
  +\beta^-(P_2)^p_m(B_3)^m_nB^n_p]\\&\qquad\qquad\times[\beta^+(B_2)^a_c(B_3)^b_aP^c_b
  +\beta^-(B_2)^c_a(B_3)^a_bP^b_c].
\end{align}
Note that the tensor structures $\mathcal{R}^{u}$, $\mathcal{R}^{s}$, and $\mathcal{R}^{t}$ include all the possible vertex structures listed in Eq.~\eqref{x7}.
With the completeness relations
\begin{align}
  \sum_{P_2}(P_2)^i_j(P_2)^k_l & = \delta^i_l\delta^k_j-\frac{1}{3}\delta^i_j\delta^k_l,\qquad \sum_{B_{2,3}}(B_{2,3})^i_j(B_{2,3})^k_l = \delta^i_l\delta^k_j-\frac{1}{3}\delta^i_j\delta^k_l,\qquad
 \sum_{V}V^i_jV^k_l = \delta^i_l\delta^k_j,
\end{align}
we obtain the rescattering amplitudes contributing to the $(1,1)$-rank amplitudes $A_1\sim A_{14}$ in the $SU(3)_F$ limit.
The results are presented in Table~\ref{tab}.

\section{Isospin symmetry and the K\"orner-Pati-Woo theorem}\label{IK}

In this section, we test isospin symmetry in the $\Xi_c\to \Xi\pi$ system in terms of the rescattering amplitudes and discuss the reliability of the K\"orner-Pati-Woo theorem.

\subsection{Test of isospin symmetry in the $\Xi_c\to \Xi\pi$ system}

In this subsection, we take the $\Xi^0_c\to \Xi^-\pi^+$, $\Xi^0_c\to \Xi^0\pi^0$, and $\Xi^+_c\to \Xi^0\pi^+$ modes as illustrative examples to test the rescattering amplitudes derived from topological amplitudes.
The decay amplitude for the $\Xi^0_c\to \Xi^-\pi^+$ mode is $\lambda_1(A_2+A_5)$.
The long-distance contributions modeled by triangle diagrams at the hadron level include
\begin{align}\label{x4}
U(A_2)_4[u,d,s,u]\,&=\, \frac{1}{2}\Delta_{\alpha^+,\gamma^-}(\Xi^0_c,\pi^+,\Xi^-,\rho^0,\pi^+,\Xi^-)
-\frac{1}{2}\Delta_{\alpha^+,\gamma^-}(\Xi^0_c,\pi^+,\Xi^-,\omega,\pi^+,\Xi^-),
\nonumber\\
U(A_2)_5[u,d,s,u]\,&=\,\Delta_{\alpha^+,\gamma^0}(\Xi^0_c,\pi^+,\Xi^-,\omega,
\pi^+,\Xi^-),\nonumber\\
U(A_2)_6[u,d,s,u]\,&=\,
-\Delta_{\alpha^+,\gamma^0}(\Xi^0_c,\pi^+,\Xi^-,\omega,\pi^+,\Xi^-),
\nonumber\\
U(A_5)[u,d,s,u]\,&=\, \frac{1}{2}\Delta_{\alpha^+,\gamma^-}(\Xi^0_c,\pi^+,\Xi^-,\rho^0,\pi^+,\Xi^-)
+\frac{1}{2}\Delta_{\alpha^+,\gamma^-}(\Xi^0_c,\pi^+,\Xi^-,\omega,\pi^+,\Xi^-).
\end{align}
To distinguish different strong coupling constants, we label the subscripts $\alpha^{\pm}, \beta^{\pm,0}, \gamma^{\pm,0}$ by the rescattering amplitudes.
In $U(A_2)_4[u,d,s,u]$, the quark constituent of the vector meson is $d\overline d$ when it couples with pseudoscalar mesons, and $u\overline u$ when it couples with octet baryons.
Since $d\overline d = -\rho^0/\sqrt 2+\omega/\sqrt 2$ and $u\overline u =  \rho^0/\sqrt 2+  \omega/\sqrt 2$, the vector propagator is $ -\rho^0/2+\omega/2$.
In $U(A_5)[u,d,s,u]$, the quark constituent of the vector meson is $u\overline u$ and the vector propagator is $ \rho^0/2+\omega/2$.
The $VPP$ couplings in $U(A_2)_4[u,d,s,u]$ and $U(A_5)[u,d,s,u]$ are $\alpha^-$ and $\alpha^+$, respectively, and the $VBB$ coupling in both is $\gamma^-$.
Given that $\alpha^-=-\alpha^+$, the vector propagators in $U(A_2)_4[u,d,s,u]$ and $U(A_5)[u,d,s,u]$ become $ \rho^0/2-\omega/2$ and $ \rho^0/2+\omega/2$, respectively.
Only the $\omega$ meson contributes to $U(A_2)_5[u,d,s,u]$ and $U(A_2)_6[u,d,s,u]$ because their $VBB$ coupling is $\gamma^0B^j_i V^k_k B^i_j$.
Amplitude $A_5$ also receives a long-distance contribution modeled by a bubble diagram,
\begin{align}
S(A_5)[u,d,s,u]\,&=\, \Theta_{\beta^-,\beta^-}(\Xi^0_c,\pi^+,\Xi^-,\Xi^0,\pi^+,\Xi^-).
\end{align}
Summing all FSI amplitudes for the $\Xi^0_c\to \Xi^-\pi^+$ decay yields
\begin{align}\label{x1}
\mathcal{A}_L(\Xi^0_c\to \Xi^-\pi^+)&= \lambda_1\{U(A_2)_4[u,d,s,u]+U(A_2)_5[u,d,s,u]+U(A_2)_6[u,d,s,u]\\&
~~~~~~+U(A_5)[u,d,s,u]+S(A_5)[u,d,s,u]\}\nonumber\\& = \lambda_1[\Delta_{\alpha^+,\gamma^-}(\Xi^0_c,\pi^+,\Xi^-,\rho^0,\pi^+,\Xi^-)
+\Theta_{\beta^-,\beta^-}(\Xi^0_c,\pi^+,\Xi^-,\Xi^0,\pi^+,\Xi^-)].
\end{align}
It should be noted that the contributions associated with the $\omega\pi\pi$ coupling cancel each other.

The decay amplitude for the $\Xi^0_c\to \Xi^0\pi^0$ decay is $\lambda_1(A_3-A_5)/\sqrt{2}$.
The long-distance contributions include
\begin{align}
U(A_3)[u,d,s,u]\,&=\, -\Delta_{\alpha^+,\gamma^-}(\Xi^0_c,\pi^+,\Xi^-,\rho^+,\pi^0,\Xi^0),\nonumber\\
T(A_3)[u,d,s,u]\,&=\, \Delta_{\beta^-,\beta^-}(\Xi^0_c,\pi^+,\Xi^-,\Xi^-,\Xi^0,\pi^0),\nonumber\\
U(A_5)[u,d,s,d]\,&=\, \Delta_{\alpha^+,\gamma^-}(\Xi^0_c,\pi^+,\Xi^-,\rho^+,\pi^0,\Xi^0),\nonumber\\
S(A_5)[u,d,s,d]\,&=\, \Theta_{\beta^-,\beta^-}(\Xi^0_c,\pi^+,\Xi^-,\Xi^0,\pi^0,\Xi^0).
\end{align}
Summing the rescattering amplitudes contributing to $A_3$ and $A_5$, we obtain
\begin{align}\label{x2}
\mathcal{A}_L(\Xi^0_c\to \Xi^0\pi^0)=&\frac{1}{\sqrt{2}}\lambda_1\{U(A_3)[u,d,s,u]+T(A_3)[u,d,s,u]
-U(A_5)[u,d,s,d]-S(A_5)[u,d,s,d]\}\nonumber\\
=& -\sqrt{2}\lambda_1\Delta_{\alpha^+,\gamma^-}(\Xi^0_c,\pi^+,\Xi^-,\rho^+,\pi^0,\Xi^0)
+\frac{1}{\sqrt{2}}\lambda_1\Delta_{\beta^-,\beta^-}
(\Xi^0_c,\pi^+,\Xi^-,\Xi^-,\Xi^0,\pi^0)
\nonumber \\&~~~~-\frac{1}{\sqrt{2}}\lambda_1\Theta_{\beta^-,\beta^-}(\Xi^0_c,\pi^+,\Xi^-,\Xi^0,\pi^0,\Xi^0).
\end{align}
The decay amplitude for the $\Xi^+_c\to \Xi^0\pi^+$ decay is $-\lambda_1(A_2+A_3)$.
The rescattering contributions include
\begin{align}\label{x5}
U(A_2)_1[d,d,s,u]\,&=\, -\frac{1}{2}\Delta_{\alpha^+,\gamma^-}(\Xi^+_c,\pi^+,\Xi^0,\rho^0,\pi^+,\Xi^0)
+\frac{1}{2}\Delta_{\alpha^+,\gamma^-}(\Xi^+_c,\pi^+,\Xi^0,\omega,\pi^+,\Xi^0),\nonumber\\
U(A_2)_5[d,d,s,u]\,&=\,
\Delta_{\alpha^+,\gamma^0}(\Xi^+_c,\pi^+,\Xi^0,\omega,\pi^+,\Xi^0),\nonumber\\
U(A_2)_6[d,d,s,u]\,&=\,
-\Delta_{\alpha^+,\gamma^0}(\Xi^+_c,\pi^+,\Xi^0,\omega,\pi^+,\Xi^0),\nonumber\\
U(A_3)[d,d,s,u]\,&=\, -\frac{1}{2}\Delta_{\alpha^+,\gamma^-}(\Xi^+_c,\pi^+,\Xi^0,\rho^0,\pi^+,\Xi^0)
-\frac{1}{2}\Delta_{\alpha^+,\gamma^-}(\Xi^+_c,\pi^+,\Xi^0,\omega,\pi^+,\Xi^0),\nonumber\\
T(A_3)[d,d,s,u]\,&=\, \Delta_{\beta^-,\beta^-}(\Xi^+_c,\pi^+,\Xi^0,\Xi^-,\Xi^0,\pi^+).
\end{align}
Summing the rescarrering amplitudes contributing to $A_2$ and $A_3$, we obtain
\begin{align}\label{x3}
\mathcal{A}_L(\Xi^+_c\to \Xi^0\pi^+)=&-\lambda_1\{U(A_2)_1[d,d,s,u]+U(A_2)_5[d,d,s,u]+U(A_2)_6[d,d,s,u]
\nonumber\\&~~~~~~~~+U(A_3)[d,d,s,u]
+T(A_3)[d,d,s,u]\}\nonumber\\
=& \lambda_1\Delta_{\alpha^+,\gamma^-}(\Xi^+_c,\pi^+,\Xi^0,\rho^0,\pi^+,\Xi^0)
-\lambda_1\Delta_{\beta^-,\beta^-}(\Xi^+_c,\pi^+,\Xi^0,\Xi^-,\Xi^0,\pi^+).
\end{align}
Again, all contributions associated with the $\omega\pi\pi$ vertex cancel each other out.

Equations~\eqref{x1}, \eqref{x2}, and \eqref{x3} can also be derived directly from the chiral Lagrangian~\eqref{x6}.
Starting from topological amplitudes, we obtain the same results.
This agreement demonstrates the consistency between the topological-amplitude and chiral-Lagrangian descriptions.
The decay amplitudes for the $\Xi^0_c\to \Xi^-\pi^+$, $\Xi^0_c\to \Xi^0\pi^0$, and $\Xi^+_c\to \Xi^0\pi^+$ modes satisfy the isospin sum rule,
\begin{align}\label{testb}
  \mathcal{A}(\Xi^0_c\to \Xi^-\pi^+)+\sqrt{2}\mathcal{A}(\Xi^0_c\to \Xi^0\pi^0)+\mathcal{A}(\Xi^+_c\to \Xi^0\pi^+)=0.
\end{align}
In the isospin limit, particles within an isospin multiplet can be treated as identical in triangle and bubble diagrams.
The isospin sum rule~\eqref{testb} is verified explicitly in terms of the $u$-, $t$-, and $s$-channel rescattering amplitudes.
Tests of isospin sum rules in other charmed baryon decay systems are presented in Appendix~\ref{exam}.

\subsection{Test of the K\"orner-Pati-Woo theorem}

In the literature, the K\"orner-Pati-Woo theorem \cite{Pati:1970fg,Korner:1970xq} plays an important role in analyzing charmed baryon weak decays.
It states that the two quarks produced by weak operators must be antisymmetric in flavor if they enter the same low-lying baryon.
A conventional argument for the K\"orner-Pati-Woo theorem proceeds as follows.
The four-quark operator for weak decay is
\begin{eqnarray}
\mathcal{O}(x)=\{\overline{q}_{\alpha,i}(x)(1-\gamma_5)q_{\beta,j}(x)\}
\{\overline{q}_{\gamma,k}(x)(1-\gamma_5)q_{\delta,l}(x)\},
\end{eqnarray}
where $\alpha$, $\beta$, $\gamma$, and $\delta$ are color indices, and $i$, $j$, $k$, and $l$ are flavor indices.
After a Fierz transformation, it can be written as
\begin{eqnarray}
\mathcal{O}(x)=\{\overline{q}_{\gamma,k}(x)(1-\gamma_5)q_{\beta,j}(x)\}
\{\overline{q}_{\alpha,i}(x)(1-\gamma_5)q_{\delta,l}(x)\}.
\end{eqnarray}
Thus, $\mathcal{O}(x)$ is symmetric under the interchange $(\alpha, i) \leftrightarrow (\gamma, k)$.
If two quarks produced by weak operators enter a final-state baryon, only the color-antisymmetric part of $\mathcal{O}(x)$ under the interchange $\alpha \leftrightarrow \gamma$ contributes, since the low-lying baryon is an $SU(3)_C$ singlet.
As a consequence, the two quarks must be antisymmetric in flavor.

By applying the K\"orner-Pati-Woo theorem to the topological diagrams, we obtain the following relations for the $(1,1)$-rank amplitudes \cite{Wang:2024ztg},
\begin{align}
  A_1=-A_3,\qquad A_5=-A_7,\qquad A_6=-A_8,\qquad A_9=-A_{10}.
\end{align}
However, these relations are not consistent with the rescattering amplitudes listed in Table~\ref{tab}, reflecting a tension between rescattering dynamics and the K\"orner-Pati-Woo theorem.

The K\"orner-Pati-Woo theorem, together with isospin symmetry, allows us to derive additional relations beyond the isospin sum rules.
The decay amplitudes for the $\Lambda_c^+\to \Sigma^+K^0$ and $\Lambda_c^+\to \Sigma^0K^+$ decays are \cite{Wang:2024ztg}
\begin{align}
 \mathcal{A}(\Lambda_c^+\to \Sigma^+K^0) &= \lambda_d(A_1+A_{11})+\lambda_s(A_7+A_{11}),\\
  \mathcal{A}(\Lambda_c^+\to \Sigma^0K^+) &= \frac{1}{\sqrt{2}}\lambda_d(-A_3+A_{11})+\frac{1}{\sqrt{2}}\lambda_s(A_7+A_{11}).
\end{align}
If the K\"orner-Pati-Woo theorem holds, we obtain
\begin{align}\label{q1}
\sqrt{2}\mathcal{A}(\Lambda_c^+\to \Sigma^0K^+) = \mathcal{A}(\Lambda_c^+\to \Sigma^+K^0)
\end{align}
in the isospin limit.
This relation was also derived in Ref.~\cite{Geng:2019xbo}.
Equation~\eqref{q1} continues to hold when $SU(3)_F$ breaking effects are included, provided isospin is preserved.
However, it is violated if the K\"orner-Pati-Woo theorem fails.
Isospin-breaking effects are naively expected to be of order $\mathcal{O}(1\%)$.
A significant violation of Eq.~\eqref{q1} would therefore signal a breakdown of the K\"orner-Pati-Woo relation.

Equation~\eqref{q1} is not preserved by rescattering dynamics.
The rescattering contributions to the $\Lambda_c^+\to \Sigma^+K^0$ decay include
\begin{align}
 T(A_1)_1[s,d,u,d]\,&=\, -\frac{1}{2}\Delta_{\beta^-,\beta^-}(\Lambda^+_c,\pi^+,n,\Sigma^0,\Sigma^+,K^0)
 +\frac{1}{6}\Delta_{\beta^-,\beta^-}(\Lambda^+_c,\pi^+,n,\Lambda^0,\Sigma^+,K^0),
 \nonumber\\
 T(A_1)_2[s,d,u,d]\,&=\, -\frac{1}{3}\Delta_{\beta^-,\beta^+}(\Lambda^+_c,\pi^+,n,\Lambda^0,\Sigma^+,K^0),
 \nonumber\\
 T(A_1)_4[s,d,u,d]\,&=\, -\frac{1}{3}\Delta_{\beta^+,\beta^+}(\Lambda^+_c,\pi^+,n,\Lambda^0,\Sigma^+,K^0),
 \nonumber\\
 U(A_7)_1[s,s,u,d]\,&=\, -\frac{1}{3}\Delta_{\alpha^+,\gamma^-}(\Lambda^+_c,K^+,\Lambda^0,\rho^+,K^0,\Sigma^+),
 \nonumber\\
 U(A_7)_2[s,s,u,d]\,&=\, -\frac{1}{3}\Delta_{\alpha^+,\gamma^+}(\Lambda^+_c,K^+,\Lambda^0,\rho^+,K^0,\Sigma^+),
 \nonumber\\
 T(A_7)_1[s,s,u,d]\,&=\, -\frac{1}{3}\Delta_{\beta^+,\beta^+}(\Lambda^+_c,K^+,\Lambda^0,\Xi^0,\Sigma^+,K^0),
 \nonumber\\
 S(A_7)_1[s,s,u,d]\,&=\, -\frac{1}{3}\Delta_{\beta^-,\beta^-}(\Lambda^+_c,K^+,\Lambda^0,p,K^0,\Sigma^+),
 \nonumber\\
 T(A_{11})[s,d,u,d]\,&=\, \frac{1}{2}\Delta_{\beta^+,\beta^-}(\Lambda^+_c,\pi^+,n,\Sigma^0,\Sigma^+,K^0)
 +\frac{1}{6}\Delta_{\beta^+,\beta^-}(\Lambda^+_c,\pi^+,n,\Lambda^0,\Sigma^+,K^0),
 \nonumber\\
 S(A_{11})[s,d,u,d]\,&=\, \Theta_{\beta^+,\beta^-}(\Lambda^+_c,\pi^+,n,p,K^0,\Sigma^+),
 \nonumber\\
 T(A_{11})[s,s,u,d]\,&=\, \frac{2}{3}\Delta_{\beta^+,\beta^-}(\Lambda^+_c,K^+,\Lambda^0,\Xi^0,\Sigma^+,K^0),
 \nonumber\\
 S(A_{11})[s,s,u,d]\,&=\, \frac{2}{3}\Theta_{\beta^+,\beta^-}(\Lambda^+_c,K^+,\Lambda^0,p,K^0,\Sigma^+).
\end{align}
The rescattering contributions to the $\Lambda_c^+\to \Sigma^0K^+$ decay include
\begin{align}
 U(A_3)[s,d,d,u]\,&=\, -\Delta_{\alpha^+,\gamma^-}(\Lambda^+_c,\pi^+,n,K^{*0},K^+,\Sigma^0),
\nonumber\\
 T(A_3)[s,d,d,u]\,&=\, \Delta_{\beta^-,\beta^-}(\Lambda^+_c,\pi^+,n,\Sigma^-,\Sigma^0,K^+),
\nonumber\\
 U(A_7)_1[s,s,u,u]\,&=\, -\frac{1}{6}\Delta_{\alpha^+,\gamma^-}(\Lambda^+_c,K^+,\Lambda^0,\rho^0,K^+,\Sigma^0)
 -\frac{1}{6}\Delta_{\alpha^+,\gamma^-}(\Lambda^+_c,K^+,\Lambda^0,\omega,K^+,\Sigma^0),
\nonumber\\
 U(A_7)_2[s,s,u,u]\,&=\, -\frac{1}{6}\Delta_{\alpha^+,\gamma^+}(\Lambda^+_c,K^+,\Lambda^0,\rho^0,K^+,\Sigma^0)
 -\frac{1}{6}\Delta_{\alpha^+,\gamma^+}(\Lambda^+_c,K^+,\Lambda^0,\omega,K^+,\Sigma^0),
\nonumber\\
 T(A_7)_1[s,s,u,u]\,&=\, -\frac{1}{3}\Delta_{\beta^+,\beta^+}(\Lambda^+_c,K^+,\Lambda^0,\Xi^-,\Sigma^0,K^+),
\nonumber\\
 S(A_7)_1[s,s,u,u]\,&=\, -\frac{1}{3}\Theta_{\beta^-,\beta^-}(\Lambda^+_c,K^+,\Lambda^0,p,K^+,\Sigma^0),
\nonumber\\
 T(A_{11})[s,d,u,u]\,&=\, \Delta_{\beta^+,\beta^-}(\Lambda^+_c,\pi^+,n,\Sigma^-,\Sigma^0,K^+),
\nonumber\\
 S(A_{11})[s,d,u,u]\,&=\, \Theta_{\beta^+,\beta^-}(\Lambda^+_c,\pi^+,n,p,K^+,\Sigma^0),
\nonumber\\
 T(A_{11})[s,s,u,u]\,&=\, \frac{2}{3}\Delta_{\beta^+,\beta^-}(\Lambda^+_c,K^+,\Lambda^0,\Xi^-,\Sigma^0,K^+),
\nonumber\\
 S(A_{11})[s,s,u,u]\,&=\, \frac{2}{3}\Theta_{\beta^+,\beta^-}(\Lambda^+_c,K^+,\Lambda^0,p,K^+,\Sigma^0),
\nonumber\\
 U(A_{19})_1[s,s,d,d]\,&=\, \frac{1}{6}\Delta_{\alpha^+,\gamma^-}(\Lambda^+_c,K^+,\Lambda^0,\rho^0,K^+,\Sigma^0)
 -\frac{1}{6}\Delta_{\alpha^+,\gamma^-}(\Lambda^+_c,K^+,\Lambda^0,\omega,K^+,\Sigma^0),
\nonumber\\
 U(A_{19})_{2}[s,s,d,d]\,&=\, \frac{1}{6}\Delta_{\alpha^+,\gamma^+}(\Lambda^+_c,K^+,\Lambda^0,\rho^0,K^+,\Sigma^0)
 -\frac{1}{6}\Delta_{\alpha^+,\gamma^+}(\Lambda^+_c,K^+,\Lambda^0,\omega,K^+,\Sigma^0),
\nonumber\\
U(A_{19})_5[s,s,u,u]\,&=\,
 -\frac{1}{3}\Delta_{\alpha^+,\gamma^0}(\Lambda^+_c,K^+,\Lambda^0,\omega,K^+,\Sigma^0),
\nonumber\\
 U(A_{19})_{6}[s,s,u,u]\,&=\,
 \frac{1}{3}\Delta_{\alpha^+,\gamma^0}(\Lambda^+_c,K^+,\Lambda^0,\phi,K^+,\Sigma^0),
 \nonumber\\
U(A_{19})_5[s,s,d,d]\,&=\,
 -\frac{1}{3}\Delta_{\alpha^+,\gamma^0}(\Lambda^+_c,K^+,\Lambda^0,\omega,K^+,\Sigma^0),
\nonumber\\
 U(A_{19})_{6}[s,s,d,d]\,&=\,
 \frac{1}{3}\Delta_{\alpha^+,\gamma^0}(\Lambda^+_c,K^+,\Lambda^0,\phi,K^+,\Sigma^0).
\end{align}
In the above formula, we include contributions from the vanishing amplitudes $A_{15}^{u}$ and $A_{15}^{d}$, where the superscripts $u$ and $d$ denote the quark components of the $\Sigma^0$ baryon.
Summing all contributions, the rescattering amplitudes contributing to the $\Lambda_c^+\to \Sigma^+K^0$ and $\Lambda_c^+\to \Sigma^0K^+$ modes are
  \begin{align} \label{q4}
  \mathcal{A}_L(\Lambda_c^+\to \Sigma^+K^0) &=-\frac{1}{2}\lambda_d\Delta_{(\beta^++\beta^-),(2\beta^+-\beta^-)}
  (\Lambda_c^+,\pi^+,n,\Lambda^0,\Sigma^+,K^0) \nonumber\\ &~~  +\frac{1}{2}\lambda_d\Delta_{(\beta^+-\beta^-),\beta^-}(\Lambda_c^+,\pi^+,n,\Sigma^0,\Sigma^+,K^0)\nonumber\\&~~ -\frac{1}{3}\lambda_s\Delta_{\alpha^+,(\gamma^++\gamma^-)}(\Lambda_c^+,K^+,\Lambda^0, \rho^{+},K^0,\Sigma^+)\nonumber\\ &~~ -\frac{1}{3}\lambda_s\Delta_{\beta^+,(\beta^+-2\beta^-)}(\Lambda_c^+,K^+,\Lambda^0, \Xi^0,\Sigma^+,K^0)\nonumber\\ &~~+\lambda_d\Theta_{\beta^+,\beta^-}(\Lambda^+_c,\pi^+,n,p,K^0,\Sigma^+)\nonumber\\ &~~
  +\frac{1}{3}\lambda_s\Theta_{(2\beta^+-\beta^-),\beta^-}(\Lambda^+_c,K^+,\Lambda^0,p,K^0,\Sigma^+),
\end{align}
\begin{align}\label{q5}
  \mathcal{A}_L(\Lambda_c^+\to \Sigma^0K^+) &=
  \frac{1}{\sqrt{2}}\lambda_d\Delta_{\alpha^+,\gamma^-}(\Lambda_c^+,\pi^+,n,\overline K^{*0},K^+,\Sigma^0) \nonumber\\ &~~
 +\frac{1}{\sqrt{2}}\lambda_d\Delta_{(\beta^+-\beta^-),\beta^-}(\Lambda_c^+,\pi^+,n,\Sigma^-,\Sigma^0,K^+) \nonumber\\&~~ -\frac{1}{3\sqrt{2}}\lambda_s\Delta_{\alpha^+,(\gamma^++\gamma^-)}(\Lambda_c^+,K^+,\Lambda^0, \rho^{0},K^+,\Sigma^0)\nonumber\\&~~-\frac{1}{3\sqrt{2}}\lambda_s\Delta_{\beta^+,(\beta^+-2\beta^-)}(\Lambda_c^+,K^+,\Lambda^0, \Xi^-,\Sigma^0,K^+)\nonumber\\&~~+\frac{1}{\sqrt{2}}\lambda_d\Theta_{\beta^+,\beta^-}
  (\Lambda^+_c,\pi^+,n,p,K^+,\Sigma^0)\nonumber\\&~~+\frac{1}{3\sqrt{2}}\lambda_s\Theta_{(2\beta^+-\beta^-),\beta^-}
  (\Lambda^+_c,K^+,\Lambda^0,p,K^+,\Sigma^0).
  \end{align}
Note that Eq.~\eqref{q1} is violated in Eqs.~\eqref{q4} and \eqref{q5}.
The two particles generated by the emission diagram $T^{SD}$ rescatter into the final state through baryon exchange in the first term of Eq.~\eqref{q4}, whereas the first term of Eq.~\eqref{q5} involves meson exchange.
The first terms in Eqs.~\eqref{q4} and \eqref{q5} include rescattering contributions from the decay amplitudes $A_1$ and $A_3$, respectively.
Thus, Eqs.~\eqref{q4} and \eqref{q5} indicate that the relation $A_1 = -A_3$ is violated in rescattering dynamics.

The final-state pseudoscalar kaons in Eq.~\eqref{q1} can be replaced by the corresponding vector mesons $K^{*+}$ and $K^{*0}$.
The branching fraction of the $\Lambda^+_c\to \Sigma^+K^{*0}$ mode is twice the branching fraction of the $\Lambda^+_c\to \Sigma^0K^{*+}$ mode according to Eq.~\eqref{q1}.
In Ref.~\cite{Jia:2024pyb}, the branching fractions for $\Lambda^+_c\to \Sigma^+K^{*0}$ and $\Lambda^+_c\to \Sigma^0K^{*+}$ were calculated to be
\begin{align}
 \mathcal{B}(\Lambda^+_c\to \Sigma^+K^{*0}) = 2.10^{+1.37}_{-0.86}\times 10^{-3},\qquad \mathcal{B}(\Lambda^+_c\to \Sigma^0K^{*+}) = 1.60^{+0.89}_{-0.62}\times 10^{-3}.
\end{align}
These results do not satisfy the relation implied by the K\"orner-Pati-Woo theorem.

\begin{figure}[t!]
  \centering
  \includegraphics[width=10cm]{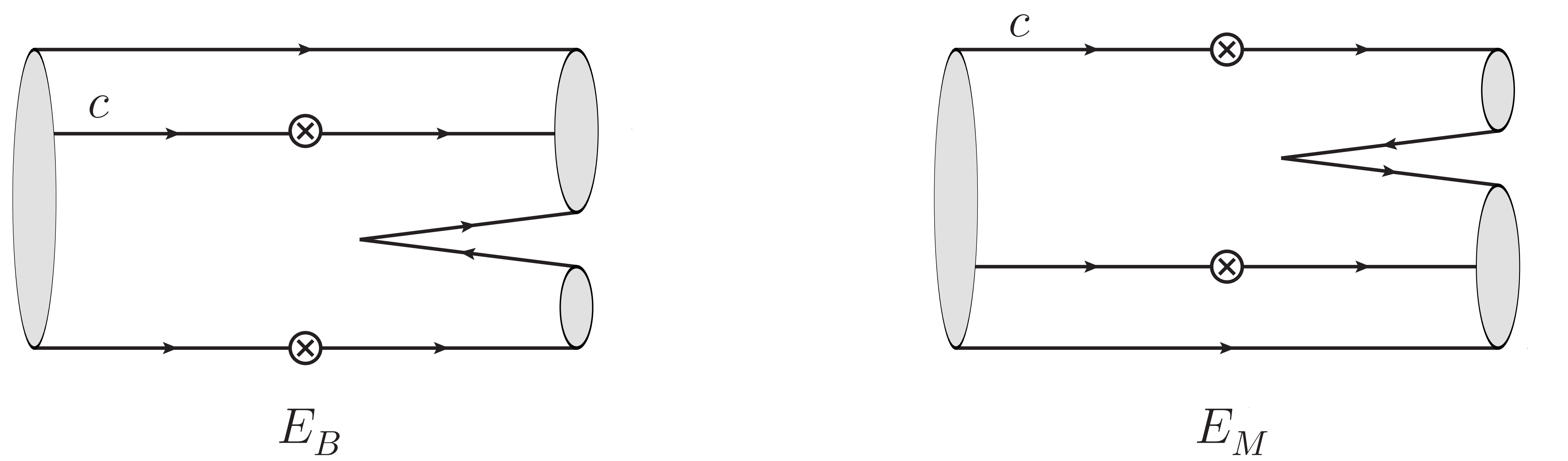}
  \caption{Two topological diagrams contributing to the $\Lambda^+_c\to \Sigma^0K^+$ and $\Lambda^+_c\to \Sigma^+K^0$ decays, respectively.}\label{f3}
\end{figure}
\begin{figure}[t!]
  \centering
  \includegraphics[width=12cm]{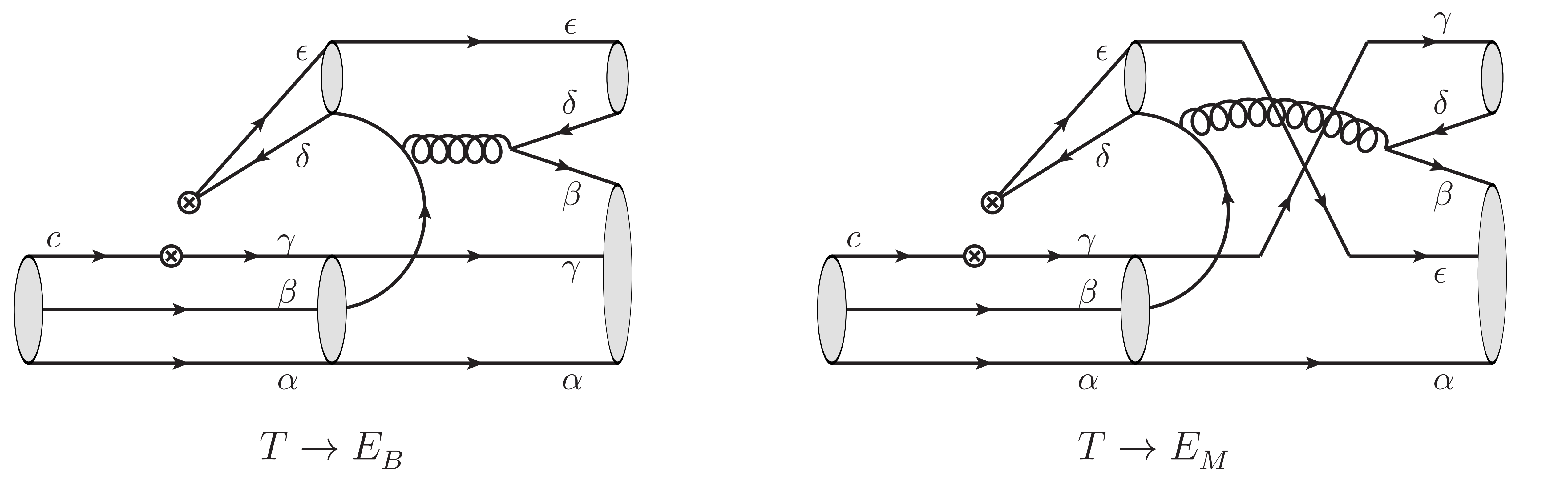}
  \caption{The long-distance contributions in the topologies $E_B$ and $E_M$, where $\alpha$, $\beta$, $\gamma$, $\delta$ and $\epsilon$ are color indices.}\label{f4}
\end{figure}
The tension between the K\"orner-Pati-Woo relation and rescattering dynamics can also be understood in terms of topological diagrams.
The two diagrams in Fig.~\ref{f3} contribute to the $\Lambda^+_c\to \Sigma^0K^+$ and $\Lambda^+_c\to \Sigma^+K^0$ decays, respectively \cite{Wang:2024ztg}.
They satisfy the relation $E_B=-E_M$ according to the K\"orner-Pati-Woo theorem.
The long-distance rescattering contributions in $E_B$ and $E_M$ are shown in Fig.~\ref{f4}.
Note that the two particles generated from the emission diagram $T^{SD}$ rescatter into final-state particles through meson exchange in $T \to E_B$ and baryon exchange in $T \to E_M$, as in Eqs.~\eqref{q4} and \eqref{q5}.
Imposing the color constraints associated with the final-state color singlets, we have $\delta=\epsilon$ in $T\to E_B$ and $\gamma=\delta=\epsilon$ in $T\to E_M$.
Thus, $E_B$ is color-favored, and $E_M$ is color-suppressed.
As a consequence, the equation $E_B=-E_M$ does not hold in this case.

Let us revisit the proof of the K\"orner-Pati-Woo theorem.
The key step assumes that the relevant color indices of $\mathcal{O}(x)$ are antisymmetric when two quarks produced by the weak operators enter the same low-lying baryon.
This argument implicitly assumes that the quark colors remain unchanged from production at the weak vertex until confinement into the final baryon.
For instance, the relation $E_B=-E_M$ is recovered if we assume that gluons cannot change the color of quarks in Fig.~\ref{f4}.
However, gluon exchange is mediated by color-octet gauge bosons.
Gluon exchange can therefore change the color carried by an individual quark.
This observation exposes a loophole in the conventional proof of the K\"orner-Pati-Woo theorem.
The proof of the K\"orner-Pati-Woo theorem \cite{Korner:1970xq,Pati:1970fg} predates the establishment of quantum chromodynamics \cite{Politzer:1973fx,Gross:1973id}.
This possibility may not have been incorporated explicitly in the original argument.
We therefore argue that the conventional K\"orner-Pati-Woo theorem warrants re-examination.

The branching fractions for the $\Lambda_c^+\to \Sigma^0K^+$ and $\Lambda_c^+\to \Sigma^+K^0_S$ modes were measured by the BESIII Collaboration as \cite{BESIII:2022wxj},
\begin{align}
\mathcal{B}(\Lambda_c^+\to \Sigma^0K^+) & = (4.7\pm 0.9\pm0.1\pm0.3)\times 10^{-4}, \nonumber\\ \mathcal{B}(\Lambda_c^+\to \Sigma^+K^0_S) & = (4.8\pm 1.4\pm0.2\pm0.3)\times 10^{-4}.
\end{align}
These measurements are compatible with the KPW relation within the current uncertainties.
However, the Belle Collaboration reported the branching fraction for the $\Lambda_c^+\to \Sigma^0K^+$ decay as \cite{Belle:2022uod}
\begin{align}
\mathcal{B}(\Lambda_c^+\to \Sigma^0K^+) & = (3.58\pm 0.19\pm0.06\pm0.19)\times 10^{-4}.
\end{align}
This value is lower than the corresponding BESIII result.
To test the K\"orner-Pati-Woo theorem, a more precise measurement of the branching fraction for the $\Lambda_c^+\to \Sigma^+K^0_S$ mode is crucial.

The K\"orner-Pati-Woo theorem can also be tested via the $\Omega^0_c\to \Sigma^+K^-$ and $\Omega^0_c\to \Sigma^0K^0_S$ modes.
If the K\"orner-Pati-Woo theorem holds, the decay amplitudes of these two modes satisfy the relation
\begin{align}
  \mathcal{A}(\Omega^0_c\to \Sigma^+K^-) = 2\mathcal{A}(\Omega^0_c\to \Sigma^0K^0_S)
\end{align}
under isospin symmetry \cite{Liu:2025hbf}.
Owing to the well-defined flavor symmetry of decuplet baryons, additional relations hold if both the KPW theorem and isospin symmetry are valid in the $B_{c\overline 3}\to B_{10}P$ decays \cite{Wang:2024nxb}:
\begin{align}\label{ax1}
\sqrt{\frac{3}{2}}\mathcal{A}(\Lambda^+_c\to \Delta^{+}\pi^0) = \sqrt{3}\mathcal{A}(\Lambda^+_c\to \Delta^{0}\pi^+)=-\mathcal{A}(\Lambda^+_c\to \Delta^{++}\pi^-),
\end{align}
\begin{align}
\mathcal{A}(\Lambda^+_c\to \Sigma^{*+} K^0) = -\sqrt{2}\mathcal{A}(\Lambda^+_c\to \Sigma^{*0}K^+),
\end{align}
\begin{align}\label{ax2}
\mathcal{A}(\Xi^+_c\to \Delta^{++} K^-) = \sqrt{3}\mathcal{A}(\Xi^+_c\to \Delta^{+} \overline K^0)=\sqrt{3}\mathcal{A}(\Xi^0_c\to \Delta^{0} \overline K^0)=\sqrt{3}\mathcal{A}(\Xi^0_c\to \Delta^{+} K^-),
\end{align}
\begin{align}
\mathcal{A}(\Xi^+_c\to \Sigma^{*0}\pi^+) = \mathcal{A}(\Xi^+_c\to \Sigma^{*+}\pi^0),
\end{align}
\begin{align}
 \sqrt{2}\mathcal{A}(\Xi^+_c\to \Sigma^{*0} K^+) = \mathcal{A}(\Xi^0_c\to \Sigma^{*-} K^+),
\end{align}
\begin{align}
\mathcal{A}(\Xi^+_c\to \Sigma^{*+} K^0)=\sqrt{2}\mathcal{A}(\Xi^0_c\to \Sigma^{*0} K^0),
\end{align}
\begin{align}
&\mathcal{A}(\Xi^+_c\to \Delta^{++}\pi^-) = \sqrt{3}\mathcal{A}(\Xi^0_c\to \Delta^{+}\pi^-),
\end{align}
\begin{align}
&\sqrt{3}\mathcal{A}(\Xi^+_c\to \Delta^{0}\pi^+) = \mathcal{A}(\Xi^0_c\to \Delta^{-}\pi^+),
\end{align}
\begin{align}
&\mathcal{A}(\Xi^+_c\to \Delta^{+}\pi^-) = \mathcal{A}(\Xi^0_c\to \Delta^{0}\pi^0).
\end{align}
These relations provide direct tests of the KPW theorem under isospin symmetry through measurements of branching fractions or decay parameters.

We can also test the K\"orner-Pati-Woo theorem under $SU(3)_F$ symmetry.
For example, the branching fractions of $\Xi^+_c\to \Sigma^{* +}\overline K^0$ and $\Xi^+_c\to \Xi^{* 0}\pi^+$ modes are zero in the $SU(3)_F$ limit if the K\"orner-Pati-Woo theorem holds.
However, these conclusions no longer hold once $SU(3)_F$-breaking effects are included.
Given the sizable $SU(3)_F$-breaking effects, testing the K\"orner-Pati-Woo theorem by imposing exact $SU(3)_F$ symmetry may not be robust.

\section{Phenomenological applications}\label{pa}

In this section, we study CP asymmetries in charmed baryon decays based on the topological amplitudes and final-state interactions.
We first extract the theoretical parameters of tree-level amplitudes from global fits to branching fractions and decay parameters in both the $SU(3)_F$ symmetric and considering  $SU(3)_F$ breaking scenarios.
Then we analyze the long-distance rescattering amplitudes contributing to quark-loop diagrams and predict the CP asymmetries in charmed baryon decays.

\subsection{Global fits to branching fractions and decay parameters}

\begin{table*}[t!]
\centering
\caption{Comparison between the experimental inputs and the theoretical results obtained from the global fit including a subset of $SU(3)_F$ breaking effects, where $\beta(\Lambda^+_c\to \Lambda^0\pi^+)$, $\gamma(\Lambda^+_c\to \Lambda^0\pi^+)$, $\beta(\Lambda^+_c\to \Lambda^0 K^+)$ and $\gamma(\Lambda^+_c\to \Lambda^0 K^+)$ are taken from \cite{Li:2025nzx}, $\mathcal{B}(\Xi^0_c\to \Lambda^0\eta)$, $\mathcal{B}(\Xi^0_c\to \Lambda^0\eta^\prime)$, and $\mathcal{B}(\Xi^0_c\to \Lambda^0\pi^0)$ are taken from \cite{Belle:2025qrq}, $\mathcal{B}(\Xi^0_c\to pK^-)$ and $\alpha(\Xi^0_c\to pK^-)$ are taken from \cite{LHCb:2026wcy}, $\beta(\Lambda^+_c\to \Sigma^0\pi^+)$, $\gamma(\Lambda^+_c\to \Sigma^0\pi^+)$, $\beta(\Lambda^+_c\to \Sigma^+\pi^0)$ and $\gamma(\Lambda^+_c\to \Sigma^+\pi^0)$ are taken from \cite{BESIII:2025zbz}, and the other data are taken from PDG \cite{PDG}.}
\label{tab:exp-theory-su3breaking}
\renewcommand{\arraystretch}{1.12}
\setlength{\tabcolsep}{3.5pt}
\resizebox{\textwidth}{!}{%
\begin{tabular}{lcccccccc}
\hline\hline
\toprule
& \multicolumn{2}{c}{$\mathcal B$}
& \multicolumn{2}{c}{$\alpha$}
& \multicolumn{2}{c}{$\beta$}
& \multicolumn{2}{c}{$\gamma$} \\\hline
\cmidrule(lr){2-3}\cmidrule(lr){4-5}\cmidrule(lr){6-7}\cmidrule(lr){8-9}
Decay mode
& Exp. & Theory
& Exp. & Theory
& Exp. & Theory
& Exp. & Theory \\\hline
\midrule

$\Lambda_c^+\to\Lambda^0\pi^+$
& $(1.32\pm0.05)\times10^{-2}$ & $(1.31\pm0.05)\times10^{-2}$
& $-0.768\pm0.015$ & $-0.779\pm0.009$
& $0.378\pm0.015$ & $0.382\pm0.014$
& $0.491\pm0.012$ & $0.497\pm0.011$ \\

$\Lambda_c^+\to\Sigma^0\pi^+$
& $(1.29\pm0.05)\times10^{-2}$ & $(1.29\pm0.04)\times10^{-2}$
& $-0.466\pm0.018$ & $-0.473\pm0.015$
& $0.704^{+0.144}_{-0.480}$ & $0.643\pm0.103$
& $-0.502^{+0.591}_{-0.304}$ & $-0.602\pm0.111$ \\

$\Lambda_c^+\to\Sigma^+\pi^0$
& $(1.27\pm0.10)\times10^{-2}$ & $(1.30\pm0.04)\times10^{-2}$
& $-0.484\pm0.027$ & $-0.472\pm0.015$
& $0.764^{+0.054}_{-0.238}$ & $0.642\pm0.103$
& $-0.262^{+0.479}_{-0.384}$ & $-0.605\pm0.110$ \\

$\Lambda_c^+\to\Sigma^+\eta$
& $(3.4\pm0.4)\times10^{-3}$ & $(3.6\pm0.4)\times10^{-3}$
& $-0.99\pm0.06$ & $-0.97\pm0.06$
& -- & -- & -- & -- \\

$\Lambda_c^+\to\Sigma^+\eta'$
& $(4.2\pm0.9)\times10^{-3}$ & $(4.1\pm0.8)\times10^{-3}$
& $-0.46\pm0.07$ & $-0.47\pm0.07$
& -- & -- & -- & -- \\

$\Lambda_c^+\to\Xi^0K^+$
& $(5.5\pm0.7)\times10^{-3}$ & $(4.6\pm0.4)\times10^{-3}$
& $0.01\pm0.16$ & $-0.19\pm0.04$
& $-0.64\pm0.70$ & $-0.96\pm0.02$
& $-0.77\pm0.59$ & $0.19\pm0.10$ \\

$\Lambda_c^+\to\Lambda^0 K^+$
& $(6.54\pm0.30)\times10^{-4}$ & $(6.61\pm0.30)\times10^{-4}$
& $-0.546\pm0.035$ & $-0.534\pm0.031$
& $0.33\pm0.08$ & $0.28\pm0.06$
& $-0.799\pm0.041$ & $-0.797\pm0.024$ \\

$\Lambda_c^+\to\Sigma^0K^+$
& $(3.76\pm0.31)\times10^{-4}$ & $(3.99\pm0.28)\times10^{-4}$
& $-0.54\pm0.20$ & $-0.49\pm0.14$
& -- & -- & -- & -- \\

$\Lambda_c^+\to\Sigma^+K_S^0$
& $(4.8\pm1.4)\times10^{-4}$ & $(2.5\pm0.5)\times10^{-4}$
& -- & -- & -- & -- & -- & -- \\

$\Lambda_c^+\to n\pi^+$
& $(6.6\pm1.3)\times10^{-4}$ & $(6.0\pm1.0)\times10^{-4}$
& -- & -- & -- & -- & -- & -- \\

$\Lambda_c^+\to p\pi^0$
& $(1.8\pm0.4)\times10^{-4}$ & $(2.0\pm0.4)\times10^{-4}$
& -- & -- & -- & -- & -- & -- \\

$\Lambda_c^+\to pK_S^0$
& $(1.63\pm0.07)\times10^{-2}$ & $(1.69\pm0.05)\times10^{-2}$
& $-0.754\pm0.010$ & $-0.753\pm0.010$
& -- & -- & -- & -- \\

$\Lambda_c^+\to p\eta$
& $(1.49\pm0.08)\times10^{-3}$ & $(1.40\pm0.08)\times10^{-3}$
& -- & -- & -- & -- & -- & -- \\

$\Lambda_c^+\to p\eta'$
& $(4.9\pm0.9)\times10^{-4}$ & $(4.3\pm0.8)\times10^{-4}$
& -- & -- & -- & -- & -- & -- \\

$\Lambda_c^+\to pK_L^0$
& $(1.67\pm0.07)\times10^{-2}$ & $(1.60\pm0.05)\times10^{-2}$
& -- & -- & -- & -- & -- & -- \\

$\Xi_c^0\to\Xi^-\pi^+$
& $(1.43\pm0.27)\times10^{-2}$ & $(1.18\pm0.20)\times10^{-2}$
& $-0.64\pm0.05$ & $-0.65\pm0.05$
& -- & -- & -- & -- \\

$\Xi_c^0\to\Xi^-K^+$
& $(3.9\pm1.1)\times10^{-4}$ & $(4.5\pm0.8)\times10^{-4}$
& -- & -- & -- & -- & -- & -- \\

$\Xi_c^0\to\Lambda^0 K_S^0$
& $(3.2\pm0.6)\times10^{-3}$ & $(3.8\pm0.6)\times10^{-3}$
& -- & -- & -- & -- & -- & -- \\

$\Xi_c^0\to\Sigma^0K_S^0$
& $(5.4\pm1.4)\times10^{-4}$ & $(6.1\pm1.4)\times10^{-4}$
& -- & -- & -- & -- & -- & -- \\

$\Xi_c^0\to\Sigma^+K^-$
& $(1.8\pm0.4)\times10^{-3}$ & $(1.2\pm0.3)\times10^{-3}$
& -- & -- & -- & -- & -- & -- \\

$\Xi_c^0\to\Xi^0\pi^0$
& $(6.9\pm1.4)\times10^{-3}$ & $(6.5\pm1.1)\times10^{-3}$
& $-0.90\pm0.27$ & $-0.61\pm0.19$
& -- & -- & -- & -- \\

$\Xi_c^0\to\Xi^0\eta$
& $(1.6\pm0.4)\times10^{-3}$ & $(1.5\pm0.4)\times10^{-3}$
& -- & -- & -- & -- & -- & -- \\

$\Xi_c^0\to\Xi^0\eta'$
& $(1.1\pm0.4)\times10^{-3}$ & $(1.4\pm0.4)\times10^{-3}$
& -- & -- & -- & -- & -- & -- \\

$\Xi_c^0\to\Lambda^0\pi^0$
& $<5.2\times10^{-4}$ & $(1.3\pm2.3)\times10^{-4}$
& -- & -- & -- & -- & -- & -- \\

$\Xi_c^0\to\Lambda^0\eta$
& $(5.95\pm1.75)\times10^{-4}$ & $(2.43\pm0.92)\times10^{-4}$
& -- & -- & -- & -- & -- & -- \\

$\Xi_c^0\to\Lambda^0\eta'$
& $(3.55\pm1.36)\times10^{-4}$ & $(2.50\pm1.12)\times10^{-4}$
& -- & -- & -- & -- & -- & -- \\

$\Xi_c^0\to pK^-$
& $(4.5\pm1.1)\times10^{-5}$ & $(5.5\pm1.0)\times10^{-5}$
& $0.32\pm0.15$ & $0.30\pm0.15$
& -- & -- & -- & -- \\

$\Xi_c^+\to\Sigma^+K_S^0$
& $(1.94\pm0.90)\times10^{-3}$ & $(3.33\pm0.80)\times10^{-3}$
& -- & -- & -- & -- & -- & -- \\

$\Xi_c^+\to\Xi^0\pi^+$
& $(7.19\pm3.23)\times10^{-3}$ & $(7.14\pm1.81)\times10^{-3}$
& -- & -- & -- & -- & -- & -- \\

$\Xi_c^+\to\Xi^0K^+$
& $(4.90\pm2.30)\times10^{-4}$ & $(7.49\pm1.50)\times10^{-4}$
& -- & -- & -- & -- & -- & -- \\

$\Xi_c^+\to pK_S^0$
& $(7.16\pm3.25)\times10^{-4}$ & $(9.48\pm2.96)\times10^{-4}$
& -- & -- & -- & -- & -- & -- \\

$\Xi_c^+\to\Lambda^0\pi^+$
& $(4.52\pm2.09)\times10^{-4}$ & $(3.05\pm0.80)\times10^{-4}$
& -- & -- & -- & -- & -- & -- \\

$\Xi_c^+\to\Sigma^0\pi^+$
& $(1.20\pm0.55)\times10^{-3}$ & $(1.96\pm0.42)\times10^{-3}$
& -- & -- & -- & -- & -- & -- \\\hline
\bottomrule
\end{tabular}%
}
\end{table*}

\begin{table*}[t!]
\centering
\caption{Best-fit parameters and propagated $1\sigma$ uncertainties, where strong phases are given in radians.}
\label{tab:fit-parameters}
\small
\renewcommand{\arraystretch}{1.08}
\setlength{\tabcolsep}{6pt}
\begin{tabular}{ccc|ccc}
\hline\hline
Parameter & $SU(3)_F$-symmetric & $SU(3)_F$-breaking
& Parameter & $SU(3)_F$-symmetric & $SU(3)_F$-breaking \\
\hline

$T_S$
& $4.264\pm2.189$
& $3.773\pm2.219$
&
$\phi_{S1}$
& $-1.254\pm0.649$
& $-1.546\pm1.327$
\\

$T_P$
& $14.878\pm7.252$
& $18.036\pm6.144$
&
$\phi_{S_3}$
& $2.647\pm0.340$
& $2.711\pm0.293$
\\

$\phi_{T_P}$
& $2.813\pm0.217$
& $2.688\pm0.672$
&
$\phi_{S_4}$
& $5.448\pm0.747$
& $5.210\pm1.573$
\\

$X_{S_1}$
& $3.992\pm1.383$
& $3.838\pm2.156$
&
$\phi_{S_5}$
& $4.633\pm1.046$
& $4.737\pm1.043$
\\

$X_{S_3}$
& $2.827\pm2.089$
& $2.318\pm2.715$
&
$\phi_{S_6}$
& $1.584\pm2.768$
& $1.377\pm5.545$
\\

$X_{S_4}$
& $-3.868\pm2.155$
& $-3.513\pm2.067$
&
$\phi_{S_7}$
& $4.355\pm0.486$
& $4.168\pm0.734$
\\

$X_{S_5}$
& $2.588\pm3.121$
& $2.632\pm3.035$
&
$\phi_{S_8}$
& $1.813\pm0.560$
& $1.658\pm0.867$
\\

$X_{S_6}$
& $1.915\pm3.237$
& $1.561\pm4.717$
&
$\phi_{P_1}$
& $5.991\pm0.362$
& $5.928\pm0.580$
\\

$X_{S_7}$
& $6.346\pm0.592$
& $6.700\pm0.563$
&
$\phi_{P_3}$
& $5.447\pm0.467$
& $5.406\pm0.889$
\\

$X_{S_8}$
& $-4.379\pm1.053$
& $-3.874\pm1.617$
&
$\phi_{P_4}$
& $2.953\pm0.353$
& $2.787\pm0.526$
\\

$X_{P_1}$
& $-10.434\pm8.241$
& $-11.943\pm7.100$
&
$\phi_{P_5}$
& $-2.354\pm0.818$
& $-2.522\pm0.936$
\\

$X_{P_3}$
& $9.549\pm6.174$
& $12.263\pm5.768$
&
$\phi_{P_6}$
& $2.836\pm1.364$
& $2.744\pm1.195$
\\

$X_{P_4}$
& $-15.270\pm6.088$
& $-18.455\pm5.353$
&
$\phi_{P_7}$
& $0.370\pm0.336$
& $0.576\pm0.528$
\\

$X_{P_5}$
& $-16.762\pm5.190$
& $-17.389\pm8.991$
&
$\phi_{P_8}$
& $-0.736\pm0.217$
& $-0.811\pm0.586$
\\

$X_{P_6}$
& $-11.359\pm11.515$
& $-12.703\pm17.337$
&
$b$
& --
& $-0.231\pm0.072$
\\

$X_{P_7}$
& $10.491\pm6.536$
& $9.054\pm6.376$
&
$b^\prime$
& --
& $-0.060\pm0.034$
\\

$X_{P_8}$
& $-20.984\pm6.458$
& $-24.124\pm5.576$
&
& & \\
\hline

$\chi^2_{\min}$
& $59.43$
& $34.76$
&
$\chi^2_{\min}/{\rm d.o.f.}$
& $59.43/23=2.58$
& $34.76/21=1.66$
\\
\hline
\end{tabular}
\end{table*}

According to Ref.~\cite{Wang:2024ztg}, there are ten tree-level $(1,1)$-rank amplitudes obtained by rearranging topological diagrams contributing to the charmed baryon decays.
The tree-induced amplitudes $A_{11} \sim A_{14}$ include contributions from quark-loop diagrams.
In the literature, quark-loop diagrams are usually classified as penguin diagrams.
$A_{11}\sim A_{14}$ are proportional to $\lambda_b=V_{cb}^*V_{ub}$ since they always appear simultaneously and  $V_{cd}^*V_{ud}+V_{cs}^*V_{us}=-V_{cb}^*V_{ub}$ in the SM.
Moreover, all penguin-induced amplitudes are proportional to $-V_{cb}^*V_{ub}$.
Due to the small $\lambda_b$, the tree-induced amplitudes $A_{11} \sim A_{14}$ and all penguin-induced amplitudes are negligible in the branching fractions and decay parameters.
Thus, only the ten tree-induced amplitudes, $A_1\sim A_{10}$, contribute to the branching fractions and decay parameters in the $SU(3)_F$ limit.
For each amplitude $A_i$, there are two partial-wave amplitudes, $\mathcal{S}_i$ and $\mathcal{P}_i$.
The decay width is expressed as
\begin{align}
  \Gamma&=\frac{p_c}{8\pi}
  \frac{(m_{B_c}+m_B)^2-m_M^2}{m_{B_c}^2}(|\mathcal{S}|^2+\kappa^2|\mathcal{P}|^2),
\end{align}
where $\mathcal{S}=\sum_i \mathcal{S}_i$, $\mathcal{P}=\sum_i \mathcal{P}_i$, and $\kappa=p_c/(E_B+m_B)=\sqrt{(E_B-m_B)/(E_B+m_B)}$.
The decay parameters $\alpha$, $\beta$, and $\gamma$ are expressed as
\begin{align}
  \alpha=\frac{2\kappa\,\mathcal{R}e(\mathcal{S}^*\mathcal{P})}{|\mathcal{S}|^2+\kappa^2|\mathcal{P}|^2},\qquad
  \beta&=\frac{2\kappa\,\mathcal{I}m(\mathcal{S}^*\mathcal{P})}{|\mathcal{S}|^2+\kappa^2|\mathcal{P}|^2},\qquad
 \gamma=\frac{|\mathcal{S}|^2-\kappa^2|\mathcal{P}|^2}{|\mathcal{S}|^2+\kappa^2|\mathcal{P}|^2}.
\end{align}
In this work, the $\eta-\eta^\prime$ mixing angle is taken to be $\phi=40.4^{\circ}$ since recent BESIII measurements give that $\phi=(40.0\pm2.0\pm0.6)^{\circ}$ \cite{BESIII:2023ajr} and $\phi=(40.2\pm2.1\pm0.7)^{\circ}$ \cite{BESIII:2023gbn}, and the value $\phi=40.4^{\circ}$ is commonly used in the literature.

Among the amplitudes $A_1\sim A_{10}$, only $A_2$ receives a contribution from the short-distance color-favored emission diagram, $T$.
In this work, we choose the $\mathcal{S}$-wave amplitude $\mathcal{S}_2$ as a real parameter $T_S$.
The $\mathcal{P}$-wave amplitude $\mathcal{P}_2$ is parameterized as $T_Pe^{i\phi_{P_T}}$.
For amplitudes $A_1$, $A_3$, $A_4$, $A_5$, $A_6$, $A_7$, and $A_8$, we parameterize the $\mathcal{S}$- and $\mathcal{P}$-wave parts as $X_{S_i}e^{i\phi_{S_i}}$ and $X_{P_i}e^{i\phi_{P_i}}$, respectively.
The amplitude $A_{9}$ is neglected in this work because its long-distance rescattering contribution is suppressed under the $SU(3)_F$ symmetry.
For amplitude $A_{10}$, we set $-A_1/3$ because its long-distance rescattering contribution is not independent, as shown in Table~\ref{tab}.
Altogether, there are 31 theoretical parameters that determine the branching fractions and decay parameters of charmed baryon decays in the $SU(3)_F$ limit.
We also consider $SU(3)_F$ breaking effects.
The $(1,1)$-rank amplitudes are obtained from the recombination of topological diagrams.
The decomposition formulas are based on the $SU(3)_F$ symmetry.
It is challenging to incorporate $SU(3)_F$ breaking effects directly into the $(1,1)$-rank amplitudes.
In this work, we introduce two real breaking parameters, $b$ for the $c\to u\overline d d$ transition and $b^\prime$ for the $c\to u\overline s s$ transition, to describe the difference between the $c\to u\overline d d$ and $c\to u\overline s s$ transitions as well as the difference between the singly Cabibbo-suppressed decays and the Cabibbo-favored decays.
Specifically, all the amplitudes of the $c\to u\overline d d$ transition are multiplied by $(1+b)$, and all the amplitudes of the $c\to u\overline s s$ transition are multiplied by $(1+b^\prime)$.
This strategy is based on the following considerations.
\begin{itemize}
  \item Compared with the $c\to u\overline d s$ and $c\to u\overline s d$ transitions, the $c\to u\overline d d$ and $c\to u\overline s s$ transitions have relatively larger and smaller energy releases, respectively.
      Thus, $SU(3)_F$ breaking effects in singly Cabibbo-suppressed modes are generally larger than those in Cabibbo-favored and the doubly Cabibbo-suppressed modes.
  \item The $SU(3)_F$ breaking components of quark-loop diagrams contribute to the branching fractions and decay parameters, enhancing the $SU(3)_F$ breaking effects in singly Cabibbo-suppressed modes.
  \item In the rescattering mechanism, the meson decay constants and baryon form factors in the short-distance $T$ diagrams differ among the four decay types.
\end{itemize}
To avoid introducing a large number of breaking parameters, we restrict these phenomenological breaking terms to singly Cabibbo-suppressed modes.

We perform global fits to the latest experimental data on branching fractions and decay parameters in both the $SU(3)_F$ symmetric and $SU(3)_F$ breaking scenarios.
We employ 54 available data points of branching fractions and decay parameters, as listed in Table~\ref{tab:exp-theory-su3breaking}.
Using these 54 observables, we fit 31 and 33 free parameters in the $SU(3)_F$ symmetric and $SU(3)_F$ breaking cases, respectively.
The numerical values are summarized in Table~\ref{tab:fit-parameters}.
A comparison between the experimental inputs and the predictions of the $SU(3)_F$ breaking fit is shown in Table~\ref{tab:exp-theory-su3breaking}.
In Appendix~\ref{amplitudes}, we present our theoretical predictions for branching fractions and decay parameters for all decay channels in both the $SU(3)_F$ symmetric and $SU(3)_F$ breaking scenarios.
The complete results are given in Tables~\ref{tab:su3-cf}, \ref{tab:su3-scs}, \ref{tab:br-cf}, and \ref{tab:br-scs}.

In our global fits, the \(\chi^2\) per degree of freedom are \(2.58\) and \(1.66\) in the $SU(3)_F$ symmetric and $SU(3)_F$ breaking scenarios, respectively, which are lower than those obtained when the K\"orner-Pati-Woo theorem is imposed.
With the latest experimental data, we performed a fit imposing both the KPW theorem and $SU(3)_F$ symmetry and found no solution with $\chi^2/{\rm d.o.f}<10.80$.
The improved fit quality supports the conclusion of the previous section that the validity of the K\"orner-Pati-Woo theorem warrants re-examination.
In addition to the \(\chi^2\) per degree of freedom, the fitting results for branching fractions and decay parameters also support this point.
The latest experimental data for $\mathcal{B}(\Xi^0_c\to pK^-)$ and $\alpha(\Xi^0_c\to pK^-)$ are not consistent with recent $SU(3)$-based predictions \cite{Cheng:2025oyr,Yang:2025orn,Zhao:2018mov,Geng:2019xbo}, whereas our results are consistent with the data within $1\sigma$.
Moreover, several branching fractions that were not well reproduced in Ref.~\cite{Cheng:2025oyr}, including $\mathcal{B}(\Xi^0_c\to \Xi^-\pi^+)$, $\mathcal{B}(\Xi^0_c\to \Xi^0\eta^\prime)$, $\mathcal{B}(\Xi^+_c\to \Lambda^0\pi^+)$, and $\mathcal{B}(\Xi^0_c\to \Lambda^0\eta^\prime)$, are accommodated by our $SU(3)_F$-symmetric fit.
To further test the KPW theorem, we suggest measuring more branching fractions and decay parameters and comparing fits with and without KPW constraints on the same dataset.
In particular, additional measurements of observables in \(\Xi_c^0\) decay channels are needed.
The decay channels of \(\Lambda_c^+\) and \(\Xi_c^+\) satisfy $U$-spin relations.
In the $SU(3)_F$ limit, the amplitudes of two $U$-spin conjugate charmed hadron decay channels are related by interchanging $\lambda_d$ and $\lambda_s$, up to an overall sign \cite{Zhang:2025jnw,Wang:2019dls}.
Measurements of \(\Xi_c^+\) modes are less constraining than those of \(\Xi_c^0\) modes because many \(\Lambda_c^+\) decay observables have already been measured.

According to Table~\ref{tab:exp-theory-su3breaking}, there is only one branching fraction, $\mathcal{B}(\Xi^0_c\to \Lambda\eta)$, for which the prediction and measurement do not overlap within \(1\sigma\).
For the decay parameters, we find that the theoretical predictions for $\alpha(\Lambda^+_c\to \Xi^0K^+)$ and $\gamma(\Lambda^+_c\to \Xi^0K^+)$ are not consistent with experimental data in the $SU(3)_F$ breaking scenario.
This discrepancy might stem from the incomplete treatment of $SU(3)_F$ breaking effects within our fit model in which only two breaking parameters are introduced.
Data on semileptonic decays of charmed baryons have revealed sizable $SU(3)_F$ breaking effects \cite{BESIII:2022ysa,Belle:2021crz,ALICE:2021bli,ALICE:2025wrq}, substantially larger than those observed in $D$-meson decays.
It is therefore plausible that sizable $SU(3)_F$-breaking effects are also present in two-body nonleptonic charmed baryon decays.
More precise measurements of the branching fraction $\mathcal{B}(\Xi^0_c\to \Lambda\eta)$ and of the decay parameters $\alpha$, $\beta$, and $\gamma$ for the $\Lambda^+_c\to \Xi^0K^+$ mode, together with measurements of currently unmeasured observables, will help improve the theoretical model.

\begin{table*}[t!]
\centering
\caption{Comparison between the theoretical and experimental results for ratios of the branching fractions of $\Xi_c^0$ decay modes to
$\mathcal B(\Xi_c^0\to\Xi^-\pi^+)$.}
\label{tab:xic0-br-ratios}
\small
\renewcommand{\arraystretch}{1.08}
\setlength{\tabcolsep}{6pt}
\begin{tabular}{lccc}
\hline\hline
Decay mode\quad
& \quad$SU(3)_F$\quad
& \quad$SU(3)_F$ breaking\quad
& \quad Experimental value \quad\\
\hline

$\Xi_c^0\to\Xi^-K^+$
& $0.0434\pm0.0053$
& $0.0380\pm0.0052$
& $0.0275\pm0.0057$
\\

$\Xi_c^0\to\Lambda^0 K_S^0$
& $0.324\pm0.067$
& $0.320\pm0.068$
& $0.225\pm0.013$
\\
$\Xi_c^0\to\Sigma^0K_S^0$
& $0.050\pm0.015$
& $0.052\pm0.015$
& $0.038\pm0.007$
\\

$\Xi_c^0\to\Sigma^+K^-$
& $0.095\pm0.036$
& $0.098\pm0.032$
& $0.123\pm0.012$
\\
$\Xi_c^0\to\Xi^0\pi^0$
& $0.58\pm0.16$
& $0.55\pm0.14$
& $0.48\pm0.04$
\\

$\Xi_c^0\to\Xi^0\eta$
& $0.14\pm0.04$
& $0.13\pm0.04$
& $0.11\pm0.01$
\\
$\Xi_c^0\to\Xi^0\eta'$
& $0.12\pm0.04$
& $0.12\pm0.04$
& $0.08\pm0.02$
\\

$\Xi_c^0\to\Lambda^0\eta$
& $0.0251\pm0.0057$
& $0.0206\pm0.0072$
& $0.0416\pm0.0094$
\\

$\Xi_c^0\to\Lambda^0\eta'$
& $0.0233\pm0.0123$
& $0.0213\pm0.0104$
& $0.0248\pm0.0083$
\\
\hline
\end{tabular}
\end{table*}

Most $\Xi^0_c$ branching fractions are inferred from measured ratios normalized to $\mathcal{B}(\Xi^0_c\to \Xi^-\pi^+)$.
The comparison between our predictions for these ratios and the corresponding experimental results is presented in Table~\ref{tab:xic0-br-ratios}.
We find that although the theoretical and experimental values of the branching fractions for the \(\Xi^0_c\to \Xi^-K^+\) mode and the \(\Xi^0_c\to \Xi^-\pi^+\) mode overlap within \(1\sigma\), the theoretical prediction for their ratio \(\mathcal{B}(\Xi^0_c\to \Xi^-K^+)/\mathcal{B}(\Xi^0_c\to \Xi^-\pi^+)\) does not overlap with the experimental measurement within \(1\sigma\).
Given the relatively large experimental uncertainty of this ratio, a more precise measurement would provide a stronger constraint on the theoretical parameters.

\begin{table}[t!]
\centering
\caption{Predictions for the $K_S^0-K_L^0$ asymmetries in charmed baryon decays.}
\label{tab:ks-kl-asymmetry}
\small
\renewcommand{\arraystretch}{1.08}
\begin{tabular}{lcc|lcc}
\hline\hline
Decay mode
& $SU(3)_F$
& $SU(3)_F$ breaking \quad& Decay mode
& $SU(3)_F$
& $SU(3)_F$ breaking \\
\hline
$\Lambda_c^+\to pK_{S,L}^0$
\quad& $0.024\pm0.005$
\quad& $0.028\pm0.006$
&
$\Xi_c^0\to\Sigma^0K_{S,L}^0$
\quad& $0.133\pm0.213$
\quad& $0.178\pm0.264$ \\

$\Xi_c^0\to\Lambda K_{S,L}^0$
& $-0.043\pm0.031$
& $-0.036\pm0.037$
&
$\Xi_c^+\to\Sigma^+K_{S,L}^0$
& $0.427\pm0.117$
& $0.362\pm0.118$ \\
\hline
\end{tabular}
\end{table}

With the theoretical parameters extracted from the global fits, we also predict the $K^0_S-K^0_L$ asymmetries in charmed baryon decays, defined as
\begin{align}
  R(B_{c\overline3}\to BK_{S,L}^0)\equiv
\frac{\mathcal B(B_{c\overline3}\to BK_S^0)-\mathcal B(B_{c\overline3}\to BK_L^0)}
{\mathcal B(B_{c\overline3}\to BK_S^0)+\mathcal B(B_{c\overline3}\to BK_L^0)}.
\end{align}
The results are shown in Table~\ref{tab:ks-kl-asymmetry}.
A notable prediction for the \(\Xi_c^+\to\Sigma^+K^0_{S,L}\) modes is a large $K_S^0-K_L^0$ asymmetry of about $40\%$
This feature can be understood through a combined $U$-spin analysis of the $\Xi_c^+\to\Sigma^+K^0_{S,L}$ and the $\Lambda_c^+\to pK^0_{S,L}$ modes \cite{Wang:2025ucg}.
The large $K^0_S-K^0_L$ asymmetry provides enhanced experimental sensitivity to the doubly Cabibbo-suppressed amplitudes.

\subsection{CP asymmetries}

\begin{table*}[t!]
\centering
\caption{Predictions for the CP-violating observables in the $SU(3)_F$-symmetric fit.}
\label{tab:cpv-penguin-su3}
\small
\renewcommand{\arraystretch}{1.08}
\begin{tabular}{lcccc}
\hline\hline
Decay mode~~~~ & ~~~~$A_{\rm CP}\;(10^{-3})$~~~~ & ~~~~$A_{\rm CP}^{\alpha}\;(10^{-3})$~~~~ & ~~~~$A_{\rm CP}^{\beta}\;(10^{-3})$~~~~ & ~~~~$A_{\rm CP}^{\gamma}\;(10^{-3})$~~~~ \\
\hline
$\Lambda_c^+\to \Lambda^0 K^+$ & $-1.227\pm0.943$ & $0.858\pm1.131$ & $0.383\pm0.774$ & $-0.435\pm0.496$ \\
$\Lambda_c^+\to \Sigma^0K^+$ & $-0.276\pm0.477$ & $-0.256\pm0.698$ & $-0.190\pm0.664$ & $0.040\pm0.510$ \\
$\Lambda_c^+\to \Sigma^+K^0$ & $-0.215\pm0.402$ & $0.076\pm0.711$ & $-0.078\pm0.972$ & $-0.318\pm0.436$ \\
$\Lambda_c^+\to p\pi^0$ & $1.543\pm1.296$ & $-0.042\pm1.532$ & $-0.201\pm1.082$ & $0.201\pm1.140$ \\
$\Lambda_c^+\to p\eta$ & $-0.744\pm0.760$ & $0.401\pm0.869$ & $0.164\pm0.652$ & $-0.285\pm0.411$ \\
$\Lambda_c^+\to p\eta'$ & $1.218\pm0.244$ & $-0.903\pm0.340$ & $0.021\pm1.024$ & $-0.998\pm0.314$ \\
$\Lambda_c^+\to n\pi^+$ & $0.927\pm0.734$ & $-0.609\pm0.488$ & $0.431\pm1.034$ & $-0.353\pm1.313$ \\
$\Xi_c^0\to \Sigma^-\pi^+$ & $-0.284\pm0.930$ & $-0.656\pm1.144$ & $-0.099\pm2.333$ & $1.832\pm2.092$ \\
$\Xi_c^0\to \Lambda^0\pi^0$ & $0.780\pm0.376$ & $-0.117\pm0.547$ & $-0.923\pm0.532$ & $-0.902\pm1.272$ \\
$\Xi_c^0\to \Lambda^0\eta$ & $0.440\pm0.476$ & $0.076\pm1.047$ & $0.475\pm1.256$ & $-1.059\pm1.090$ \\
$\Xi_c^0\to \Lambda^0\eta'$ & $-0.671\pm0.639$ & $0.436\pm0.968$ & $0.042\pm0.716$ & $-0.208\pm0.834$ \\
$\Xi_c^0\to \Sigma^0\pi^0$ & $-0.594\pm0.859$ & $-0.136\pm1.824$ & $-0.319\pm2.913$ & $2.093\pm0.793$ \\
$\Xi_c^0\to \Sigma^0\eta$ & $-0.846\pm1.635$ & $-0.451\pm1.638$ & $1.581\pm1.259$ & $-0.494\pm3.136$ \\
$\Xi_c^0\to \Sigma^0\eta'$ & $1.591\pm1.095$ & $-0.490\pm2.055$ & $0.165\pm1.640$ & $0.545\pm1.555$ \\
$\Xi_c^0\to n\overline K^0$ & $-0.614\pm0.352$ & $-0.141\pm0.616$ & $0.147\pm0.364$ & $0.373\pm1.362$ \\
$\Xi_c^0\to \Xi^-K^+$ & $-0.138\pm0.801$ & $0.396\pm1.482$ & $0.422\pm2.469$ & $-1.893\pm1.489$ \\
$\Xi_c^0\to \Xi^0K^0$ & $0.434\pm0.462$ & $-0.027\pm0.748$ & $-0.152\pm0.235$ & $-0.429\pm1.094$ \\
$\Xi_c^0\to \Sigma^+\pi^-$ & $0.596\pm2.665$ & $1.348\pm4.139$ & $-1.800\pm3.827$ & $-0.069\pm11.293$ \\
$\Xi_c^0\to pK^-$ & $-0.614\pm4.723$ & $-1.642\pm5.468$ & $2.202\pm3.946$ & $0.103\pm16.904$ \\
$\Xi_c^+\to \Lambda^0\pi^+$ & $-0.324\pm0.460$ & $1.985\pm0.307$ & $-0.043\pm0.212$ & $0.150\pm1.113$ \\
$\Xi_c^+\to \Sigma^0\pi^+$ & $0.958\pm0.901$ & $-0.552\pm1.102$ & $-0.359\pm0.632$ & $0.187\pm0.486$ \\
$\Xi_c^+\to \Sigma^+\pi^0$ & $0.955\pm1.114$ & $-0.641\pm1.419$ & $-0.316\pm0.570$ & $0.239\pm0.486$ \\
$\Xi_c^+\to \Sigma^+\eta$ & $-1.118\pm0.525$ & $0.225\pm0.949$ & $0.483\pm0.213$ & $0.730\pm0.521$ \\
$\Xi_c^+\to \Sigma^+\eta'$ & $-0.719\pm0.153$ & $0.645\pm0.198$ & $-0.490\pm0.328$ & $0.348\pm0.327$ \\
$\Xi_c^+\to p\overline K^0$ & $0.104\pm0.553$ & $-0.137\pm0.684$ & $0.106\pm0.743$ & $0.214\pm0.304$ \\
$\Xi_c^+\to \Xi^0K^+$ & $-0.807\pm0.326$ & $0.572\pm0.235$ & $-0.288\pm1.114$ & $0.244\pm0.899$ \\
\hline
\end{tabular}
\end{table*}

\begin{table*}[t!]
\centering
\caption{Predictions for the CP-violating observables in the $SU(3)_F$-breaking fit. }
\label{tab:cpv-penguin-breaking}
\small
\renewcommand{\arraystretch}{1.08}
\begin{tabular}{lcccc}
\hline\hline
Decay mode~~~~ & ~~~~$A_{\rm CP}\;(10^{-3})$~~~~ & ~~~~$A_{\rm CP}^{\alpha}\;(10^{-3})$~~~~ & ~~~~$A_{\rm CP}^{\beta}\;(10^{-3})$~~~~ & ~~~~$A_{\rm CP}^{\gamma}\;(10^{-3})$~~~~ \\
\hline
$\Lambda_c^+\to \Lambda^0 K^+$ & $-1.375\pm0.810$ & $0.801\pm1.698$ & $0.815\pm2.104$ & $-0.111\pm0.495$ \\
$\Lambda_c^+\to \Sigma^0K^+$ & $-0.220\pm0.617$ & $-0.169\pm1.028$ & $-0.149\pm0.797$ & $0.063\pm0.354$ \\
$\Lambda_c^+\to \Sigma^+K^0$ & $-0.169\pm0.305$ & $0.107\pm0.580$ & $0.049\pm1.359$ & $-0.262\pm0.216$ \\
$\Lambda_c^+\to p\pi^0$ & $4.754\pm6.295$ & $-1.116\pm7.384$ & $-0.418\pm3.577$ & $0.521\pm1.756$ \\
$\Lambda_c^+\to p\eta$ & $-0.743\pm0.703$ & $0.474\pm1.150$ & $0.341\pm1.198$ & $-0.170\pm0.198$ \\
$\Lambda_c^+\to p\eta'$ & $1.397\pm0.444$ & $-0.991\pm1.070$ & $-0.434\pm1.288$ & $-1.151\pm0.611$ \\
$\Lambda_c^+\to n\pi^+$ & $2.290\pm2.598$ & $-1.209\pm2.982$ & $0.265\pm4.950$ & $-1.691\pm3.784$ \\
$\Xi_c^0\to \Sigma^-\pi^+$ & $-0.220\pm0.768$ & $-0.590\pm0.480$ & $-0.156\pm1.173$ & $0.990\pm1.501$ \\
$\Xi_c^0\to \Lambda^0\pi^0$ & $0.912\pm1.049$ & $-0.423\pm0.492$ & $-0.905\pm1.540$ & $-1.098\pm1.407$ \\
$\Xi_c^0\to \Lambda^0\eta$ & $0.745\pm0.532$ & $0.358\pm0.915$ & $0.334\pm1.356$ & $-1.314\pm1.298$ \\
$\Xi_c^0\to \Lambda^0\eta'$ & $-0.602\pm0.810$ & $0.527\pm1.114$ & $0.179\pm0.986$ & $-0.060\pm1.129$ \\
$\Xi_c^0\to \Sigma^0\pi^0$ & $-0.471\pm0.463$ & $-0.384\pm1.120$ & $-0.373\pm1.070$ & $1.122\pm1.346$ \\
$\Xi_c^0\to \Sigma^0\eta$ & $-1.225\pm2.075$ & $-0.432\pm4.100$ & $3.197\pm3.452$ & $-0.464\pm5.727$ \\
$\Xi_c^0\to \Sigma^0\eta'$ & $0.798\pm0.917$ & $-0.508\pm1.150$ & $0.242\pm1.156$ & $0.225\pm1.689$ \\
$\Xi_c^0\to n\overline K^0$ & $-0.745\pm0.237$ & $-0.214\pm0.399$ & $0.004\pm1.530$ & $0.485\pm2.103$ \\
$\Xi_c^0\to \Xi^-K^+$ & $-0.052\pm1.401$ & $1.028\pm1.889$ & $0.614\pm3.259$ & $-2.691\pm2.103$ \\
$\Xi_c^0\to \Xi^0K^0$ & $0.427\pm0.629$ & $-0.017\pm0.896$ & $-0.165\pm0.268$ & $-0.458\pm1.901$ \\
$\Xi_c^0\to \Sigma^+\pi^-$ & $0.411\pm3.221$ & $1.845\pm7.032$ & $-0.963\pm6.150$ & $-0.252\pm6.819$ \\
$\Xi_c^0\to pK^-$ & $-0.412\pm1.910$ & $-1.582\pm3.733$ & $1.473\pm2.050$ & $0.152\pm13.148$ \\
$\Xi_c^+\to \Lambda^0\pi^+$ & $-0.170\pm0.257$ & $1.175\pm0.275$ & $-0.124\pm0.048$ & $-0.002\pm0.588$ \\
$\Xi_c^+\to \Sigma^0\pi^+$ & $0.941\pm1.191$ & $-0.599\pm1.571$ & $-0.261\pm0.894$ & $0.274\pm0.582$ \\
$\Xi_c^+\to \Sigma^+\pi^0$ & $0.831\pm1.195$ & $-0.647\pm1.702$ & $-0.183\pm0.796$ & $0.306\pm0.552$ \\
$\Xi_c^+\to \Sigma^+\eta$ & $-1.138\pm0.906$ & $0.216\pm1.393$ & $0.130\pm0.256$ & $0.904\pm0.264$ \\
$\Xi_c^+\to \Sigma^+\eta'$ & $-0.424\pm0.158$ & $0.369\pm0.165$ & $-0.312\pm0.231$ & $0.066\pm0.303$ \\
$\Xi_c^+\to p\overline K^0$ & $0.118\pm0.449$ & $-0.236\pm0.732$ & $0.020\pm1.844$ & $0.294\pm0.210$ \\
$\Xi_c^+\to \Xi^0K^+$ & $-0.458\pm0.310$ & $0.342\pm0.238$ & $-0.220\pm0.816$ & $0.117\pm0.827$ \\
\hline
\end{tabular}
\end{table*}

The decay amplitude for a singly Cabibbo-suppressed charmed hadron decay mode is written as $\mathcal{A}=\lambda_d\mathcal{A}_d+\lambda_s\mathcal{A}_s$.
The corresponding direct CP asymmetry can be written approximately as
\begin{align}
A_{\rm CP}\simeq -2\,\frac{{ Im}(\lambda_d^*\lambda_s)}{|\lambda_d|^2} \frac{{ Im}(\mathcal{A}_d^*\mathcal{A}_s)}{|\mathcal{A}_d-\mathcal{A}_s|^2}.
\end{align}
In charm meson decays, quark-loop diagrams provide important sources of CP asymmetries.
Quark-loop contributions in charm decays are difficult to extract from branching-fraction fits or to calculate reliably using QCD-inspired approaches.
The rescattering mechanism provides a feasible strategy for estimating the long-distance contributions to quark-loop diagrams.
Our previous study showed that, in the $SU(3)$-symmetric limit, the long-distance contribution to the quark-loop topology $P$ is equivalent to that of the $W$-exchange topology $E$, $P_L\simeq E_L$ \cite{Wang:2021rhd}.
This finding provides a dynamical basis for understanding CP violation in charm meson decays \cite{Cheng:2012xb,LHCb:2019hro}.

In charmed baryon decays, the amplitudes $A_{11}\sim A_{14}$ are also important sources of CP asymmetries.
As in $D$-meson decays, the long-distance quark-loop contributions in charmed-baryon decays can be related to the corresponding contributions from tree topologies.
Table~\ref{tab} shows that the rescattering contributions to $A_{11}$ coincide with those to $A_{6}$, whereas those to $A_{12}$ can be expressed in terms of several tree-level amplitudes,
\begin{align}\label{key}
  (A_{11})_L\simeq (A_{6})_L,\qquad (A_{12})_L\simeq -(A_{3})_L-(A_{6})_L-3(A_{8})_L.
\end{align}
According to Eq.~\eqref{key}, the decay amplitudes containing quark-loop contributions can be estimated from the tree-level amplitudes when the corresponding short-distance contributions are neglected.
The resulting amplitudes can then be used to predict CP asymmetries in charmed-baryon decays.
As argued in Ref.~\cite{Jia:2024pyb}, the $s$-channel rescattering amplitudes for charmed hadron decays are suppressed by the finite widths of the intermediate resonances.
In this work, we neglect the long-distance contribution to $A_{14}$.
The rescattering contributions to $A_{11}$ and $A_{12}$ are comparable in magnitude to those contributing to tree amplitudes.
It indicates that the CP asymmetries in some charmed baryon decay channels could reach $\mathcal{O}(10^{-3})$.

In Refs.~\cite{Yang:2025orn,He:2024unv,Cheng:2025oyr}, the long-distance quark-loop contributions depend on additional nonperturbative parameters $r_-$ and $\rho_+$.
By contrast, the relations in Eq.~\eqref{key} are compact and introduce no additional nonperturbative parameters.
The rescattering amplitudes derived from the topological diagrams follow directly from the topology-rescattering matching procedure and are consistent with those obtained independently from the chiral Lagrangian.

In charmed baryon decays, CP-violating observables can be defined in terms of the decay width and the decay parameters $\alpha$, $\beta$, and $\gamma$,
\begin{align}
 A_{\rm CP} = \frac{\Gamma-\overline \Gamma}{\Gamma+\overline \Gamma},\qquad
 A^\alpha_{\rm CP} = \frac{\alpha+\overline \alpha}{2},\qquad
 A^\beta_{\rm CP} = \frac{\beta+\overline \beta}{2},\qquad
 A^\gamma_{\rm CP} = \frac{\gamma-\overline \gamma}{2}.
\end{align}
Using the fit results from the previous subsection together with Eq.~\eqref{key}, we predict the CP-violating observables associated with the decay width $\Gamma$ and the decay parameters $\alpha$, $\beta$, and $\gamma$ for singly Cabibbo-suppressed decays.
The results for the $SU(3)_F$ symmetric and $SU(3)_F$ breaking scenarios are presented in Tables~\ref{tab:cpv-penguin-su3} and \ref{tab:cpv-penguin-breaking}, respectively.
As shown in Tables~\ref{tab:cpv-penguin-su3} and \ref{tab:cpv-penguin-breaking}, the predicted CP observables generally carry sizable uncertainties.
Consequently, it is difficult to identify a channel with a clear experimental advantage for observing CP violation in the charmed-baryon sector.
Among the channels considered, the CP asymmetry in $\Lambda^+_c\to p\eta^\prime$ exhibits both a relatively large central value and a comparatively small uncertainty.
Thus, the $\Lambda^+_c\to p\eta^\prime$ mode is a promising benchmark channel for CP violation searches in the charmed baryon sector.
These large uncertainties arise because the CP asymmetries are sensitive to the strong phases of the $S$- and $P$-wave amplitudes.
However, the available data predominantly constrain CP-even combinations of amplitude moduli.
The relative strong phases therefore remain weakly constrained.
The relatively small uncertainty of $A_{\rm CP}(\Lambda^+_c\to p\eta^\prime)$ arises because $\eta-\eta^\prime$ mixing leads to fewer cancellations in the amplitude combination, and the relevant parameter combination lies close to a well-constrained direction of the fit covariance matrix rather than to a nearly flat direction.

\begin{table*}[t!]
\centering
\caption{Ranking of experimental observables according to their impact on the theoretical uncertainties of CP asymmetries, assuming prospective precisions of $\sigma_{\mathcal{B}}/\mathcal{B}=10\%$ and $\sigma_\alpha=\sigma_\beta=\sigma_\gamma=0.10$. }
\label{tab:rank}
\small
\renewcommand{\arraystretch}{1.08}
\begin{tabular}{c| l c | l c}
\hline\hline
\quad Rank\quad\quad
& \qquad $SU(3)_F$\quad
& \quad Average reduction\quad
& $SU(3)_F$ breaking\quad
& \quad Average reduction \quad \\
\hline

1
& $\beta(\Xi_c^0\to\Sigma^+K^-)$
& $39.8\%$
& $\gamma(\Xi_c^0\to\Xi^0\eta)$
& $38.6\%$ \\

2
& $\gamma(\Xi_c^0\to\Sigma^+K^-)$
& $38.3\%$
& $\mathcal B(\Xi_c^0\to\Lambda\pi^0)$
& $36.7\%$ \\

3
& $\gamma(\Xi_c^0\to\Sigma^0K_S^0)$
& $37.4\%$
& $\beta(\Xi_c^0\to\Sigma^+K^-)$
& $36.4\%$ \\

4
& $\alpha(\Xi_c^0\to\Xi^0\eta)$
& $36.9\%$
& $\alpha(\Xi_c^0\to\Lambda\pi^0)$
& $35.6\%$ \\

5
& $\mathcal B(\Xi_c^0\to\Lambda\pi^0)$
& $36.8\%$
& $\mathcal B(\Xi_c^0\to\Sigma^0\pi^0)$
& $35.3\%$ \\

6
& $\gamma(\Xi_c^0\to\Sigma^0K_L^0)$
& $36.1\%$
& $\gamma(\Xi_c^0\to\Sigma^0K_S^0)$
& $35.2\%$ \\

7
& $\mathcal B(\Xi_c^0\to\Sigma^0\pi^0)$
& $35.6\%$
& $\gamma(\Xi_c^0\to\Sigma^0K_L^0)$
& $35.1\%$ \\

8
& $\gamma(\Xi_c^0\to\Xi^0\eta)$
& $34.6\%$
& $\gamma(\Xi_c^0\to\Sigma^+K^-)$
& $35.0\%$ \\

9
& $\alpha(\Xi_c^0\to\Sigma^0K_L^0)$
& $31.1\%$
& $\alpha(\Xi_c^0\to\Xi^0\eta')$
& $34.2\%$ \\

10
& $\alpha(\Xi_c^0\to\Sigma^0K_S^0)$
& $30.3\%$
& $\gamma(\Xi_c^0\to\Lambda\eta')$
& $34.2\%$ \\

11
& $\beta(\Xi_c^0\to\Sigma^0\pi^0)$
& $28.2\%$
& $\beta(\Xi_c^0\to n\overline K^0)$
& $33.3\%$ \\

12
& $\alpha(\Xi_c^0\to\Lambda\pi^0)$
& $28.1\%$
& $\beta(\Xi_c^0\to\Sigma^0\pi^0)$
& $33.3\%$ \\

13
& $\beta(\Xi_c^0\to\Sigma^0K_S^0)$
& $27.8\%$
& $\alpha(\Xi_c^0\to\Sigma^0\pi^0)$
& $33.0\%$ \\

14
& $\alpha(\Xi_c^0\to\Sigma^0\pi^0)$
& $27.7\%$
& $\alpha(\Xi_c^0\to\Lambda\eta')$
& $32.6\%$ \\

15
& $\alpha(\Xi_c^0\to\Lambda K_L^0)$
& $26.7\%$
& $\alpha(\Xi_c^0\to\Lambda K_L^0)$
& $32.6\%$ \\

16
& $\beta(\Xi_c^0\to\Xi^0K^0)$
& $26.1\%$
& $\alpha(\Xi_c^0\to\Lambda K_S^0)$
& $31.1\%$ \\

17
& $\beta(\Xi_c^0\to\Lambda\eta)$
& $26.1\%$
& $\alpha(\Xi_c^0\to\Xi^0K^0)$
& $30.9\%$ \\

18
& $\beta(\Xi_c^0\to n\overline K^0)$
& $25.9\%$
& $\beta(\Xi_c^0\to\Lambda K_L^0)$
& $30.8\%$ \\

19
& $\beta(\Xi_c^0\to\Sigma^0K_L^0)$
& $25.8\%$
& $\beta(\Xi_c^0\to\Sigma^0K_S^0)$
& $30.7\%$ \\

20
& $\beta(\Xi_c^0\to\Lambda K_L^0)$
& $25.8\%$
& $\beta(\Xi_c^0\to\Lambda K_S^0)$
& $30.5\%$ \\

21
& $\beta(\Xi_c^0\to\Lambda K_S^0)$
& $25.6\%$
& $\beta(\Xi_c^0\to\Sigma^0K_L^0)$
& $30.4\%$ \\

22
& $\alpha(\Xi_c^0\to\Lambda K_S^0)$
& $24.3\%$
& $\gamma(\Xi_c^0\to\Xi^0\eta')$
& $29.8\%$ \\

23
& $\alpha(\Xi_c^0\to\Lambda\eta)$
& $23.9\%$
& $\beta(\Xi_c^0\to\Lambda\eta)$
& $29.6\%$ \\

24
& $\mathcal B(\Xi_c^0\to n\overline K^0)$
& $23.7\%$
& $\alpha(\Xi_c^0\to\Xi^0\eta)$
& $28.0\%$ \\

25
& $\beta(\Xi_c^0\to\Xi^0\eta')$
& $22.5\%$
& $\mathcal B(\Xi_c^0\to\Xi^0K^0)$
& $27.7\%$ \\

26
& $\beta(\Xi_c^0\to\Lambda\pi^0)$
& $20.9\%$
& $\beta(\Xi_c^0\to\Xi^0K^0)$
& $27.6\%$ \\

27
& $\beta(\Xi_c^0\to\Xi^0\eta)$
& $20.1\%$
& $\alpha(\Xi_c^0\to\Lambda\eta)$
& $26.3\%$ \\

28
& $\alpha(\Xi_c^0\to n\overline K^0)$
& $19.8\%$
& $\beta(\Xi_c^0\to\Xi^0\eta')$
& $26.3\%$ \\

29
& $\alpha(\Xi_c^0\to\Xi^0\eta')$
& $19.6\%$
& $\gamma(\Xi_c^0\to\Xi^0K^0)$
& $25.5\%$ \\

30
& $\beta(\Xi_c^+\to\Sigma^+K_S^0)$
& $19.4\%$
& $\beta(\Xi_c^0\to\Xi^0\eta)$
& $21.7\%$ \\

\hline
\end{tabular}%
\end{table*}

To reduce the theoretical uncertainties of CP observables, additional measurements of decay parameters are needed to better constrain the parameter space and weaken parameter correlations.
To identify which currently unmeasured observables would most effectively reduce the theoretical uncertainties of CP asymmetries, we add each candidate observable individually to the fit and quantify the resulting uncertainty reduction.
The average fractional reduction in the theoretical uncertainties of the 26 direct CP asymmetries is defined as
\begin{align}
  {\cal I}
=
\frac{1}{26}\sum_f
\left(
1-\frac{\sigma^\prime_{A_{\rm CP}(f)}}
{\sigma_{A_{\rm CP}(f)}}
\right),
\end{align}
where $\sigma^\prime_{A_{\rm CP}(f)}$ is the uncertainty of $A_{\rm CP}(f)$ when one additional observable is included in the global fit, and $\sigma_{A_{\rm CP}(f)}$ is the original uncertainty reported in Tables~\ref{tab:cpv-penguin-su3} and \ref{tab:cpv-penguin-breaking}.
For comparison, we assume a common prospective precision for every candidate observable: the uncertainty of a branching fraction is taken to be $10\%$ of its theoretical central value, whereas the uncertainties of the decay parameters are set to $\sigma_\alpha=\sigma_\beta=\sigma_\gamma=0.10$.
The 30 observables that produce the largest average uncertainty reductions in the $SU(3)_F$-symmetric and $SU(3)_F$-breaking scenarios are listed in Table~\ref{tab:rank}.
The results exhibit a clear pattern: the highest-ranked observables are dominated by \(\Xi_c^0\) modes, and decay parameters generally rank ahead of branching fractions.
Thus, we suggest measuring additional decay parameters in $\Xi^0_c$ modes to reduce theoretical uncertainties of CP asymmetries.

\section{Summary}\label{sum}

In this work, we develop a framework that connects quark-level topological amplitudes with hadron-level final-state interactions.
The (1,1)-rank amplitudes, which are linear combinations of topological amplitudes, provide the interface to the chiral Lagrangian.
The rescattering amplitudes derived from topological diagrams are consistent with those obtained from the chiral Lagrangian.
The preservation of the complete set of isospin sum rules provides a nontrivial consistency check of the framework.

We find that hadronic rescattering violates the amplitude relations implied by the K\"orner-Pati-Woo theorem
Moreover, a loophole in the conventional proof of the K\"orner-Pati-Woo theorem is identified: it assumes that the color assignments of the quarks are preserved from the weak vertex until their confinement into the final baryon, whereas gluon exchanges can change these colors.
We suggest a precise measurement of the branching fraction of the $\Lambda^+_c\to \Sigma^+K^0_S$ mode at Belle II to test the K\"orner-Pati-Woo theorem.

As a phenomenological application, we perform global fits to the latest data on branching fractions and decay parameters in the $SU(3)_F$ symmetric and $SU(3)_F$ breaking scenarios.
The \(\chi^2\) per degree of freedom are \(2.58\) and \(1.66\), respectively.
Many branching fractions that were not well reproduced in previous analyses are accommodated by our fits, further supporting the conclusion that the validity of the K\"orner-Pati-Woo theorem warrants re-examination.
Notably, the recent measurements of both the branching fraction and decay parameter of \(\Xi_c^0\to pK^-\), which are not consistent with the theoretical predictions in the literature, are described within \(1\sigma\) in our framework.
In addition, a large $K^0_S-K^0_L$ asymmetry in the \(\Xi_c^+\to\Sigma^+K^0_{S,L}\) modes is predicted.

We derive the $u$-, $t$-, and $s$-channel rescattering contributions to each (1,1)-rank amplitude in the $SU(3)_F$ limit through tensor contractions.
We find that the rescattering amplitudes contributing to quark-loop diagrams can be expressed in terms of those contributing to tree diagrams without introducing additional hadronic parameters.
Using the tree-level decay amplitudes extracted from the $SU(3)_F$-symmetric and $SU(3)_F$-breaking global fits, we predict the CP asymmetries in charmed-baryon decays.
Due to the weak constraints of available experimental data on the theoretical parameters, the predicted uncertainties of CP asymmetries are substantial.
We suggest measuring additional observables in $\Xi^0_c$ decay modes to reduce these theoretical uncertainties.
We also identify $\Lambda^+_c\to p\eta^\prime$ as a promising channel for CP-violation searches in the charmed-baryon sector, because $A_{\rm CP}(\Lambda^+_c\to p\eta^\prime)$ can reach the $10^{-3}$ level with a relatively small uncertainty.

\begin{acknowledgements}

We are grateful to Fu-Sheng Yu, Zhu-Ding Duan, and Long-Ke Li for useful discussions.
This work was supported in part by the Scientific Research Fund of Hunan Provincial Department under No. 25B0090 and 24B0435 and the National Natural Science Foundation of China under No. 32502553.

\end{acknowledgements}

\begin{appendix}

\section{Decay amplitudes}\label{amplitudes}

\begin{table*}[htp!]
\caption{Decay amplitudes of the Cabibbo-favored and doubly Cabibbo-suppressed $B_{c\overline{3}}\to B_{8}M$ decays.}\label{ta7cf}
\centering
\setlength{\tabcolsep}{3pt}
\footnotesize
\begin{tabular}{|l|c|l|c|}
\hline\hline
Channel & Amplitude & Channel & Amplitude \\ \hline
$\Lambda_c^+\to\Lambda^0\pi^+$ & $\frac{\lambda_1}{\sqrt6}(-2A_2+A_7+A_8)$
& $\Lambda_c^+\to\Sigma^0\pi^+$ & $\frac{\lambda_1}{\sqrt2}(A_7-A_8)$ \\ \hline
$\Lambda_c^+\to\Sigma^+\pi^0$ & $\frac{\lambda_1}{\sqrt2}(-A_7+A_8)$
& $\Lambda_c^+\to\Sigma^+\eta$ & $\lambda_1\left[\frac{\cos\phi}{\sqrt2}(2A_{10}+A_7+A_8)-\sin\phi(A_1+A_{10})\right]$ \\ \hline
$\Lambda_c^+\to\Sigma^+\eta'$ & $\lambda_1\left[\frac{\sin\phi}{\sqrt2}(2A_{10}+A_7+A_8)+\cos\phi(A_1+A_{10})\right]$
& $\Lambda_c^+\to pK_S^0$ & $\frac{1}{\sqrt2}\big[\lambda_2(A_1+A_4)-\lambda_1(A_4+A_7)\big]$ \\ \hline
$\Lambda_c^+\to pK_L^0$ & $\frac{1}{\sqrt2}\big[\lambda_2(A_1+A_4)+\lambda_1(A_4+A_7)\big]$
& $\Lambda_c^+\to\Xi^0K^+$ & $\lambda_1(A_3+A_8)$ \\ \hline
$\Xi_c^0\to\Xi^-\pi^+$ & $\lambda_1(A_2+A_5)$
& $\Xi_c^0\to\Xi^0\pi^0$ & $\frac{\lambda_1}{\sqrt2}(A_3-A_5)$ \\ \hline
$\Xi_c^0\to\Xi^0\eta$ & $\lambda_1\left[\frac{\cos\phi}{\sqrt2}(A_3+A_5+2A_9)-\sin\phi(A_6+A_9)\right]$
& $\Xi_c^0\to\Xi^0\eta'$ & $\lambda_1\left[\frac{\sin\phi}{\sqrt2}(A_3+A_5+2A_9)+\cos\phi(A_6+A_9)\right]$ \\ \hline
$\Xi_c^0\to\Lambda^0K_S^0$ & $\frac{1}{\sqrt{12}}\big[\lambda_2(A_4+A_5-2A_6)-\lambda_1(A_4-2A_5+A_6)\big]$
& $\Xi_c^0\to\Lambda^0K_L^0$ & $\frac{1}{\sqrt{12}}\big[\lambda_2(A_4+A_5-2A_6)+\lambda_1(A_4-2A_5+A_6)\big]$ \\ \hline
$\Xi_c^0\to\Sigma^+K^-$ & $\lambda_1(A_1+A_6)$
& $\Xi_c^0\to\Sigma^0K_S^0$ & $\frac12\big[\lambda_2(A_4-A_5)-\lambda_1(A_4-A_6)\big]$ \\ \hline
$\Xi_c^0\to\Sigma^0K_L^0$ & $\frac12\big[\lambda_2(A_4-A_5)+\lambda_1(A_4-A_6)\big]$
& $\Xi_c^+\to\Xi^0\pi^+$ & $-\lambda_1(A_2+A_3)$ \\ \hline
$\Xi_c^+\to\Sigma^+K_S^0$ & $\frac{1}{\sqrt2}\big[\lambda_1(A_1+A_4)-\lambda_2(A_4+A_7)\big]$
& $\Xi_c^+\to\Sigma^+K_L^0$ & $-\frac{1}{\sqrt2}\big[\lambda_1(A_1+A_4)+\lambda_2(A_4+A_7)\big]$ \\ \toprule[1.2pt]
$\Lambda_c^+\to nK^+$ & $\lambda_2(A_2+A_3)$
& $\Xi_c^0\to\Sigma^-K^+$ & $\lambda_2(A_2+A_5)$ \\ \hline
$\Xi_c^0\to p\pi^-$ & $\lambda_2(A_1+A_6)$
& $\Xi_c^0\to n\pi^0$ & $\frac{\lambda_2}{\sqrt2}(A_3-A_6)$ \\ \hline
$\Xi_c^0\to n\eta$ & $\lambda_2\left[\frac{\cos\phi}{\sqrt2}(A_3+A_6+2A_9)-\sin\phi(A_5+A_9)\right]$
& $\Xi_c^0\to n\eta'$ & $\lambda_2\left[\frac{\sin\phi}{\sqrt2}(A_3+A_6+2A_9)+\cos\phi(A_5+A_9)\right]$ \\ \hline
$\Xi_c^+\to\Lambda^0K^+$ & $-\frac{\lambda_2}{\sqrt6}(A_2+A_7-2A_8)$
& $\Xi_c^+\to\Sigma^0K^+$ & $\frac{\lambda_2}{\sqrt2}(A_2-A_7)$ \\ \hline
$\Xi_c^+\to p\pi^0$ & $\frac{\lambda_2}{\sqrt2}(A_1-A_8)$
& $\Xi_c^+\to p\eta$ & $-\lambda_2\left[\frac{\cos\phi}{\sqrt2}(A_1+2A_{10}+A_8)-\sin\phi(A_{10}+A_7)\right]$ \\ \hline
$\Xi_c^+\to p\eta'$ & $-\lambda_2\left[\frac{\sin\phi}{\sqrt2}(A_1+2A_{10}+A_8)+\cos\phi(A_{10}+A_7)\right]$
& $\Xi_c^+\to n\pi^+$ & $-\lambda_2(A_3+A_8)$ \\ \hline
\end{tabular}
\end{table*}

\begin{table*}[htp!]
\caption{Decay amplitudes of the singly Cabibbo-suppressed $\Lambda^+_c$ and $\Xi^+_c$ decays.}\label{ta8a}
\footnotesize
\centering
\begin{tabular}{|l|c|}
\hline\hline
Channel & Amplitude \\ \hline
$\Lambda_c^+\to\Lambda^0K^+$
& $\displaystyle -\frac{\lambda_b}{\sqrt6}\left(A_3^P-2A_4^P+A_5^P-2A_6^P+A_{11}^P-2A_{12}^P-6A_{15}^P+3A_{17}^P\right)$
\\[-1mm]& $\displaystyle \quad +\frac{\lambda_d}{\sqrt6}\left(A_3+A_{11}-2A_{12}\right)
+\frac{\lambda_s}{\sqrt6}\left(-2A_2-2A_3+A_7-2A_8+A_{11}-2A_{12}\right)$ \\
\hline

$\Lambda_c^+\to\Sigma^0K^+$
& $\displaystyle -\frac{\lambda_b}{\sqrt2}\left(A_3^P+A_5^P+A_{11}^P+3A_{17}^P\right)
+\frac{\lambda_d}{\sqrt2}\left(-A_3+A_{11}\right)
+\frac{\lambda_s}{\sqrt2}\left(A_7+A_{11}\right)$ \\
\hline

$\Lambda_c^+\to\Sigma^+K^0$
& $\displaystyle -\lambda_b\left(A_3^P+A_5^P+A_{11}^P+3A_{17}^P\right)
+\lambda_d\left(A_1+A_{11}\right)+\lambda_s\left(A_7+A_{11}\right)$ \\
\hline

$\Lambda_c^+\to p\pi^0$
& $\displaystyle -\frac{\lambda_b}{\sqrt2}\left(A_4^P+A_6^P+A_{12}^P+3A_{15}^P\right)
+\frac{\lambda_d}{\sqrt2}\left(-A_4+A_{12}\right)
+\frac{\lambda_s}{\sqrt2}\left(A_8+A_{12}\right)$ \\
\hline

$\Lambda_c^+\to p\eta$
& $\displaystyle -\lambda_b\Bigg[\frac{\cos\phi}{\sqrt2}\left(2A_2^P+A_4^P+A_6^P+2A_9^P+A_{12}^P+2A_{13}^P+3A_{15}^P+6A_{16}^P\right)$
\\[-1mm]& $\displaystyle \qquad-\sin\phi\left(A_2^P+A_3^P+A_5^P+A_9^P+A_{11}^P+A_{13}^P+3A_{16}^P+3A_{17}^P\right)\Bigg]$
\\[-1mm]& $\displaystyle \quad+\lambda_d\Bigg[\frac{\cos\phi}{\sqrt2}\left(A_4+A_{12}+2A_{13}\right)-\sin\phi\left(A_{11}+A_{13}\right)\Bigg]$
\\[-1mm]& $\displaystyle \quad+\lambda_s\Bigg[\frac{\cos\phi}{\sqrt2}\left(2A_{10}+A_8+A_{12}+2A_{13}\right)-\sin\phi\left(A_1+A_4+A_7+A_{10}+A_{11}+A_{13}\right)\Bigg]$ \\
\hline

$\Lambda_c^+\to p\eta'$
& $\displaystyle -\lambda_b\Bigg[\frac{\sin\phi}{\sqrt2}\left(2A_2^P+A_4^P+A_6^P+2A_9^P+A_{12}^P+2A_{13}^P+3A_{15}^P+6A_{16}^P\right)$
\\[-1mm]& $\displaystyle \qquad+\cos\phi\left(A_2^P+A_3^P+A_5^P+A_9^P+A_{11}^P+A_{13}^P+3A_{16}^P+3A_{17}^P\right)\Bigg]$
\\[-1mm]& $\displaystyle \quad+\lambda_d\Bigg[\frac{\sin\phi}{\sqrt2}\left(A_4+A_{12}+2A_{13}\right)+\cos\phi\left(A_{11}+A_{13}\right)\Bigg]$
\\[-1mm]& $\displaystyle \quad+\lambda_s\Bigg[\frac{\sin\phi}{\sqrt2}\left(2A_{10}+A_8+A_{12}+2A_{13}\right)+\cos\phi\left(A_1+A_4+A_7+A_{10}+A_{11}+A_{13}\right)\Bigg]$ \\
\hline

$\Lambda_c^+\to n\pi^+$
& $\displaystyle -\lambda_b\left(A_4^P+A_6^P+A_{12}^P+3A_{15}^P\right)
+\lambda_d\left(A_2+A_{12}\right)+\lambda_s\left(A_8+A_{12}\right)$ \\
\hline

$\Xi_c^+\to\Lambda^0\pi^+$
& $\displaystyle \frac{\lambda_b}{\sqrt6}\left(A_3^P+A_4^P+A_5^P+A_6^P+A_{11}^P+A_{12}^P+3A_{15}^P+3A_{17}^P\right)$
\\[-1mm]& $\displaystyle \quad-\frac{\lambda_d}{\sqrt6}\left(A_2+A_3+A_7+A_8+A_{11}+A_{12}\right)
-\frac{\lambda_s}{\sqrt6}\left(-2A_3+A_{11}+A_{12}\right)$ \\
\hline

$\Xi_c^+\to\Sigma^0\pi^+$
& $\displaystyle \frac{\lambda_b}{\sqrt2}\left(A_3^P-A_4^P+A_5^P-A_6^P+A_{11}^P-A_{12}^P-3A_{15}^P+3A_{17}^P\right)$
\\[-1mm]& $\displaystyle \quad+\frac{\lambda_d}{\sqrt2}\left(A_2+A_3-A_7+A_8-A_{11}+A_{12}\right)
+\frac{\lambda_s}{\sqrt2}\left(-A_{11}+A_{12}\right)$ \\
\hline

$\Xi_c^+\to\Sigma^+\pi^0$
& $\displaystyle -\frac{\lambda_b}{\sqrt2}\left(A_3^P-A_4^P+A_5^P-A_6^P+A_{11}^P-A_{12}^P-3A_{15}^P+3A_{17}^P\right)$
\\[-1mm]& $\displaystyle \quad+\frac{\lambda_d}{\sqrt2}\left(A_1+A_4+A_7-A_8+A_{11}-A_{12}\right)
+\frac{\lambda_s}{\sqrt2}\left(A_{11}-A_{12}\right)$ \\
\hline

$\Xi_c^+\to\Sigma^+\eta$
& $\displaystyle \lambda_b\Bigg[\frac{\cos\phi}{\sqrt2}\left(2A_2^P+A_3^P+A_4^P+A_5^P+A_6^P+2A_9^P+A_{11}^P+A_{12}^P+2A_{13}^P+3A_{15}^P+6A_{16}^P+3A_{17}^P\right)$
\\[-1mm]& $\displaystyle -\sin\phi\left(A_2^P+A_9^P+A_{13}^P+3A_{16}^P\right)\Bigg]$
\\[-1mm]& $\displaystyle \quad-\lambda_d\Bigg[\frac{\cos\phi}{\sqrt2}\left(A_1+A_4+A_7+A_8+2A_{10}+A_{11}+A_{12}+2A_{13}\right)-\sin\phi\left(A_{10}+A_{13}\right)\Bigg]$
\\[-1mm]& $\displaystyle \quad-\lambda_s\Bigg[\frac{\cos\phi}{\sqrt2}\left(A_{11}+A_{12}+2A_{13}\right)-\sin\phi\left(A_4+A_{13}\right)\Bigg]$ \\
\hline

$\Xi_c^+\to\Sigma^+\eta'$
& $\displaystyle \lambda_b\Bigg[\frac{\sin\phi}{\sqrt2}\left(2A_2^P+A_3^P+A_4^P+A_5^P+A_6^P+2A_9^P+A_{11}^P+A_{12}^P+2A_{13}^P+3A_{15}^P+6A_{16}^P+3A_{17}^P\right)
$
\\[-1mm]& $\displaystyle +\cos\phi\left(A_2^P+A_9^P+A_{13}^P+3A_{16}^P\right)\Bigg]$
\\[-1mm]& $\displaystyle \quad-\lambda_d\Bigg[\frac{\sin\phi}{\sqrt2}\left(A_1+A_4+A_7+A_8+2A_{10}+A_{11}+A_{12}+2A_{13}\right)+\cos\phi\left(A_{10}+A_{13}\right)\Bigg]$
\\[-1mm]& $\displaystyle \quad-\lambda_s\Bigg[\frac{\sin\phi}{\sqrt2}\left(A_{11}+A_{12}+2A_{13}\right)+\cos\phi\left(A_4+A_{13}\right)\Bigg]$ \\
\hline

$\Xi_c^+\to p\overline K^0$
& $\displaystyle \lambda_b\left(A_3^P+A_5^P+A_{11}^P+3A_{17}^P\right)
-\lambda_d\left(A_7+A_{11}\right)-\lambda_s\left(A_1+A_{11}\right)$ \\
\hline

$\Xi_c^+\to\Xi^0K^+$
& $\displaystyle \lambda_b\left(A_4^P+A_6^P+A_{12}^P+3A_{15}^P\right)
-\lambda_d\left(A_8+A_{12}\right)-\lambda_s\left(A_2+A_{12}\right)$ \\\hline
\end{tabular}
\end{table*}

\begin{table*}[htp!]
\caption{Decay amplitudes of the singly Cabibbo-suppressed $\Xi^0_c$ decays.}\label{ta8b}
\footnotesize
\centering
\begin{tabular}{|l|c|}
\hline\hline
Channel & Amplitude \\ \hline

$\Xi_c^0\to\Sigma^-\pi^+$
& $\displaystyle -\lambda_b\left(A_4^P+A_6^P+A_7^P+A_8^P+A_{12}^P+A_{14}^P+3A_{15}^P+3A_{18}^P\right)$
\\[-1mm]& $\displaystyle \quad+\lambda_d\left(A_2+A_5+A_{12}+A_{14}\right)
+\lambda_s\left(A_{12}+A_{14}\right)$ \\
\hline

$\Xi_c^0\to\Lambda^0\pi^0$
& $\displaystyle -\frac{\lambda_b}{2\sqrt3}\left(A_3^P+A_4^P+A_5^P+A_6^P+A_{11}^P+A_{12}^P+3A_{15}^P+3A_{17}^P\right)$
\\[-1mm]& $\displaystyle \quad+\frac{\lambda_d}{2\sqrt3}\left(A_3-A_4-A_5-A_6+A_{11}+A_{12}\right)
+\frac{\lambda_s}{2\sqrt3}\left(A_{11}+A_{12}-2A_3\right)$ \\
\hline

$\Xi_c^0\to\Lambda^0\eta$
& $\displaystyle -\lambda_b\Bigg[\frac{\cos\phi}{2\sqrt3}\left(2A_2^P+A_3^P+A_4^P+A_5^P+A_6^P+2A_7^P+2A_8^P+2A_9^P+A_{11}^P+A_{12}^P+2A_{13}^P+2A_{14}^P\right.$
\\[-1mm]& $\displaystyle \quad\left.+3A_{15}^P+6A_{16}^P+3A_{17}^P+6A_{18}^P\right)-\frac{\sin\phi}{\sqrt6}\left(A_2^P-2A_7^P-2A_8^P+A_9^P+A_{13}^P-2A_{14}^P+3A_{16}^P-6A_{18}^P\right)\Bigg]$
\\[-1mm]& $\displaystyle \quad+\lambda_d\Bigg[\frac{\cos\phi}{2\sqrt3}\left(A_3+A_4+A_5+A_6+2A_9+A_{11}+A_{12}+2A_{13}+2A_{14}\right)-\frac{\sin\phi}{\sqrt6}\left(A_9+A_{13}-2A_{14}\right)\Bigg]$
\\[-1mm]& $\displaystyle +\lambda_s\Bigg[\frac{\cos\phi}{2\sqrt3}\left(-2A_3-4A_9+A_{11}+A_{12}+2A_{13}+2A_{14}\right)-\frac{\sin\phi}{\sqrt6}\left(A_4-2A_5-2A_6-2A_9+A_{13}-2A_{14}\right)\Bigg]$ \\
\hline

$\Xi_c^0\to\Lambda^0\eta'$
& $\displaystyle -\lambda_b\Bigg[\frac{\sin\phi}{2\sqrt3}\left(2A_2^P+A_3^P+A_4^P+A_5^P+A_6^P+2A_7^P+2A_8^P+2A_9^P+A_{11}^P+A_{12}^P+2A_{13}^P+2A_{14}^P\right.$
\\[-1mm]& $\displaystyle \left.\qquad+3A_{15}^P+6A_{16}^P+3A_{17}^P+6A_{18}^P\right)+\frac{\cos\phi}{\sqrt6}\left(A_2^P-2A_7^P-2A_8^P+A_9^P+A_{13}^P-2A_{14}^P+3A_{16}^P-6A_{18}^P\right)\Bigg]$
\\[-1mm]& $\displaystyle \quad+\lambda_d\Bigg[\frac{\sin\phi}{2\sqrt3}\left(A_3+A_4+A_5+A_6+2A_9+A_{11}+A_{12}+2A_{13}+2A_{14}\right)+\frac{\cos\phi}{\sqrt6}\left(A_9+A_{13}-2A_{14}\right)\Bigg]$
\\[-1mm]& $\displaystyle +\lambda_s\Bigg[\frac{\sin\phi}{2\sqrt3}\left(-2A_3-4A_9+A_{11}+A_{12}+2A_{13}+2A_{14}\right)+\frac{\cos\phi}{\sqrt6}\left(A_4-2A_5-2A_6-2A_9+A_{13}-2A_{14}\right)\Bigg]$ \\
\hline

$\Xi_c^0\to\Sigma^0\pi^0$
& $\displaystyle -\frac{\lambda_b}{2}\left(A_3^P+A_4^P+A_5^P+A_6^P+2A_7^P+2A_8^P+A_{11}^P+A_{12}^P+2A_{14}^P+3A_{15}^P+3A_{17}^P+6A_{18}^P\right)$
\\[-1mm]& $\displaystyle \quad+\frac{\lambda_d}{2}\left(-A_3-A_4+A_5+A_6+A_{11}+A_{12}+2A_{14}\right)
+\frac{\lambda_s}{2}\left(A_{11}+A_{12}+2A_{14}\right)$ \\
\hline

$\Xi_c^0\to\Sigma^0\eta$
& $\displaystyle -\lambda_b\Bigg[\frac{\cos\phi}{2}\left(2A_2^P+A_3^P+A_4^P+A_5^P+A_6^P+2A_9^P+A_{11}^P+A_{12}^P+2A_{13}^P+3A_{15}^P+6A_{16}^P+3A_{17}^P\right)$
\\[-1mm]& $\displaystyle -\frac{\sin\phi}{\sqrt2}\left(A_2^P+A_9^P+A_{13}^P+3A_{16}^P\right)\Bigg]$
\\[-1mm]& $\displaystyle \quad+\lambda_d\Bigg[\frac{\cos\phi}{2}\left(-A_3+A_4-A_5-A_6-2A_9+A_{11}+A_{12}+2A_{13}\right)-\frac{\sin\phi}{\sqrt2}\left(A_{13}-A_9\right)\Bigg]$
\\[-1mm]& $\displaystyle \quad+\lambda_s\Bigg[\frac{\cos\phi}{2}\left(A_{11}+A_{12}+2A_{13}\right)-\frac{\sin\phi}{\sqrt2}\left(A_4+A_{13}\right)\Bigg]$ \\
\hline

$\Xi_c^0\to\Sigma^0\eta'$
& $\displaystyle -\lambda_b\Bigg[\frac{\sin\phi}{2}\left(2A_2^P+A_3^P+A_4^P+A_5^P+A_6^P+2A_9^P+A_{11}^P+A_{12}^P+2A_{13}^P+3A_{15}^P+6A_{16}^P+3A_{17}^P\right)$
\\[-1mm]& $\displaystyle +\frac{\cos\phi}{\sqrt2}\left(A_2^P+A_9^P+A_{13}^P+3A_{16}^P\right)\Bigg]$
\\[-1mm]& $\displaystyle \quad+\lambda_d\Bigg[\frac{\sin\phi}{2}\left(-A_3+A_4-A_5-A_6-2A_9+A_{11}+A_{12}+2A_{13}\right)+\frac{\cos\phi}{\sqrt2}\left(A_{13}-A_9\right)\Bigg]$
\\[-1mm]& $\displaystyle \quad+\lambda_s\Bigg[\frac{\sin\phi}{2}\left(A_{11}+A_{12}+2A_{13}\right)+\frac{\cos\phi}{\sqrt2}\left(A_4+A_{13}\right)\Bigg]$ \\
\hline

$\Xi_c^0\to n\overline K^0$
& $\displaystyle -\lambda_b\left(A_7^P+A_8^P+A_{14}^P+3A_{18}^P\right)
+\lambda_d\left(A_5+A_{14}\right)+\lambda_s\left(A_6+A_{14}\right)$ \\
\hline

$\Xi_c^0\to\Xi^-K^+$
& $\displaystyle -\lambda_b\left(A_4^P+A_6^P+A_7^P+A_8^P+A_{12}^P+A_{14}^P+3A_{15}^P+3A_{18}^P\right)$
\\[-1mm]& $\displaystyle \quad+\lambda_d\left(A_{12}+A_{14}\right)
+\lambda_s\left(A_2+A_5+A_{12}+A_{14}\right)$ \\
\hline

$\Xi_c^0\to\Xi^0K^0$
& $\displaystyle -\lambda_b\left(A_7^P+A_8^P+A_{14}^P+3A_{18}^P\right)
+\lambda_d\left(A_6+A_{14}\right)+\lambda_s\left(A_5+A_{14}\right)$ \\
\hline

$\Xi_c^0\to\Sigma^+\pi^-$
& $\displaystyle -\lambda_b\left(A_3^P+A_5^P+A_7^P+A_8^P+A_{11}^P+A_{14}^P+3A_{17}^P+3A_{18}^P\right)$
\\[-1mm]& $\displaystyle \quad+\lambda_d\left(A_1+A_6+A_{11}+A_{14}\right)
+\lambda_s\left(A_{11}+A_{14}\right)$ \\
\hline

$\Xi_c^0\to pK^-$
& $\displaystyle -\lambda_b\left(A_3^P+A_5^P+A_7^P+A_8^P+A_{11}^P+A_{14}^P+3A_{17}^P+3A_{18}^P\right)$
\\[-1mm]& $\displaystyle \quad+\lambda_d\left(A_{11}+A_{14}\right)
+\lambda_s\left(A_1+A_6+A_{11}+A_{14}\right)$ \\
\hline
\end{tabular}
\end{table*}

\section{Rescattering structures}\label{scatter}

$A_2\to A_1$:
\begin{align}
T(A_1)_1[i,j,k,l]\,=\,B_{c\overline 3}^qH_{pm}^n \contraction[1mm]{}{P}{_n^pB_q^m\,\cdot\,}{P}P_n^p
\contraction[2mm]{}{B}{_q^m\,\cdot\,P_p^nB_k^p B_j^k\,\cdot\,}{B}B_q^m\,\cdot\,P_p^n\contraction[1mm]{}{B}{_k^p B_j^k\,\cdot\,B_m^q}{B}B_k^p B_j^k\,\cdot\,B_m^qB_l^m P_i^l\,\cdot\,
\delta_{kp}\delta_{lm}\delta_{jn}\,\cdot\,\delta_{iq},
\end{align}
\begin{align}
T(A_1)_2[i,j,k,l]\,=\, B_{c\overline 3}^qH_{pm}^n\contraction[1mm]{}{P}{_n^pB_q^m\,\cdot\,}{P}P_n^p
\contraction[2mm]{}{B}{_q^m\,\cdot\,P_p^nB_k^p B_j^k\,\cdot\,}{B}B_q^m\,\cdot\,P_p^n\contraction[1mm]{}{B}{_k^p B_j^k\,\cdot\,B_m^q}{B}B_k^p B_j^k\,\cdot\,B_m^qB_q^i P_i^l\,\cdot\,
\delta_{kp}\delta_{lm}\delta_{jn}\,\cdot\,\delta_{iq},
\end{align}
\begin{align}
T(A_1)_3[i,j,k,l]\,=\, B_{c\overline 3}^qH_{pm}^n\contraction[1mm]{}{P}{_n^pB_q^m\,\cdot\,}{P}P_n^p
\contraction[2mm]{}{B}{_q^m\,\cdot\,P_p^nB_n^j B_j^k\,\cdot\,}{B}B_q^m\,\cdot\,P_p^n
\contraction[1mm]{}{B}{_n^j B_j^k\,\cdot\,B_m^q}{B}B_n^j B_j^k\,\cdot\,B_m^qB_l^m P_i^l\,\cdot\,
\delta_{kp}\delta_{lm}\delta_{jn}\,\cdot\,\delta_{iq},
\end{align}
\begin{align}
T(A_1)_4[i,j,k,l]\,=\, B_{c\overline 3}^qH_{pm}^n\contraction[1mm]{}{P}{_n^pB_q^m\,\cdot\,}{P}P_n^p
\contraction[2mm]{}{B}{_q^m\,\cdot\,P_p^nB_n^j B_j^k\,\cdot\,}{B}B_q^m\,\cdot\,P_p^n
\contraction[1mm]{}{B}{_n^j B_j^k\,\cdot\,B_m^q}{B}B_n^j B_j^k\,\cdot\,B_m^qB_q^i P_i^l\,\cdot\,
\delta_{kp}\delta_{lm}\delta_{jn}\,\cdot\,\delta_{iq}.
\end{align}
$A_2\to A_2$:
\begin{align}
U(A_2)_1[i,j,k,l]\,=\, B_{c\overline 3}^qH_{pm}^n\contraction[1mm]{}{P}{_n^pB_q^m\,\cdot\,}{P}P_n^p
\contraction[2mm]{}{B}{_q^m\,\cdot\,P_p^nV_l^p P_j^l\,\cdot\,}{B}B_q^m\,\cdot\,P_p^n
\contraction[1mm]{}{V}{_l^p P_j^l\,\cdot\,B_m^q}{V}V_l^p P_j^l\,\cdot\,B_m^qV_k^m B_i^k\,\cdot\,
\delta_{lp}\delta_{km}\delta_{jn}\,\cdot\,\delta_{iq},
\end{align}
\begin{align}
U(A_2)_2[i,j,k,l]\,=\, B_{c\overline 3}^qH_{pm}^n\contraction[1mm]{}{P}{_n^pB_q^m\,\cdot\,}{P}P_n^p
\contraction[2mm]{}{B}{_q^m\,\cdot\,P_p^nV_l^p P_j^l\,\cdot\,}{B}B_q^m\,\cdot\,P_p^n
\contraction[1mm]{}{V}{_l^p P_j^l\,\cdot\,B_m^q}{V}V_l^p P_j^l\,\cdot\,B_m^qV_q^i B_i^k\,\cdot\,
\delta_{lp}\delta_{km}\delta_{jn}\,\cdot\,\delta_{iq},
\end{align}
\begin{align}
U(A_2)_3[i,j,k,l]\,=\, B_{c\overline 3}^qH_{pm}^n\contraction[1mm]{}{P}{_n^pB_q^m\,\cdot\,}{P}P_n^p
\contraction[2mm]{}{B}{_q^m\,\cdot\,P_p^nV_n^j P_j^l\,\cdot\,}{B}B_q^m\,\cdot\,P_p^n
\contraction[1mm]{}{V}{_n^j P_j^l\,\cdot\,B_m^q}{V}V_n^j P_j^l\,\cdot\,B_m^qV_k^m B_i^k\,\cdot\,
\delta_{lp}\delta_{km}\delta_{jn}\,\cdot\,\delta_{iq},
\end{align}
\begin{align}
U(A_2)_4[i,j,k,l]\,=\, B_{c\overline 3}^qH_{pm}^n\contraction[1mm]{}{P}{_n^pB_q^m\,\cdot\,}{P}P_n^p
\contraction[2mm]{}{B}{_q^m\,\cdot\,P_p^nV_n^j P_j^l\,\cdot\,}{B}B_q^m\,\cdot\,P_p^n
\contraction[1mm]{}{V}{_n^j P_j^l\,\cdot\,B_m^q}{V}V_n^j P_j^l\,\cdot\,B_m^qV_q^i B_i^k\,\cdot\,
\delta_{lp}\delta_{km}\delta_{jn}\,\cdot\,\delta_{iq},
\end{align}
\begin{align}
U(A_2)_5[i,j,k,l]\,=\, B_{c\overline 3}^qH_{pm}^n\contraction[1mm]{}{P}{_n^pB_q^m\,\cdot\,}{P}P_n^p
\contraction[2mm]{}{B}{_q^m\,\cdot\,P_p^nV_l^p P_j^l\,\cdot\,}{B}B_q^m\,\cdot\,P_p^n
\contraction[1mm]{}{V}{_l^p P_j^l\,\cdot\,B_m^q}{V}V_l^p P_j^l\,\cdot\,B_m^qV_r^r B_i^k\,\cdot\,
\delta_{lp}\delta_{km}\delta_{jn}\,\cdot\,\delta_{iq},
\end{align}
\begin{align}
U(A_2)_6[i,j,k,l]\,=\, B_{c\overline 3}^qH_{pm}^n\contraction[1mm]{}{P}{_n^pB_q^m\,\cdot\,}{P}P_n^p
\contraction[2mm]{}{B}{_q^m\,\cdot\,P_p^nV_n^j P_j^l\,\cdot\,}{B}B_q^m\,\cdot\,P_p^n
\contraction[1mm]{}{V}{_n^j P_j^l\,\cdot\,B_m^q}{V}V_n^j P_j^l\,\cdot\,B_m^qV_r^r B_i^k\,\cdot\,
\delta_{lp}\delta_{km}\delta_{jn}\,\cdot\,\delta_{iq}.
\end{align}
$A_2\to A_3$:
\begin{align}
U(A_3)[i,j,k,l]\,=\, B_{c\overline 3}^qH_{pm}^n
\contraction[1mm]{}{P}{_n^pB_q^m\,\cdot\,}{P}P_n^p
\contraction[2mm]{}{B}{_q^m\,\cdot\,P_p^nV_n^i P_i^l\,\cdot\,}{B}B_q^m\,\cdot\,P_p^n
\contraction[1mm]{}{V}{_n^i P_i^l\,\cdot\,B_m^q}{V}V_n^i P_i^l\,\cdot\,B_m^qV_i^n B_j^k\,\cdot\,
\delta_{lp}\delta_{km}\delta_{jn}\,\cdot\,\delta_{iq},
\end{align}
\begin{align}
T(A_3)[i,j,k,l]\,=\, B_{c\overline 3}^qH_{pm}^n\contraction[1mm]{}{P}{_n^pB_q^m\,\cdot\,}{P}P_n^p
\contraction[2mm]{}{B}{_q^m\,\cdot\,P_p^nB_k^p B_j^k\,\cdot\,}{B}B_q^m\,\cdot\,P_p^n
\contraction[1mm]{}{B}{_k^p B_j^k\,\cdot\,B_m^q}{B}B_k^p B_j^k\,\cdot\,B_m^qB_p^k P_i^l\,\cdot\,
\delta_{lp}\delta_{km}\delta_{jn}\,\cdot\,\delta_{iq}.
\end{align}
$A_2\to A_4$:
\begin{align}
U(A_4)[i,j,k,l]\,=\,B_{c\overline 3}^qH_{pm}^n\contraction[1mm]{}{P}{_n^pB_q^m\,\cdot\,}{P}P_n^p
\contraction[2mm]{}{B}{_q^m\,\cdot\,P_p^nV_l^p P_j^l\,\cdot\,}{B}B_q^m\,\cdot\,P_p^n
\contraction[1mm]{}{V}{_l^p P_j^l\,\cdot\,B_m^q}{V}V_l^p P_j^l\,\cdot\,B_m^qV_p^l B_i^k\,\cdot\,
\delta_{kp}\delta_{lm}\delta_{jn}\,\cdot\,\delta_{iq},
\end{align}
\begin{align}
T(A_4)[i,j,k,l]\,=\,B_{c\overline 3}^qH_{pm}^n\contraction[1mm]{}{P}{_n^pB_q^m\,\cdot\,}{P}P_n^p
\contraction[2mm]{}{B}{_q^m\,\cdot\,P_p^nB_n^i B_i^k\,\cdot\,}{B}B_q^m\,\cdot\,P_p^n
\contraction[1mm]{}{B}{_n^i B_i^k\,\cdot\,B_m^q}{B}B_n^i B_i^k\,\cdot\,B_m^qB_i^n P_j^l\,\cdot\,
\delta_{kp}\delta_{lm}\delta_{jn}\,\cdot\,\delta_{iq}.
\end{align}
$A_2\to A_6$:
\begin{align}
T(A_6)[i,j,k,l]\,=\,B_{c\overline 3}^qH_{pm}^n\contraction[1mm]{}{P}{_n^pB_q^m\,\cdot\,}{P}P_n^p
\contraction[2mm]{}{B}{_q^m\,\cdot\,P_p^nB_k^q B_j^k\,\cdot\,}{B}B_q^m\,\cdot\,P_p^n
\contraction[1mm]{}{B}{_k^q B_j^k\,\cdot\,B_m^q}{B}B_k^q B_j^k\,\cdot\,B_m^qB_q^k P_k^l\,\cdot\,
\delta_{ip}\delta_{lm}\delta_{jn}\,\cdot\,\delta_{iq},
\end{align}
\begin{align}
S(A_6)[i,j,k,l]\,=\,B_{c\overline 3}^qH_{pm}^n\contraction[1mm]{}{P}{_n^pB_q^m\,\cdot\,}{P}P_n^p
\contraction[2mm]{}{B}{_q^m\,\cdot\,P_p^n}{B}B_q^m\,\cdot\,P_p^nB_m^q
\contraction[1mm]{}{B}{_n^m\,\cdot\,}{B}B_n^m\,\cdot\,B_l^j P_k^l B_j^k\,\cdot\,
\delta_{ip}\delta_{lm}\delta_{jn}\,\cdot\,\delta_{iq}.
\end{align}
$A_2\to A_7$:
\begin{align}
U(A_7)_1[i,j,k,l]\,=\,B_{c\overline 3}^qH_{pm}^n\contraction[1mm]{}{P}{_n^pB_q^m\,\cdot\,}{P}P_n^p
\contraction[2mm]{}{B}{_q^m\,\cdot\,P_p^nV_l^p P_j^l\,\cdot\,}{B}B_q^m\,\cdot\,P_p^n
\contraction[1mm]{}{V}{_l^p P_j^l\,\cdot\,B_k^r}{V}V_l^p P_j^l\,\cdot\,B_k^rV_r^l B_l^k\,\cdot\,
\delta_{kp}\delta_{im}\delta_{jn}\,\cdot\,\delta_{iq}\,\cdot\,\delta_{kr},
\end{align}
\begin{align}
U(A_7)_2[i,j,k,l]\,=\,B_{c\overline 3}^qH_{pm}^n\contraction[1mm]{}{P}{_n^pB_q^m\,\cdot\,}{P}P_n^p
\contraction[2mm]{}{B}{_q^m\,\cdot\,P_p^nV_l^p P_j^l\,\cdot\,}{B}B_q^m\,\cdot\,P_p^n
\contraction[1mm]{}{V}{_l^p P_j^l\,\cdot\,B_r^l}{V}V_l^p P_j^l\,\cdot\,B_r^lV_k^r B_l^k\,\cdot\,
\delta_{kp}\delta_{im}\delta_{jn}\,\cdot\,\delta_{iq}\,\cdot\,\delta_{lr},
\end{align}
\begin{align}
T(A_7)_1[i,j,k,l]\,=\,B_{c\overline 3}^qH_{pm}^n\contraction[1mm]{}{P}{_n^pB_q^m\,\cdot\,}{P}P_n^p
\contraction[2mm]{}{B}{_q^m\,\cdot\,P_p^nB_n^l B_l^k\,\cdot\,}{B}B_q^m\,\cdot\,P_p^n
\contraction[1mm]{}{B}{_n^l B_l^k\,\cdot\,B_l^r}{B}B_n^l B_l^k\,\cdot\,B_l^rB_r^j P_j^l\,\cdot\,
\delta_{kp}\delta_{im}\delta_{jn}\,\cdot\,\delta_{iq}\,\cdot\,\delta_{lr},
\end{align}
\begin{align}
T(A_7)_2[i,j,k,l]\,=\, B_{c\overline 3}^qH_{pm}^n\contraction[1mm]{}{P}{_n^pB_q^m\,\cdot\,}{P}P_n^p
\contraction[2mm]{}{B}{_q^m\,\cdot\,P_p^nB_n^l B_l^k\,\cdot\,}{B}B_q^m\,\cdot\,P_p^n
\contraction[1mm]{}{B}{_n^l B_l^k\,\cdot\,B_r^j}{B}B_n^l B_l^k\,\cdot\,B_r^jB_l^r P_j^l\,\cdot\,
\delta_{kp}\delta_{im}\delta_{jn}\,\cdot\,\delta_{iq}\,\cdot\,\delta_{jr},
\end{align}
\begin{align}
S(A_{7})_1[i,j,k,l]\,=\,B_{c\overline 3}^qH_{pm}^n\contraction[1mm]{}{P}{_n^pB_q^m\,\cdot\,}{P}P_n^p
\contraction[2mm]{}{B}{_q^m\,\cdot\,P_p^n}{B}B_q^m\,\cdot\,P_p^nB_k^p
\contraction[1mm]{}{B}{_k^p\,\cdot\,}{B}B_j^k\,\cdot\,B_k^j P_j^l B_l^k\,\cdot\,
\delta_{kp}\delta_{im}\delta_{jn}\,\cdot\,\delta_{iq},
\end{align}
\begin{align}
S(A_{7})_2[i,j,k,l]\,=\,B_{c\overline 3}^qH_{pm}^n\contraction[1mm]{}{P}{_n^pB_q^m\,\cdot\,}{P}P_n^p
\contraction[2mm]{}{B}{_q^m\,\cdot\,P_p^n}{B}B_q^m\,\cdot\,P_p^nB_n^j
\contraction[1mm]{}{B}{_n^m\,\cdot\,}{B}B_j^k\,\cdot\,B_k^j P_j^l B_l^k\,\cdot\,
\delta_{kp}\delta_{im}\delta_{jn}\,\cdot\,\delta_{iq}.
\end{align}
$A_2\to A_8$:
\begin{align}
U(A_8)_1[i,j,k,l]\,=\,B_{c\overline 3}^qH_{pm}^n\contraction[1mm]{}{P}{_n^pB_q^m\,\cdot\,}{P}P_n^p
\contraction[2mm]{}{B}{_q^m\,\cdot\,P_p^nV_n^k P_k^l\,\cdot\,}{B}B_q^m\,\cdot\,P_p^n
\contraction[1mm]{}{V}{_n^k P_k^l\,\cdot\,B_k^r}{V}V_n^k P_k^l\,\cdot\,B_k^rV_r^j B_j^k\,\cdot\,
\delta_{lp}\delta_{im}\delta_{jn}\,\cdot\,\delta_{iq}\,\cdot\,\delta_{kr},
\end{align}
\begin{align}
U(A_8)_2[i,j,k,l]\,=\,B_{c\overline 3}^qH_{pm}^n\contraction[1mm]{}{P}{_n^pB_q^m\,\cdot\,}{P}P_n^p
\contraction[2mm]{}{B}{_q^m\,\cdot\,P_p^nV_n^k P_k^l\,\cdot\,}{B}B_q^m\,\cdot\,P_p^n
\contraction[1mm]{}{V}{_n^k P_k^l\,\cdot\,B_r^j}{V}V_n^k P_k^l\,\cdot\,B_r^jV_k^r B_j^k\,\cdot\,
\delta_{lp}\delta_{im}\delta_{jn}\,\cdot\,\delta_{iq}\,\cdot\,\delta_{jr},
\end{align}
\begin{align}
T(A_8)_1[i,j,k,l]\,=\,B_{c\overline 3}^qH_{pm}^n\contraction[1mm]{}{P}{_n^pB_q^m\,\cdot\,}{P}P_n^p
\contraction[2mm]{}{B}{_q^m\,\cdot\,P_p^nB_k^p B_j^k\,\cdot\,}{B}B_q^m\,\cdot\,P_p^n
\contraction[1mm]{}{B}{_k^p B_j^k\,\cdot\,B_l^r}{B}B_k^p B_j^k\,\cdot\,B_l^rB_r^k P_k^l\,\cdot\,
\delta_{lp}\delta_{im}\delta_{jn}\,\cdot\,\delta_{iq}\,\cdot\,\delta_{lr},
\end{align}
\begin{align}
T(A_8)_2[i,j,k,l]\,=\,B_{c\overline 3}^qH_{pm}^n\contraction[1mm]{}{P}{_n^pB_q^m\,\cdot\,}{P}P_n^p
\contraction[2mm]{}{B}{_q^m\,\cdot\,P_p^nB_k^p B_j^k\,\cdot\,}{B}B_q^m\,\cdot\,P_p^n
\contraction[1mm]{}{B}{_k^p B_j^k\,\cdot\,B_r^k}{B}B_k^p B_j^k\,\cdot\,B_r^kB_l^r P_k^l\,\cdot\,
\delta_{lp}\delta_{im}\delta_{jn}\,\cdot\,\delta_{iq}\,\cdot\,\delta_{kr},
\end{align}
\begin{align}
S(A_{8})_1[i,j,k,l]\,=\,B_{c\overline 3}^qH_{pm}^n\contraction[1mm]{}{P}{_n^pB_q^m\,\cdot\,}{P}P_n^p
\contraction[2mm]{}{B}{_q^m\,\cdot\,P_p^n}{B}B_q^m\,\cdot\,P_p^nB_l^p
\contraction[1mm]{}{B}{_q^p\,\cdot\,}{B}B_j^l\,\cdot\,B_l^j P_k^l B_j^k\,\cdot\,
\delta_{lp}\delta_{im}\delta_{jn}\,\cdot\,\delta_{iq},
\end{align}
\begin{align}
S(A_{8})_2[i,j,k,l]\,=\,B_{c\overline 3}^qH_{pm}^n\contraction[1mm]{}{P}{_n^pB_q^m\,\cdot\,}{P}P_n^p
\contraction[2mm]{}{B}{_q^m\,\cdot\,P_p^n}{B}B_q^m\,\cdot\,P_p^nB_n^j
\contraction[1mm]{}{B}{_n^m\,\cdot\,}{B}B_j^l\,\cdot\,B_l^j P_k^l B_j^k\,\cdot\,
\delta_{lp}\delta_{im}\delta_{jn}\,\cdot\,\delta_{iq}.
\end{align}
$A_2\to A_{10}$:
\begin{align}
T(A_{10})_1[i,j,k,l]\,=\, B_{c\overline 3}^qH_{pm}^n
\contraction[1mm]{}{P}{_n^pB_q^m\,\cdot\,}{P}P_n^p
\contraction[2mm]{}{B}{_q^m\,\cdot\,P_p^nB_k^p B_j^k\,\cdot\,}{B}B_q^m\,\cdot\,P_p^n
\contraction[1mm]{}{B}{_k^p B_j^k\,\cdot\,B_m^q}{V}B_k^p B_j^k\,\cdot\,B_l^sB_s^r P_r^l\,\cdot\,
\delta_{kp}\delta_{im}\delta_{jn}\,\cdot\,\delta_{iq}\,\cdot\,\delta_{lr}\,\cdot\,\delta_{ls},
\end{align}
\begin{align}
T(A_{10})_2[i,j,k,l]\,=\, B_{c\overline 3}^qH_{pm}^n
\contraction[1mm]{}{P}{_n^pB_q^m\,\cdot\,}{P}P_n^p
\contraction[2mm]{}{B}{_q^m\,\cdot\,P_p^nB_k^p B_j^k\,\cdot\,}{B}B_q^m\,\cdot\,P_p^n
\contraction[1mm]{}{B}{_k^p B_j^k\,\cdot\,B_m^q}{V}B_k^p B_j^k\,\cdot\,B_r^sB_l^r P_s^l\,\cdot\,
\delta_{kp}\delta_{im}\delta_{jn}\,\cdot\,\delta_{iq}\,\cdot\,\delta_{lr}\,\cdot\,\delta_{ls},
\end{align}
\begin{align}
T(A_{10})_3[i,j,k,l]\,=\, B_{c\overline 3}^qH_{pm}^n
\contraction[1mm]{}{P}{_n^pB_q^m\,\cdot\,}{P}P_n^p
\contraction[2mm]{}{B}{_q^m\,\cdot\,P_p^nB_n^j B_j^k\,\cdot\,}{B}B_q^m\,\cdot\,P_p^n
\contraction[1mm]{}{B}{_n^j B_j^k\,\cdot\,B_m^q}{V}B_n^j B_j^k\,\cdot\,B_l^sB_s^r P_r^l\,\cdot\,
\delta_{kp}\delta_{im}\delta_{jn}\,\cdot\,\delta_{iq}\,\cdot\,\delta_{lr}\,\cdot\,\delta_{ls},
\end{align}
\begin{align}
T(A_{10})_4[i,j,k,l]\,=\, B_{c\overline 3}^qH_{pm}^n
\contraction[1mm]{}{P}{_n^pB_q^m\,\cdot\,}{P}P_n^p
\contraction[2mm]{}{B}{_q^m\,\cdot\,P_p^nB_n^j B_j^k\,\cdot\,}{B}B_q^m\,\cdot\,P_p^n
\contraction[1mm]{}{B}{_n^j B_j^k\,\cdot\,B_m^q}{V}B_n^j B_j^k\,\cdot\,B_r^sB_l^r P_s^l\,\cdot\,
\delta_{kp}\delta_{im}\delta_{jn}\,\cdot\,\delta_{iq}\,\cdot\,\delta_{lr}\,\cdot\,\delta_{ls}.
\end{align}
$A_2\to A_{11}$:
\begin{align}
T(A_{11})[i,j,k,l]\,=\,B_{c\overline 3}^qH_{pm}^n\contraction[1mm]{}{P}{_n^pB_q^m\,\cdot\,}{P}P_n^p
\contraction[2mm]{}{B}{_q^m\,\cdot\,P_p^nB_n^l B_l^k\,\cdot\,}{B}B_q^m\,\cdot\,P_p^n
\contraction[1mm]{}{B}{_n^l B_l^k\,\cdot\,B_m^q}{B}B_n^l B_l^k\,\cdot\,B_m^qB_l^m P_i^l\,\cdot\,
\delta_{kp}\delta_{jm}\delta_{jn}\,\cdot\,\delta_{iq},
\end{align}
\begin{align}
S(A_{11})_1[i,j,k,l]\,=\,B_{c\overline 3}^qH_{pm}^n\contraction[1mm]{}{P}{_n^pB_q^m\,\cdot\,}{P}P_n^p
\contraction[2mm]{}{B}{_q^m\,\cdot\,P_p^n}{B}B_q^m\,\cdot\,P_p^nB_m^q
\contraction[1mm]{}{B}{_q^p\,\cdot\,}{B}B_q^p\,\cdot\,B_k^i P_i^l B_l^k\,\cdot\,
\delta_{kp}\delta_{jm}\delta_{jn}\,\cdot\,\delta_{iq}.
\end{align}
$A_2\to A_{12}$:
\begin{align}
U(A_{12})[i,j,k,l]\,=\,B_{c\overline 3}^qH_{pm}^n\contraction[1mm]{}{P}{_n^pB_q^m\,\cdot\,}{P}P_n^p
\contraction[2mm]{}{B}{_q^m\,\cdot\,P_p^nV_n^k P_k^l\,\cdot\,}{B}B_q^m\,\cdot\,P_p^n
\contraction[1mm]{}{V}{_n^k P_k^l\,\cdot\,B_m^q}{V}V_n^k P_k^l\,\cdot\,B_m^qV_k^m B_i^k\,\cdot\,
\delta_{lp}\delta_{jm}\delta_{jn}\,\cdot\,\delta_{iq},
\end{align}
\begin{align}
S(A_{12})_1[i,j,k,l]\,=\,B_{c\overline 3}^qH_{pm}^n\contraction[1mm]{}{P}{_n^pB_q^m\,\cdot\,}{P}P_n^p
\contraction[2mm]{}{B}{_q^m\,\cdot\,P_p^n}{B}B_q^m\,\cdot\,P_p^nB_m^q
\contraction[1mm]{}{B}{_q^p\,\cdot\,}{B}B_q^p\,\cdot\,B_l^i P_k^l B_i^k\,\cdot\,
\delta_{lp}\delta_{jm}\delta_{jn}\,\cdot\,\delta_{iq}.
\end{align}
$A_2\to A_{14}$:
\begin{align}
S(A_{14})_1[i,j,k,l]\,=\,B_{c\overline 3}^qH_{pm}^n\contraction[1mm]{}{P}{_n^pB_q^m\,\cdot\,}{P}P_n^p
\contraction[2mm]{}{B}{_q^m\,\cdot\,P_p^n}{B}B_q^m\,\cdot\,P_p^nB_m^q
\contraction[1mm]{}{B}{_q^p\,\cdot\,}{B}B_n^m\,\cdot\,B_r^l P_k^r B_l^k\,\cdot\,
\delta_{ip}\delta_{jm}\delta_{jn}\,\cdot\,\delta_{iq}\,\cdot\,\delta_{lr},
\end{align}
\begin{align}
S(A_{14})_2[i,j,k,l]\,=\,B_{c\overline 3}^qH_{pm}^n\contraction[1mm]{}{P}{_n^pB_q^m\,\cdot\,}{P}P_n^p
\contraction[2mm]{}{B}{_q^m\,\cdot\,P_p^n}{B}B_q^m\,\cdot\,P_p^nB_m^q
\contraction[1mm]{}{B}{_q^p\,\cdot\,}{B}B_n^m\,\cdot\,B_k^r P_r^l B_l^k\,\cdot\,
\delta_{ip}\delta_{jm}\delta_{jn}\,\cdot\,\delta_{iq}\,\cdot\,\delta_{kr},
\end{align}
\begin{align}
S(A_{14})_3[i,j,k,l]\,=\,B_{c\overline 3}^qH_{pm}^n\contraction[1mm]{}{P}{_n^pB_q^m\,\cdot\,}{P}P_n^p
\contraction[2mm]{}{B}{_q^m\,\cdot\,P_p^n}{B}B_q^m\,\cdot\,P_p^nB_m^q
\contraction[1mm]{}{B}{_q^p\,\cdot\,}{B}B_q^p\,\cdot\,B_r^l P_k^r B_l^k\,\cdot\,
\delta_{ip}\delta_{jm}\delta_{jn}\,\cdot\,\delta_{iq}\,\cdot\,\delta_{lr},
\end{align}
\begin{align}
S(A_{14})_4[i,j,k,l]\,=\,B_{c\overline 3}^qH_{pm}^n\contraction[1mm]{}{P}{_n^pB_q^m\,\cdot\,}{P}P_n^p
\contraction[2mm]{}{B}{_q^m\,\cdot\,P_p^n}{B}B_q^m\,\cdot\,P_p^nB_m^q
\contraction[1mm]{}{B}{_q^p\,\cdot\,}{B}B_q^p\,\cdot\,B_k^r P_r^l B_l^k\,\cdot\,
\delta_{ip}\delta_{jm}\delta_{jn}\,\cdot\,\delta_{iq}\,\cdot\,\delta_{kr}.
\end{align}
$A_2\to A_{19}$:
\begin{align}
U(A_{19})_1[i,j,k,l]\,=\,B_{c\overline 3}^qH_{pm}^n\contraction[1mm]{}{P}{_n^pB_q^m\,\cdot\,}{P}P_n^p
\contraction[2mm]{}{B}{_q^m\,\cdot\,P_p^nV_l^p P_j^l\,\cdot\,}{B}B_q^m\,\cdot\,P_p^n
\contraction[1mm]{}{V}{_l^p P_j^l\,\cdot\,B_k^r}{V}V_k^p P_j^k\,\cdot\,B_r^sV_l^r B_s^l\,\cdot\,
\delta_{kp}\delta_{im}\delta_{jn}\,\cdot\,\delta_{iq}\,\cdot\,\delta_{lr}\,\cdot\,\delta_{ls},
\end{align}
\begin{align}
U(A_{19})_2[i,j,k,l]\,=\,B_{c\overline 3}^qH_{pm}^n\contraction[1mm]{}{P}{_n^pB_q^m\,\cdot\,}{P}P_n^p
\contraction[2mm]{}{B}{_q^m\,\cdot\,P_p^nV_l^p P_j^l\,\cdot\,}{B}B_q^m\,\cdot\,P_p^n
\contraction[1mm]{}{V}{_l^p P_j^l\,\cdot\,B_k^r}{V}V_k^p P_j^k\,\cdot\,B_l^rV_r^s B_s^l\,\cdot\,
\delta_{kp}\delta_{im}\delta_{jn}\,\cdot\,\delta_{iq}\,\cdot\,\delta_{lr}\,\cdot\,\delta_{ls},
\end{align}
\begin{align}
U(A_{19})_3[i,j,k,l]\,=\,B_{c\overline 3}^qH_{pm}^n\contraction[1mm]{}{P}{_n^pB_q^m\,\cdot\,}{P}P_n^p
\contraction[2mm]{}{B}{_q^m\,\cdot\,P_p^nV_l^p P_j^l\,\cdot\,}{B}B_q^m\,\cdot\,P_p^n
\contraction[1mm]{}{V}{_l^p P_j^l\,\cdot\,B_k^r}{V}V_n^j P_j^k\,\cdot\,B_r^sV_l^r B_s^l\,\cdot\,
\delta_{kp}\delta_{im}\delta_{jn}\,\cdot\,\delta_{iq}\,\cdot\,\delta_{lr}\,\cdot\,\delta_{ls},
\end{align}
\begin{align}
U(A_{19})_4[i,j,k,l]\,=\,B_{c\overline 3}^qH_{pm}^n\contraction[1mm]{}{P}{_n^pB_q^m\,\cdot\,}{P}P_n^p
\contraction[2mm]{}{B}{_q^m\,\cdot\,P_p^nV_l^p P_j^l\,\cdot\,}{B}B_q^m\,\cdot\,P_p^n
\contraction[1mm]{}{V}{_l^p P_j^l\,\cdot\,B_k^r}{V}V_n^j P_j^k\,\cdot\,B_l^rV_r^s B_s^l\,\cdot\,
\delta_{kp}\delta_{im}\delta_{jn}\,\cdot\,\delta_{iq}\,\cdot\,\delta_{lr}\,\cdot\,\delta_{ls},
\end{align}
\begin{align}
U(A_{19})_5[i,j,k,l]\,=\,B_{c\overline 3}^qH_{pm}^n\contraction[1mm]{}{P}{_n^pB_q^m\,\cdot\,}{P}P_n^p
\contraction[2mm]{}{B}{_q^m\,\cdot\,P_p^nV_l^p P_j^l\,\cdot\,}{B}B_q^m\,\cdot\,P_p^n
\contraction[1mm]{}{V}{_l^p P_j^l\,\cdot\,B_k^r}{V}V_k^p P_j^k\,\cdot\,B_l^rV_s^s B_r^l\,\cdot\,
\delta_{kp}\delta_{im}\delta_{jn}\,\cdot\,\delta_{iq}\,\cdot\,\delta_{lr},
\end{align}
\begin{align}\label{lx}
U(A_{19})_6[i,j,k,l]\,=\,B_{c\overline 3}^qH_{pm}^n\contraction[1mm]{}{P}{_n^pB_q^m\,\cdot\,}{P}P_n^p
\contraction[2mm]{}{B}{_q^m\,\cdot\,P_p^nV_l^p P_j^l\,\cdot\,}{B}B_q^m\,\cdot\,P_p^n
\contraction[1mm]{}{V}{_l^p P_j^l\,\cdot\,B_k^r}{V}V_n^j P_j^k\,\cdot\,B_l^rV_s^s B_r^l\,\cdot\,
\delta_{kp}\delta_{im}\delta_{jn}\,\cdot\,\delta_{iq}\,\cdot\,\delta_{lr}.
\end{align}

\section{Test of isospin symmetry}\label{exam}

\subsection{The $\Lambda_c\to  \Sigma\pi$ system}

The decay amplitude for the $\Lambda^+_c\to \Sigma^0\pi^+$ mode is $\lambda_1(A_7-A_8)/\sqrt{2}$.
The rescattering amplitudes contributing to the $\Lambda^+_c\to \Sigma^0\pi^+$ decay include
\begin{align}\label{xx1}
U(A_7)_1[s,d,u,u]\,&=\, -\frac{1}{6}\Delta_{\alpha^+,\gamma^-}(\Lambda^+_c,\pi^+,\Lambda^0,\rho^0,\pi^+,\Sigma^0)
-\frac{1}{6}\Delta_{\alpha^+,\gamma^-}(\Lambda^+_c,\pi^+,\Lambda^0,\omega,\pi^+,\Sigma^0),
\nonumber\\
U(A_7)_2[s,d,u,u]\,&=\, -\frac{1}{6}\Delta_{\alpha^+,\gamma^+}(\Lambda^+_c,\pi^+,\Lambda^0,\rho^0,\pi^+,\Sigma^0)
-\frac{1}{6}\Delta_{\alpha^+,\gamma^+}(\Lambda^+_c,\pi^+,\Lambda^0,\omega,\pi^+,\Sigma^0),
\nonumber\\
T(A_7)_1[s,d,u,u]\,&=\, -\frac{1}{3}\Delta_{\beta^+,\beta^+}(\Lambda^+_c,\pi^+,\Lambda^0,\Sigma^-,\Sigma^0,\pi^+),
\nonumber\\
T(A_7)_2[s,d,u,u]\,&=\, -\frac{1}{3}\Delta_{\beta^+,\beta^-}(\Lambda^+_c,\pi^+,\Lambda^0,\Sigma^-,\Sigma^0,\pi^+),
\nonumber\\
S(A_7)_1[s,d,u,u]\,&=\, -\frac{1}{3}\Theta_{\beta^-,\beta^-}(\Lambda^+_c,\pi^+,\Lambda^0,\Sigma^+,\pi^+,\Sigma^0),
\nonumber\\
S(A_7)_2[s,d,u,u]\,&=\, -\frac{1}{3}\Theta_{\beta^+,\beta^-}(\Lambda^+_c,\pi^+,\Lambda^0,\Sigma^+,\pi^+,\Sigma^0),
\nonumber\\
U(A_8)_1[s,d,d,u]\,&=\, \frac{1}{6}\Delta_{\alpha^+,\gamma^-}(\Lambda^+_c,\pi^+,\Lambda^0,\rho^0,\pi^+,\Sigma^0)
+\frac{1}{6}\Delta_{\alpha^+,\gamma^-}(\Lambda^+_c,\pi^+,\Lambda^0,\omega,\pi^+,\Sigma^0),
\nonumber\\
U(A_8)_2[s,d,d,u]\,&=\, \frac{1}{6}\Delta_{\alpha^+,\gamma^+}(\Lambda^+_c,\pi^+,\Lambda^0,\rho^0,\pi^+,\Sigma^0)
+\frac{1}{6}\Delta_{\alpha^+,\gamma^+}(\Lambda^+_c,\pi^+,\Lambda^0,\omega,\pi^+,\Sigma^0),
\nonumber\\
T(A_8)_1[s,d,d,u]\,&=\, -\frac{1}{3}\Delta_{\beta^-,\beta^+}(\Lambda^+_c,\pi^+,\Lambda^0,\Sigma^-,\Sigma^0,\pi^+),
\nonumber\\
T(A_8)_2[s,d,d,u]\,&=\, -\frac{1}{3}\Delta_{\beta^-,\beta^-}(\Lambda^+_c,\pi^+,\Lambda^0,
\Sigma^-,\Sigma^0,\pi^+),\nonumber\\
S(A_8)_1[s,d,d,u]\,&=\, -\frac{1}{3}\Theta_{\beta^-,\beta^+}(\Lambda^+_c,\pi^+,\Lambda^0,\Sigma^+,\pi^+,\Sigma^0),
\nonumber\\
S(A_8)_2[s,d,d,u]\,&=\, -\frac{1}{3}\Theta_{\beta^+,\beta^+}(\Lambda^+_c,\pi^+,\Lambda^0,\Sigma^+,\pi^+,
\Sigma^0),\nonumber\\
U(A_{19})_1[s,d,u,d]\,&=\, \frac{1}{6}\Delta_{\alpha^+,\gamma^-}(\Lambda^+_c,\pi^+,\Lambda^0,\rho^0,\pi^+,\Sigma^0)
-\frac{1}{6}\Delta_{\alpha^+,\gamma^-}(\Lambda^+_c,\pi^+,\Lambda^0,\omega,\pi^+,\Sigma^0),
\nonumber\\
U(A_{19})_2[s,d,u,d]\,&=\, \frac{1}{6}\Delta_{\alpha^+,\gamma^+}(\Lambda^+_c,\pi^+,\Lambda^0,\rho^0,\pi^+,\Sigma^0)
-\frac{1}{6}\Delta_{\alpha^+,\gamma^+}(\Lambda^+_c,\pi^+,\Lambda^0,\omega,\pi^+,\Sigma^0),
\nonumber\\
U(A_{19})_3[s,d,u,u]\,&=\, -\frac{1}{6}\Delta_{\alpha^+,\gamma^-}(\Lambda^+_c,\pi^+,\Lambda^0,\rho^0,\pi^+,\Sigma^0)
+\frac{1}{6}\Delta_{\alpha^+,\gamma^-}(\Lambda^+_c,\pi^+,\Lambda^0,\omega,\pi^+,\Sigma^0),
\nonumber\\
U(A_{19})_4[s,d,u,u]\,&=\, -\frac{1}{6}\Delta_{\alpha^+,\gamma^+}(\Lambda^+_c,\pi^+,\Lambda^0,\rho^0,\pi^+,\Sigma^0)
+\frac{1}{6}\Delta_{\alpha^+,\gamma^+}(\Lambda^+_c,\pi^+,\Lambda^0,\omega,\pi^+,\Sigma^0),
\nonumber\\
U(A_{19})_5[s,d,u,u]\,&=\,
-\frac{1}{3}\Delta_{\alpha^+,\gamma^0}(\Lambda^+_c,\pi^+,\Lambda^0,\omega,\pi^+,\Sigma^0),
\nonumber\\
U(A_{19})_6[s,d,u,u]\,&=\,
\frac{1}{3}\Delta_{\alpha^+,\gamma^0}(\Lambda^+_c,\pi^+,\Lambda^0,\omega,\pi^+,\Sigma^0),
\nonumber\\
U(A_{19})_5[s,d,u,d]\,&=\,
-\frac{1}{3}\Delta_{\alpha^+,\gamma^0}(\Lambda^+_c,\pi^+,\Lambda^0,\omega,\pi^+,\Sigma^0),
\nonumber\\
U(A_{19})_6[s,d,u,d]\,&=\,
\frac{1}{3}\Delta_{\alpha^+,\gamma^0}(\Lambda^+_c,\pi^+,\Lambda^0,\omega,\pi^+,\Sigma^0).
\end{align}
Summing all long-distance contributions, we have
\begin{align}\label{xx3}
  \mathcal{A}_L(\Lambda^+_c\to \Sigma^0\pi^+)&=-\frac{\sqrt{2}}{3}\lambda_1\Delta_{\alpha^+,(\gamma^++\gamma^-)}
  (\Lambda^+_c,\pi^+,\Lambda^0,\rho^0,\pi^+,\Sigma^0)\nonumber\\
  &~~-\frac{\sqrt{2}}{6}\lambda_1\Delta_{(\beta^+-\beta^-),(\beta^++\beta^-)}
  (\Lambda^+_c,\pi^+,\Lambda^0,\Sigma^-,\Sigma^0,\pi^+)
  \nonumber\\
  &~~+\frac{\sqrt{2}}{6}\lambda_1\Theta_{(\beta^++\beta^-),(\beta^+-\beta^-)}
  (\Lambda^+_c,\pi^+,\Lambda^0,\Sigma^+,\pi^+,\Sigma^0).
\end{align}
The decay amplitude for the $\Lambda^+_c\to \Sigma^+\pi^0$ mode is $-\lambda_1(A_7-A_8)/\sqrt{2}$.
The rescattering amplitudes contributing to the $\Lambda^+_c\to \Sigma^+\pi^0$ decay include
\begin{align}\label{xx2}
U(A_7)_1[s,d,u,d]\,&=\, -\frac{1}{3}\Delta_{\alpha^+,\gamma^-}(\Lambda^+_c,\pi^+,\Lambda^0,\rho^+,\pi^0,\Sigma^+),
\nonumber\\
U(A_7)_2[s,d,u,d]\,&=\, -\frac{1}{3}\Delta_{\alpha^+,\gamma^+}(\Lambda^+_c,\pi^+,\Lambda^0,\rho^+,\pi^0,\Sigma^+),
\nonumber\\
T(A_7)_1[s,d,u,d]\,&=\, -\frac{1}{6}\Delta_{\beta^+,\beta^+}(\Lambda^+_c,\pi^+,\Lambda^0,\Sigma^0,\Sigma^+,\pi^0)
-\frac{1}{18}\Delta_{\beta^+,\beta^+}(\Lambda^+_c,\pi^+,\Lambda^0,\Lambda^0,\Sigma^+,\pi^0),
\nonumber\\
T(A_7)_2[s,d,u,d]\,&=\, -\frac{1}{6}\Delta_{\beta^+,\beta^-}(\Lambda^+_c,\pi^+,\Lambda^0,\Sigma^0,\Sigma^+,\pi^0)
-\frac{1}{18}\Delta_{\beta^+,\beta^-}(\Lambda^+_c,\pi^+,\Lambda^0,\Lambda^0,\Sigma^+,\pi^0),
\nonumber\\
S(A_7)_1[s,d,u,d]\,&=\, -\frac{1}{3}\Theta_{\beta^-,\beta^-}(\Lambda^+_c,\pi^+,\Lambda^0,\Sigma^+,\pi^0,\Sigma^+),
\nonumber\\
S(A_7)_2[s,d,u,d]\,&=\, -\frac{1}{3}\Theta_{\beta^+,\beta^-}(\Lambda^+_c,\pi^+,\Lambda^0,\Sigma^+,\pi^0,\Sigma^+),
\nonumber\\
U(A_8)_1[s,d,u,u]\,&=\, \frac{1}{3}\Delta_{\alpha^+,\gamma^-}(\Lambda^+_c,\pi^+,\Lambda^0,\rho^+,\pi^0,\Sigma^+),
\nonumber\\
U(A_8)_2[s,d,u,u]\,&=\, \frac{1}{3}\Delta_{\alpha^+,\gamma^+}(\Lambda^+_c,\pi^+,\Lambda^0,\rho^+,\pi^0,\Sigma^+),
\nonumber\\
T(A_8)_1[s,d,u,u]\,&=\, -\frac{1}{6}\Delta_{\beta^-,\beta^+}(\Lambda^+_c,\pi^+,\Lambda^0,\Sigma^0,\Sigma^+,\pi^0)
-\frac{1}{18}\Delta_{\beta^-,\beta^+}(\Lambda^+_c,\pi^+,\Lambda^0,\Lambda^0,\Sigma^+,\pi^0),
\nonumber\\
T(A_8)_2[s,d,u,u]\,&=\, -\frac{1}{6}\Delta_{\beta^-,\beta^-}(\Lambda^+_c,\pi^+,\Lambda^0,\Sigma^0,\Sigma^+,\pi^0)
-\frac{1}{18}\Delta_{\beta^-,\beta^-}(\Lambda^+_c,\pi^+,\Lambda^0,\Lambda^0,\Sigma^+,\pi^0),
\nonumber\\
S(A_8)_1[s,d,u,u]\,&=\, -\frac{1}{3}\Theta_{\beta^-,\beta^+}(\Lambda^+_c,\pi^+,\Lambda^0,\Sigma^+,\pi^0,\Sigma^+),
\nonumber\\
S(A_8)_2[s,d,u,u]\,&=\, -\frac{1}{3}\Theta_{\beta^+,\beta^+}(\Lambda^+_c,\pi^+,\Lambda^0,\Sigma^+,\pi^0,
\Sigma^+),\nonumber\\
T(A_{10})_1[s,d,u,d]\,&=\, \frac{1}{6}\Delta_{\beta^-,\beta^+}(\Lambda^+_c,\pi^+,\Lambda^0,\Sigma^0,\Sigma^+,\pi^0)
-\frac{1}{18}\Delta_{\beta^-,\beta^+}(\Lambda^+_c,\pi^+,\Lambda^0,\Lambda^0,\Sigma^+,\pi^0),
\nonumber\\
T(A_{10})_2[s,d,u,d]\,&=\, \frac{1}{6}\Delta_{\beta^-,\beta^-}(\Lambda^+_c,\pi^+,\Lambda^0,\Sigma^0,\Sigma^+,\pi^0)
-\frac{1}{18}\Delta_{\beta^-,\beta^-}(\Lambda^+_c,\pi^+,\Lambda^0,\Lambda^0,\Sigma^+,\pi^0),
\nonumber\\
T(A_{10})_3[s,d,u,u]\,&=\, \frac{1}{6}\Delta_{\beta^+,\beta^+}(\Lambda^+_c,\pi^+,\Lambda^0,\Sigma^0,\Sigma^+,\pi^0)
-\frac{1}{18}\Delta_{\beta^+,\beta^+}(\Lambda^+_c,\pi^+,\Lambda^0,\Lambda^0,\Sigma^+,\pi^0),
\nonumber\\
T(A_{10})_{4}[s,d,u,u]\,&=\, \frac{1}{6}\Delta_{\beta^+,\beta^-}(\Lambda^+_c,\pi^+,\Lambda^0,\Sigma^0,\Sigma^+,\pi^0)
-\frac{1}{18}\Delta_{\beta^+,\beta^-}(\Lambda^+_c,\pi^+,\Lambda^0,\Lambda^0,\Sigma^+,
\pi^0).
\end{align}
Summing all long-distance contributions, we have
\begin{align}\label{xx4}
  \mathcal{A}_L(\Lambda^+_c\to \Sigma^+\pi^0)&=\frac{\sqrt{2}}{3}\lambda_1\Delta_{\alpha^+,(\gamma^++\gamma^-)}
  (\Lambda^+_c,\pi^+,\Lambda^0,\rho^+,\pi^0,\Sigma^+)\nonumber\\
  &~~+\frac{\sqrt{2}}{6}\lambda_1\Delta_{(\beta^+-\beta^-),(\beta^++\beta^-)}
  (\Lambda^+_c,\pi^+,\Lambda^0,\Sigma^0,\Sigma^+,\pi^0)  \nonumber\\
  &~~-\frac{\sqrt{2}}{6}\lambda_1\Theta_{(\beta^++\beta^-),(\beta^+-\beta^-)}
  (\Lambda^+_c,\pi^+,\Lambda^0,\Sigma^+,\pi^0,\Sigma^+).
\end{align}
Comparing Eqs.~\eqref{xx3} and \eqref{xx4}, one finds that the isospin sum rule
\begin{align}
\mathcal{A}(\Lambda^+_c\to \Sigma^+\pi^0)+\mathcal{A}(\Lambda^+_c\to \Sigma^0\pi^+)=0
\end{align}
 is satisfied.

\subsection{The $\Xi_c\to \Sigma K $ system (1)}

The decay amplitude for the $\Xi_{c}^{0}\to \Sigma^+K^-$ mode is $\lambda_1(A_1+A_6)$.
The rescattering amplitudes contributing to the $\Xi_{c}^{0}\to \Sigma^+K^-$ decay include
\begin{align}
  T(A_6)[u,d,u,s]\,&=\, \frac{1}{2}\Delta_{\beta^-,\beta^+}(\Xi^0_c,\pi^+,\Xi^-,\Sigma^0,\Sigma^+,K^-)
+\frac{1}{6}\Delta_{\beta^-,\beta^+}(\Xi^0_c,\pi^+,\Xi^-,\Lambda^0,\Sigma^+,K^-),
\nonumber\\
  T(A_1)_1[u,d,u,s]\,&=\,
-\frac{1}{3}\Delta_{\beta^-,\beta^-}(\Xi^0_c,\pi^+,\Xi^-,\Lambda^0,\Sigma^+,K^-),
\nonumber\\
  T(A_1)_3[u,d,u,s]\,&=\,
-\frac{1}{3}\Delta_{\beta^+,\beta^-}(\Xi^0_c,\pi^+,\Xi^-,\Lambda^0,\Sigma^+,K^-),
\nonumber\\
  T(A_1)_4[u,d,u,s]\,&=\,-\frac{1}{2}\Delta_{\beta^+,\beta^+}
  (\Xi^0_c,\pi^+,\Xi^-,\Sigma^0,\Sigma^+,K^-)
+\frac{1}{6}\Delta_{\beta^+,\beta^+}(\Xi^0_c,\pi^+,\Xi^-,\Lambda^0,\Sigma^+,K^-),
\nonumber\\
S(A_6)[u,d,u,s]\,&=\,\Theta_{\beta^-,\beta^+}(\Xi^0_c,\pi^+,\Xi^-,\Xi^0,K^-,\Sigma^+).
\end{align}
Summing all the long-distance contributions, we have
\begin{align}\label{xx5}
  \mathcal{A}_L(\Xi_{c}^{0}\to \Sigma^+K^-)&=-\frac{1}{2}\lambda_1\Delta_{(\beta^+-\beta^-),\beta^+}
  (\Xi^0_c,\pi^+,\Xi^-,\Sigma^0,\Sigma^+,K^-)\nonumber\\
  &~~+\frac{1}{6}\lambda_1\Delta_{(\beta^++\beta^-),(\beta^+-2\beta^-)}
  (\Xi^0_c,\pi^+,\Xi^-,\Lambda^0,\Sigma^+,K^-)\nonumber\\
  &~~+\lambda_1\Theta_{\beta^-,\beta^+}(\Xi^0_c,\pi^+,\Xi^-,\Xi^0,K^-,\Sigma^+).
\end{align}
The decay amplitude for the $\Xi_{c}^{0}\to \Sigma^0\overline{K}^0$ mode is $\lambda_1(A_4-A_6)/\sqrt{2}$.
The rescattering amplitudes contributing to the $\Xi_{c}^{0}\to \Sigma^0\overline{K}^0$ decay include
\begin{align}
  T(A_6)[u,d,u,s]\,&=\, \Delta_{\beta^-,\beta^+}(\Xi^0_c,\pi^+,\Xi^-,\Sigma^-,\Sigma^0,\overline K^0),
\nonumber\\
  U(A_4)[u,d,u,s]\,&=\, \Delta_{\alpha^+,\gamma^+}(\Xi^0_c,\pi^+,\Xi^-,K^{*+},\overline K^0,\Sigma^0),
\nonumber\\
  T(A_4)[u,d,u,s]\,&=\, \Delta_{\beta^+,\beta^+}(\Xi^0_c,\pi^+,\Xi^-,\Sigma^-,\Sigma^0,\overline K^0),
\nonumber\\
S(A_6)[u,d,u,s]\,&=\,\Theta_{\beta^-,\beta^+}(\Xi^0_c,\pi^+,\Xi^-,\Xi^0,\overline K^0,\Sigma^0).
\end{align}
Summing all long-distance contributions, we have
\begin{align}\label{xx6}
  \mathcal{A}_L(\Xi_{c}^{0}\to \Sigma^0\overline K^0)&=\frac{1}{\sqrt{2}}\lambda_1\Delta_{(\beta^+-\beta^-),\beta^+}
  (\Xi^0_c,\pi^+,\Xi^-,\Sigma^-,\Sigma^0,\overline K^0)\nonumber\\
  &~~+\frac{1}{\sqrt{2}}\lambda_1\Delta_{\alpha^+,\gamma^+}(\Xi^0_c,\pi^+,\Xi^-,K^{*+},\overline K^0,\Sigma^0)\nonumber\\
  &~~-\frac{1}{\sqrt{2}}\lambda_1\Theta_{\beta^-,\beta^+}(\Xi^0_c,\pi^+,\Xi^-,\Xi^0,\overline K^0,\Sigma^0).
\end{align}
The decay amplitude for the $\Xi_{c}^{+}\to \Sigma^+\overline{K}^0$ mode is $-\lambda_1(A_1+A_4)$.
The rescattering amplitudes contributing to the $\Xi_{c}^{+}\to \Sigma^+\overline{K}^0$ decay include
\begin{align}
  U(A_4)[d,d,u,s]\,&=\, \Delta_{\alpha^+,\gamma^+}(\Xi^+_c,\pi^+,\Xi^0,K^{*+},\overline K^0,\Sigma^+),
\nonumber\\
  T(A_4)[d,d,u,s]\,&=\, \frac{1}{2}\Delta_{\beta^+,\beta^+}(\Xi^+_c,\pi^+,\Xi^0,\Sigma^0,\Sigma^+,\overline K^0)+\frac{1}{6}\Delta_{\beta^+,\beta^+}(\Xi^+_c,\pi^+,\Xi^0,\Lambda^0,\Sigma^+,\overline K^0),
\nonumber\\
  T(A_1)_1[d,d,u,s]\,&=\,
-\frac{1}{3}\Delta_{\beta^-,\beta^-}(\Xi^+_c,\pi^+,\Xi^0,\Lambda^0,\Sigma^+,\overline K^0),  \nonumber\\
  T(A_1)_2[d,d,u,s]\,&=\,
-\frac{1}{2}\Delta_{\beta^-,\beta^+}(\Xi^+_c,\pi^+,\Xi^0,\Sigma^0,\Sigma^+,\overline K^0)+\frac{1}{6}\Delta_{\beta^-,\beta^+}(\Xi^+_c,\pi^+,\Xi^0,\Lambda^0,\Sigma^+,\overline K^0),
\nonumber\\
  T(A_1)_3[d,d,u,s]\,&=\,
-\frac{1}{3}\Delta_{\beta^+,\beta^-}(\Xi^+_c,\pi^+,\Xi^0,\Lambda^0,\Sigma^+,\overline K^0).
\end{align}
Summing all long-distance contributions, we have
\begin{align}\label{xx7}
  \mathcal{A}_L(\Xi_{c}^{+}\to \Sigma^+\overline K^0)&=-\frac{1}{2}\lambda_1\Delta_{(\beta^+-\beta^-),\beta^+}
  (\Xi^+_c,\pi^+,\Xi^0,\Sigma^0,\Sigma^+,\overline K^0)\nonumber\\
  &~~-\frac{1}{6}\lambda_1\Delta_{(\beta^++\beta^-),(\beta^+-2\beta^-)}
  (\Xi^0_c,\pi^+,\Xi^0,\Lambda^0,\Sigma^+,\overline K^0)\nonumber\\
  &~~-\lambda_1\Delta_{\alpha^+,\gamma^+}(\Xi^+_c,\pi^+,\Xi^0,K^{*+},\overline K^0,\Sigma^+).
\end{align}
One can verify that the isospin sum rule
\begin{align}
 \mathcal{A}(\Xi^0_c\to \Sigma^+K^-)+\sqrt{2}\,\mathcal{A}(\Xi^0_c\to \Sigma^0\overline K^0)+\mathcal{A}(\Xi^+_c\to \Sigma^+\overline K^0)=0
\end{align}
is satisfied.

\subsection{The $\Xi_c\to \Sigma K $ system (2)}

The decay amplitude for the $\Xi_{c}^{0}\to \Sigma^-K^+$ mode is $\lambda_2(A_2+A_5)$.
The rescattering amplitudes contributing to the $\Xi_{c}^{0}\to \Sigma^-K^+$ decay include
\begin{align}
  U(A_2)_1[u,s,d,u]\,&=\, -\frac{1}{2}\Delta_{\alpha^+,\gamma^+}(\Xi^0_c,K^+,\Sigma^-,\rho^0, K^+,\Sigma^-)+\frac{1}{2}\Delta_{\alpha^+,\gamma^+}(\Xi^0_c,K^+,\Sigma^-,\omega, K^+,\Sigma^-),
\nonumber\\
  U(A_2)_5[u,s,d,u]\,&=\, \Delta_{\alpha^+,\gamma^0}(\Xi^0_c,K^+,\Sigma^-,\omega, K^+,\Sigma^-),
\nonumber\\
  U(A_2)_6[u,s,d,u]\,&=\, -\Delta_{\alpha^+,\gamma^0}(\Xi^0_c,K^+,\Sigma^-,\phi, K^+,\Sigma^-),
\nonumber\\
  U(A_5)[u,s,d,u]\,&=\, \frac{1}{2}\Delta_{\alpha^+,\gamma^-}(\Xi^0_c,K^+,\Sigma^-,\rho^0, K^+,\Sigma^-)+\frac{1}{2}\Delta_{\alpha^+,\gamma^-}(\Xi^0_c,K^+,\Sigma^-,\omega, K^+,\Sigma^-),
\nonumber\\
  S(A_5)[u,s,d,u]\,&=\, \Theta_{\beta^-,\beta^-}(\Xi^0_c,K^+,\Sigma^-,n, K^+,\Sigma^-).
\end{align}
Summing all long-distance contributions, we have
\begin{align}\label{xx8}
  \mathcal{A}_L(\Xi_{c}^{0}\to \Sigma^-K^+)&=-\frac{1}{2}\lambda_2\Delta_{\alpha^+,(\gamma^+-\gamma^-)}(\Xi^0_c,K^+,\Sigma^-,\rho^0, K^+,\Sigma^-)\nonumber\\&~~+\frac{1}{2}\lambda_2\Delta_{\alpha^+,(\gamma^+-\gamma^-)}(\Xi^0_c,K^+,\Sigma^-,\omega, K^+,\Sigma^-)\nonumber\\&~~+\lambda_2\Delta_{\alpha^+,\gamma^-}(\Xi^0_c,K^+,\Sigma^-,\phi, K^+,\Sigma^-)\nonumber\\&~~+\lambda_2\Theta_{\beta^-,\beta^-}(\Xi^0_c,K^+,\Sigma^-,n, K^+,\Sigma^-).
\end{align}
The decay amplitude for the $\Xi_{c}^{0}\to \Sigma^0K^0$ mode is $\lambda_2(A_4-A_5)/\sqrt{2}$.
The rescattering amplitudes contributing to the $\Xi_{c}^{0}\to \Sigma^0K^0$ decay include
\begin{align}
  U(A_4)[u,s,u,d]\,&=\, \Delta_{\alpha^+,\gamma^+}(\Xi^0_c,K^+,\Sigma^-,\rho^+, K^0,\Sigma^0),
\nonumber\\
  T(A_4)[u,s,u,d]\,&=\, \Delta_{\beta^+,\beta^+}(\Xi^0_c,K^+,\Sigma^-,\Xi^-, \Sigma^0, K^0),
\nonumber\\
  U(A_5)[u,s,d,d]\,&=\, \Delta_{\alpha^+,\gamma^-}(\Xi^0_c,K^+,\Sigma^-,\rho^+, K^0,\Sigma^0),
\nonumber\\
  S(A_5)[u,s,d,d]\,&=\, \Theta_{\beta^-,\beta^-}(\Xi^0_c,K^+,\Sigma^-,n, K^0,\Sigma^0).
\end{align}
Summing all long-distance contributions, we have
\begin{align}\label{xx9}
  \mathcal{A}_L(\Xi_{c}^{0}\to \Sigma^0K^0)&=\frac{1}{\sqrt{2}}\lambda_2\Delta_{\alpha^+,(\gamma^+-\gamma^-)}(\Xi^0_c,K^+,\Sigma^-,\rho^+, K^0,\Sigma^0)\nonumber\\&~~+\frac{1}{\sqrt{2}}\lambda_2\Delta_{\beta^+,\beta^+}(\Xi^0_c,K^+,\Sigma^-,\Xi^-, \Sigma^0, K^0)\nonumber\\&~~-\frac{1}{\sqrt{2}}\lambda_2\Theta_{\beta^-,\beta^-}(\Xi^0_c,K^+,\Sigma^-,n, K^0,\Sigma^0).
\end{align}
The decay amplitude for the $\Xi_{c}^{+}\to \Sigma^0K^+$ mode is $\lambda_2(A_2-A_7)/\sqrt{2}$.
The rescattering amplitudes contributing to the $\Xi_{c}^{+}\to \Sigma^0K^+$ decay include
\begin{align}
  U(A_2)_1[d,s,d,u]\,&=\, -\frac{1}{4}\Delta_{\alpha^+,\gamma^+}(\Xi^+_c,K^+,\Sigma^0,\rho^0, K^+,\Sigma^0)+\frac{1}{4}\Delta_{\alpha^+,\gamma^+}(\Xi^+_c,K^+,\Sigma^0,\omega, K^+,\Sigma^0)\nonumber\\&~~~~-\frac{1}{12}\Delta_{\alpha^+,\gamma^+}
  (\Xi^+_c,K^+,\Lambda^0,\rho^0, K^+,\Sigma^0)+\frac{1}{12}\Delta_{\alpha^+,\gamma^+}(\Xi^+_c,K^+,\Lambda^0,\omega, K^+,\Sigma^0),
\nonumber\\
  U(A_2)_2[d,s,d,u]\,&=\, -\frac{1}{4}\Delta_{\alpha^+,\gamma^-}(\Xi^+_c,K^+,\Sigma^0,\rho^0, K^+,\Sigma^0)+\frac{1}{4}\Delta_{\alpha^+,\gamma^-}(\Xi^+_c,K^+,\Sigma^0,\omega, K^+,\Sigma^0)\nonumber\\&~~~~-\frac{1}{12}\Delta_{\alpha^+,\gamma^-}
  (\Xi^+_c,K^+,\Lambda^0,\rho^0, K^+,\Sigma^0)+\frac{1}{12}\Delta_{\alpha^+,\gamma^-}(\Xi^+_c,K^+,\Lambda^0,\omega, K^+,\Sigma^0),
\nonumber\\
  U(A_2)_5[d,s,d,u]\,&=\, \frac{1}{2}\Delta_{\alpha^+,\gamma^0}(\Xi^+_c,K^+,\Sigma^0,\omega, K^+,\Sigma^0)+\frac{1}{6}\Delta_{\alpha^+,\gamma^0}(\Xi^+_c,K^+,\Lambda^0,\omega, K^+,\Sigma^0),
\nonumber\\
  U(A_2)_6[d,s,d,u]\,&=\, -\frac{1}{2}\Delta_{\alpha^+,\gamma^0}(\Xi^+_c,K^+,\Sigma^0,\phi, K^+,\Sigma^0)-\frac{1}{6}\Delta_{\alpha^+,\gamma^0}(\Xi^+_c,K^+,\Lambda^0,\phi, K^+,\Sigma^0),
  \nonumber\\
   U(A_7)_1[d,s,u,u]\,&=\, -\frac{1}{4}\Delta_{\alpha^+,\gamma^-}(\Xi^+_c,K^+,\Sigma^0,\rho^0, K^+,\Sigma^0)-\frac{1}{4}\Delta_{\alpha^+,\gamma^-}(\Xi^+_c,K^+,\Sigma^0,\omega,
    K^+,\Sigma^0)\nonumber\\&
   ~~~~+\frac{1}{12}\Delta_{\alpha^+,\gamma^-}(\Xi^+_c,K^+,\Lambda^0,\rho^0, K^+,\Sigma^0)+\frac{1}{12}\Delta_{\alpha^+,\gamma^-}(\Xi^+_c,K^+,\Lambda^0,
   \omega, K^+,\Sigma^0),
   \nonumber\\
    U(A_7)_2[d,s,u,u]\,&=\, -\frac{1}{4}\Delta_{\alpha^+,\gamma^+}(\Xi^+_c,K^+,\Sigma^0,\rho^0, K^+,\Sigma^0)-\frac{1}{4}\Delta_{\alpha^+,\gamma^+}(\Xi^+_c,K^+,\Sigma^0,
   \omega, K^+,\Sigma^0)\nonumber\\&
   ~~~~+\frac{1}{12}\Delta_{\alpha^+,\gamma^+}(\Xi^+_c,K^+,\Lambda^0,\rho^0, K^+,\Sigma^0)+\frac{1}{12}\Delta_{\alpha^+,\gamma^+}(\Xi^+_c,K^+,\Lambda^0,
   \omega, K^+,\Sigma^0),
   \nonumber\\
T(A_7)_{1}[d,s,u,u]\,&=\, -\frac{1}{2}\Delta_{\beta^+,\beta^+}(\Xi^+_c,K^+,\Sigma^0,
   \Xi^-,\Sigma^0,K^+)+\frac{1}{6}\Delta_{\beta^+,\beta^+}(\Xi^+_c,K^+,\Lambda^0,
   \Xi^-, \Sigma^0,K^+),
   \nonumber\\
T(A_7)_{2}[d,s,u,u]\,&=\, -\frac{1}{3}\Delta_{\beta^+,\beta^-}(\Xi^+_c,K^+,\Lambda^0,
   \Xi^-, \Sigma^0,K^+),\nonumber\\
   S(A_7)_{1}[d,s,u,u]\,&=\, -\frac{1}{2}\Theta_{\beta^-,\beta^-}(\Xi^+_c,K^+,\Sigma^0,
   p, K^+,\Sigma^0)+\frac{1}{6}\Theta_{\beta^-,\beta^-}(\Xi^+_c,K^+,\Lambda^0,
   p, K^+,\Sigma^0),
   \nonumber\\
   S(A_7)_{2}[d,s,u,u]\,&=\, -\frac{1}{3}\Theta_{\beta^+,\beta^-}(\Xi^+_c,K^+,\Lambda^0,
   p, K^+,\Sigma^0),
   \nonumber\\
   U(A_{19})_5[d,s,u,u]\,&=\, -\frac{1}{2}\Delta_{\alpha^+,\gamma^0}(\Xi^+_c,K^+,\Sigma^0,
   \omega, K^+,\Sigma^0)+\frac{1}{6}\Delta_{\alpha^+,\gamma^0}(\Xi^+_c,K^+,\Lambda^0,
   \omega, K^+,\Sigma^0),
   \nonumber\\
    U(A_{19})_{6}[d,s,u,u]\,&=\, \frac{1}{2}\Delta_{\alpha^+,\gamma^0}(\Xi^+_c,K^+,\Sigma^0,
   \phi, K^+,\Sigma^0)-\frac{1}{6}\Delta_{\alpha^+,\gamma^0}(\Xi^+_c,K^+,\Lambda^0,
   \phi, K^+,\Sigma^0).
\end{align}
Summing all long-distance contributions, we have
\begin{align}\label{xx10}
  \mathcal{A}_L(\Xi_{c}^{+}\to \Sigma^0K^+)&=\frac{1}{2\sqrt{2}}\lambda_2\Delta_{\alpha^+,(\gamma^+-\gamma^-)}(\Xi^+_c,K^+,\Sigma^0,\omega, K^+,\Sigma^0)\nonumber\\&~~-\frac{1}{6\sqrt{2}}\lambda_2\Delta_{\alpha^+,(\gamma^++\gamma^-)}(\Xi^+_c,K^+,\Lambda^0,\rho^0, K^+,\Sigma^0)\nonumber\\&~~+\frac{1}{\sqrt{2}}\lambda_2\Delta_{\alpha^+,\gamma^-}(\Xi^+_c,K^+,\Sigma^0,\phi, K^+,\Sigma^0)\nonumber\\&~~+\frac{1}{2\sqrt{2}}\lambda_2\Delta_{\beta^+,\beta^+}(\Xi^+_c,K^+,\Sigma^0,
   \Xi^-,\Sigma^0,K^+)\nonumber\\&~~~~-\frac{1}{6\sqrt{2}}\lambda_2\Delta_{\beta^+,(\beta^+-2\beta^-)}(\Xi^+_c,K^+,\Lambda^0,
   \Xi^-, \Sigma^0,K^+)\nonumber\\&~~+\frac{1}{2\sqrt{2}}\lambda_2\Theta_{\beta^-,\beta^-}(\Xi^+_c,K^+,\Sigma^0,
   p, K^+,\Sigma^0)\nonumber\\&~~+\frac{1}{6\sqrt{2}}\lambda_2\Theta_{(2\beta^+-\beta^-),\beta^-}(\Xi^+_c,K^+,\Lambda^0,
   p, K^+,\Sigma^0).
\end{align}
The decay amplitude for the $\Xi_{c}^{+}\to \Sigma^+K^0$ mode is $-\lambda_2(A_4+A_7)$.
The rescattering amplitudes contributing to the $\Xi_{c}^{+}\to \Sigma^+K^0$ decay include
\begin{align}
  U(A_4)[d,s,u,d]\,&=\, \frac{1}{2}\Delta_{\alpha^+,\gamma^+}(\Xi^+_c,K^+,\Sigma^0,\rho^+, K^0,\Sigma^+)+\frac{1}{6}\Delta_{\alpha^+,\gamma^+}(\Xi^+_c,K^+,\Lambda^0,\rho^+, K^0,\Sigma^+),
\nonumber\\
  T(A_4)[d,s,u,d]\,&=\, \frac{1}{2}\Delta_{\beta^+,\beta^+}(\Xi^+_c,K^+,\Sigma^0,\Xi^-, \Sigma^+,K^0)+\frac{1}{6}\Delta_{\beta^+,\beta^+}(\Xi^+_c,K^+,\Lambda^0,\Xi^-, \Sigma^+,K^0),
\nonumber\\
  U(A_7)_1[d,s,u,d]\,&=\,-\frac{1}{2}\Delta_{\alpha^+,\gamma^-}(\Xi^+_c,K^+,\Sigma^0,\rho^+, K^0,\Sigma^+)+\frac{1}{6}\Delta_{\alpha^+,\gamma^-}(\Xi^+_c,K^+,\Lambda^0,\rho^+, K^0, \Sigma^+)\nonumber\\
  T(A_7)_2[d,s,u,d]\,&=\, -\frac{1}{3}\Delta_{\beta^+,\beta^-}(\Xi^+_c,K^+,\Lambda^0,\Xi^-, \Sigma^+,K^0),
   \nonumber\\
   S(A_7)_{1}[d,s,u,d]\,&=\, -\frac{1}{2}\Theta_{\beta^-,\beta^-}(\Xi^+_c,K^+,\Sigma^0,
   p, K^0,\Sigma^+)+\frac{1}{6}\Theta_{\beta^-,\beta^-}(\Xi^+_c,K^+,\Lambda^0,
   p, K^0,\Sigma^+),
   \nonumber\\
   S(A_7)_{2}[d,s,u,d]\,&=\, -\frac{1}{3}\Theta_{\beta^+,\beta^-}(\Xi^+_c,K^+,\Lambda^0,
   p, K^0,\Sigma^+).
\end{align}
Summing all long-distance contributions, we have
\begin{align}
  \mathcal{A}_L(\Xi_{c}^{+}\to \Sigma^+K^0)&=-\frac{1}{2}\lambda_2\Delta_{\alpha^+,(\gamma^+-\gamma^-)}(\Xi^+_c,K^+,\Sigma^0,\rho^+, K^0,\Sigma^+)\nonumber\\&~~-\frac{1}{6}\lambda_2\Delta_{\alpha^+,(\gamma^++\gamma^-)}(\Xi^+_c,K^+,\Lambda^0,\rho^+, K^0,\Sigma^+)\nonumber\\&~~-\frac{1}{2}\lambda_2\Delta_{\beta^+,\beta^+}(\Xi^+_c,K^+,\Sigma^0,\Xi^-, \Sigma^+,K^0)\nonumber\\&~~-\frac{1}{6}\lambda_2\Delta_{\beta^+,(\beta^+-2\beta^-)}(\Xi^+_c,K^+,\Lambda^0,\Xi^-, \Sigma^+,K^0)\nonumber\\&~~+\frac{1}{2}\lambda_2\Theta_{\beta^-,\beta^-}(\Xi^+_c,K^+,\Sigma^0,
   p, K^0,\Sigma^+)\nonumber\\&~~+\frac{1}{6}\lambda_2\Theta_{(2\beta^+-\beta^-),\beta^-}(\Xi^+_c,K^+,\Lambda^0,
   p, K^0,\Sigma^+).
\end{align}
One can verify that the isospin sum rule
\begin{align}
\sqrt{2}\,\mathcal{A}( \Xi^0_c\to \Sigma^0K^0)+\mathcal{A}(\Xi^0_c\to \Sigma^-K^+)
+\mathcal{A}(\Xi^+_c\to \Sigma^+K^0)-\sqrt{2}\,\mathcal{A}(\Xi^+_c\to \Sigma^0K^+)=0
\end{align}
is satisfied.

\subsection{The $\Xi_c\to N\pi$ system}

The decay amplitude for the $\Xi_{c}^{0}\to p\pi^-$ mode is $\lambda_2(A_1+A_6)$.
The rescattering amplitudes contributing to the $\Xi_{c}^{0}\to p\pi^-$ decay include
\begin{align}
  T(A_1)_1[u,s,u,d]\,&=\, -\frac{1}{2}\Delta_{\beta^-,\beta^-}(\Xi^0_c,K^+,\Sigma^-,\Sigma^0, p,\pi^-)+\frac{1}{6}\Delta_{\beta^-,\beta^-}(\Xi^0_c,K^+,\Sigma^-,\Lambda^0, p,\pi^-),\nonumber\\
 T(A_1)_3[u,s,u,d]\,&=\, -\frac{1}{3}\Delta_{\beta^+,\beta^-}(\Xi^0_c,K^+,\Sigma^-,\Lambda^0, p,\pi^-),\nonumber\\
  T(A_1)_4[u,s,u,d]\,&=\, -\frac{1}{3}\Delta_{\beta^+,\beta^+}(\Xi^0_c,K^+,\Sigma^-,\Lambda^0, p,\pi^-),\nonumber\\
  T(A_6)[u,s,u,d]\,&=\frac{1}{2}\Delta_{\beta^-,\beta^+}(\Xi^0_c,K^+,\Sigma^-,\Sigma^0, p,\pi^-)+\frac{1}{6}\Delta_{\beta^-,\beta^+}(\Xi^0_c,K^+,\Sigma^-,\Lambda^0, p,\pi^-),\nonumber\\
   S(A_6)[u,s,u,d]\,&=\Theta_{\beta^-,\beta^+}(\Xi^0_c,K^+,\Sigma^-,n, p,\pi^-).
\end{align}
Summing all long-distance contributions, we have
\begin{align}
  \mathcal{A}_L(\Xi_{c}^{0}\to p\pi^-)&=\frac{1}{2}\lambda_2\Delta_{\beta^-,(\beta^+-\beta^-)}(\Xi^0_c,K^+,\Sigma^-,\Sigma^0, p,\pi^-)\nonumber\\&~~+\frac{1}{6}\lambda_2\Delta_{(\beta^--2\beta^+),(\beta^++\beta^-)}(\Xi^0_c,K^+,\Sigma^-,\Lambda^0, p,\pi^-)\nonumber\\&~~+\lambda_2\Theta_{\beta^-,\beta^+}(\Xi^0_c,K^+,\Sigma^-,n, p,\pi^-).
\end{align}
The decay amplitude for the $\Xi_{c}^{0}\to n\pi^0$ mode is $\frac{1}{\sqrt{2}}\lambda_2(A_3-A_6)$.
The rescattering amplitudes contributing to the $\Xi_{c}^{0}\to n\pi^0$ decay include
\begin{align}
  U(A_3)[d,s,d,u]\,&=\, -\Delta_{\alpha^+,\gamma^-}(\Xi^0_c,K^+,\Sigma^-,K^{*+}, \pi^0,n),\nonumber\\
  T(A_3)[d,s,d,u]\,&=\, \Delta_{\beta^-,\beta^-}(\Xi^0_c,K^+,\Sigma^-,\Sigma^-, n, \pi^0),\nonumber\\
    T(A_6)[d,s,d,d]\,&=\, \Delta_{\beta^-,\beta^+}(\Xi^0_c,K^+,\Sigma^-,\Sigma^-, n, \pi^0),\nonumber\\
    S(A_6)[u,s,d,d]\,&=\Theta_{\beta^-,\beta^+}(\Xi^0_c,K^+,\Sigma^-,n, n,\pi^0).
\end{align}
Summing all long-distance contributions, we have
\begin{align}
  \mathcal{A}_L(\Xi_{c}^{0}\to n\pi^0)&=\, -\frac{1}{\sqrt{2}}\lambda_2\Delta_{\alpha^+,\gamma^-}(\Xi^0_c,K^+,\Sigma^-,K^{*+}, \pi^0,n)\nonumber\\&~~-\frac{1}{\sqrt{2}}\lambda_2\Delta_{\beta^-,(\beta^+-\beta^-)}(\Xi^0_c,K^+,\Sigma^-,\Sigma^0, p,\pi^-)\nonumber\\&~~-\frac{1}{\sqrt{2}}\lambda_2\Theta_{\beta^-,\beta^+}(\Xi^0_c,K^+,\Sigma^-,n, n,\pi^0).
\end{align}
The decay amplitude for the $\Xi_{c}^+\to p\pi^0$ mode is $\frac{1}{\sqrt{2}}\lambda_2(A_1-A_8)$.
The rescattering amplitudes contributing to the $\Xi_{c}^+\to p\pi^0$ decay include
\begin{align}
  T(A_1)_1[d,s,u,d]\,&=\, -\frac{1}{4}\Delta_{\beta^-,\beta^-}(\Xi^+_c,K^+,\Sigma^0,\Sigma^0, p,\pi^0)+\frac{1}{12}\Delta_{\beta^-,\beta^-}(\Xi^+_c,K^+,\Sigma^0,\Lambda^0, p,\pi^0)\nonumber\\
  &~~~~-\frac{1}{12}\Delta_{\beta^-,\beta^-}(\Xi^+_c,K^+,\Lambda^0,\Sigma^0, p,\pi^0)+\frac{1}{36}\Delta_{\beta^-,\beta^-}(\Xi^+_c,K^+,\Lambda^0,\Lambda^0, p,\pi^0),\nonumber\\
  T(A_1)_2[d,s,u,d]\,&=\, -\frac{1}{4}\Delta_{\beta^-,\beta^+}(\Xi^+_c,K^+,\Sigma^0,\Sigma^0, p,\pi^0)+\frac{1}{12}\Delta_{\beta^-,\beta^+}(\Xi^+_c,K^+,\Sigma^0,\Lambda^0, p,\pi^0)\nonumber\\
&~~~~-\frac{1}{12}\Delta_{\beta^-,\beta^+}(\Xi^+_c,K^+,\Lambda^0,\Sigma^0, p,\pi^0)+\frac{1}{36}\Delta_{\beta^-,\beta^+}(\Xi^+_c,K^+,\Lambda^0,\Lambda^0, p,\pi^0),\nonumber\\
   T(A_1)_3[d,s,u,d]\,&=\, -\frac{1}{6}\Delta_{\beta^+,\beta^-}(\Xi^+_c,K^+,\Sigma^0,\Lambda^0, p,\pi^0)-\frac{1}{18}\Delta_{\beta^+,\beta^-}(\Xi^+_c,K^+,\Lambda^0,\Lambda^0, p,\pi^0),\nonumber\\
   T(A_1)_4[d,s,u,d]\,&=\, -\frac{1}{6}\Delta_{\beta^+,\beta^+}(\Xi^+_c,K^+,\Sigma^0,\Lambda^0, p,\pi^0)-\frac{1}{18}\Delta_{\beta^+,\beta^+}(\Xi^+_c,K^+,\Lambda^0,\Lambda^0, p,\pi^0),\nonumber\\
   U(A_8)_1[d,s,u,u]\,&=\, \frac{1}{2}\Delta_{\alpha^+,\gamma^-}(\Xi^+_c,K^+,\Sigma^0,K^{*+}, \pi^0, p)-\frac{1}{6}\Delta_{\alpha^+,\gamma^-}(\Xi^+_c,K^+,\Lambda^0,K^{*+}, \pi^0, p),\nonumber\\
   U(A_8)_2[d,s,u,u]\,&=\, \frac{1}{3}\Delta_{\alpha^+,\gamma^+}(\Xi^+_c,K^+,\Lambda^0,K^{*+}, \pi^0, p),\nonumber\\
   T(A_8)_1[d,s,u,u]\,&=\, -\frac{1}{4}\Delta_{\beta^-,\beta^+}(\Xi^+_c,K^+,\Sigma^0,\Sigma^0, p, \pi^0)-\frac{1}{12}\Delta_{\beta^-,\beta^+}(\Xi^+_c,K^+,\Sigma^0,\Lambda^0, p, \pi^0),\nonumber\\
  &~~~~+\frac{1}{12}\Delta_{\beta^-,\beta^+}(\Xi^+_c,K^+,\Lambda^0,\Sigma^0, p, \pi^0)+\frac{1}{36}\Delta_{\beta^-,\beta^+}(\Xi^+_c,K^+,\Lambda^0,\Lambda^0, p, \pi^0),\nonumber\\
   T(A_8)_2[d,s,u,u]\,&=\, -\frac{1}{4}\Delta_{\beta^-,\beta^-}(\Xi^+_c,K^+,\Sigma^0,\Sigma^0, p, \pi^0)-\frac{1}{12}\Delta_{\beta^-,\beta^-}(\Xi^+_c,K^+,\Sigma^0,\Lambda^0, p, \pi^0),\nonumber\\
  &~~~~+\frac{1}{12}\Delta_{\beta^-,\beta^-}(\Xi^+_c,K^+,\Lambda^0,\Sigma^0, p, \pi^0)+\frac{1}{36}\Delta_{\beta^-,\beta^-}(\Xi^+_c,K^+,\Lambda^0,\Lambda^0, p, \pi^0),\nonumber\\
   S(A_8)_1[d,s,u,u]\,&=\, -\frac{1}{2}\Theta_{\beta^-,\beta^+}(\Xi^+_c,K^+,\Sigma^0,p, \pi^0, p)+\frac{1}{6}\Theta_{\beta^-,\beta^+}(\Xi^+_c,K^+,\Lambda^0,p, \pi^0, p)
,\nonumber\\
   S(A_8)_2[d,s,u,u]\,&=\, -\frac{1}{3}\Theta_{\beta^+,\beta^+}(\Xi^+_c,K^+,\Lambda^0,p, \pi^0, p),
\nonumber\\
   T(A_{10})_3[d,s,u,u]\,&=\, \frac{1}{6}\Delta_{\beta^+,\beta^+}(\Xi^+_c,K^+,\Sigma^0,\Lambda^0, p, \pi^0)-\frac{1}{18}\Delta_{\beta^+,\beta^+}(\Xi^+_c,K^+,\Lambda^0,\Lambda^0, p, \pi^0),\nonumber\\
   T(A_{10})_4[d,s,u,u]\,&=\, \frac{1}{6}\Delta_{\beta^+,\beta^-}(\Xi^+_c,K^+,\Sigma^0,\Lambda^0, p, \pi^0)-\frac{1}{18}\Delta_{\beta^+,\beta^-}(\Xi^+_c,K^+,\Lambda^0,\Lambda^0, p, \pi^0).
\end{align}
Summing all long-distance contributions, we have
\begin{align}
  \mathcal{A}_L(\Xi_{c}^+\to p\pi^0)&=-\frac{1}{2\sqrt{2}}\lambda_2\Delta_{\alpha^+,\gamma^-}(\Xi^+_c,K^+,\Sigma^0,K^{*+}, \pi^0, p)\nonumber\\&~~+\frac{1}{6\sqrt{2}}\lambda_2\Delta_{\alpha^+,(\gamma^--2\gamma^+)}(\Xi^+_c,K^+,\Lambda^0,K^{*+}, \pi^0, p)\nonumber\\&~~+\frac{1}{6\sqrt{2}}\lambda_2\Delta_{(\beta^--2\beta^+),(\beta^++\beta^-)}(\Xi^+_c,K^+,\Sigma^0,\Lambda^0, p,\pi^0)\nonumber\\
  &~~-\frac{1}{6\sqrt{2}}\lambda_2\Delta_{\beta^-,(\beta^++\beta^-)}(\Xi^+_c,K^+,\Lambda^0,\Sigma^0, p,\pi^0)\nonumber\\
  &~~+\frac{1}{2\sqrt{2}}\lambda_2\Theta_{\beta^-,\beta^+}(\Xi^+_c,K^+,\Sigma^0,p, \pi^0, p)\nonumber\\
  &~~+\frac{1}{6\sqrt{2}}\lambda_2\Theta_{(2\beta^+-\beta^-),\beta^+}(\Xi^+_c,K^+,\Lambda^0,p, \pi^0, p).
\end{align}
The decay amplitude for the $\Xi_{c}^+\to n\pi^+$ mode is $-\lambda_2(A_3+A_8)$.
The rescattering amplitudes contributing to the $\Xi_{c}^+\to n\pi^+$  decay include
\begin{align}
  U(A_3)[d,s,d,u]\,&=\, -\frac{1}{2}\Delta_{\alpha^+,\gamma^-}(\Xi^+_c,K^+,\Sigma^0,K^{*0}, \pi^+,n)-\frac{1}{6}\Delta_{\alpha^+,\gamma^-}(\Xi^+_c,K^+,\Lambda^0,K^{*0}, \pi^+,n),\nonumber\\
  T(A_3)[d,s,d,u]\,&=\, \frac{1}{2}\Delta_{\beta^-,\beta^-}(\Xi^+_c,K^+,\Sigma^0,\Sigma^+, n,\pi^+)+\frac{1}{6}\Delta_{\beta^-,\beta^-}(\Xi^+_c,K^+,\Lambda^0,\Sigma^+, n,\pi^+),\nonumber\\
  U(A_8)_2[d,s,d,u]\,&=\, \frac{1}{3}\Delta_{\alpha^+,\gamma^+}(\Xi^+_c,K^+,\Lambda^0,K^{*0}, \pi^+,n),\nonumber\\
 T(A_8)_1[d,s,d,u]\,&=\, -\frac{1}{2}\Delta_{\beta^-,\beta^+}(\Xi^+_c,K^+,\Sigma^0,\Sigma^+, n,\pi^+)+\frac{1}{6}\Delta_{\beta^-,\beta^+}(\Xi^+_c,K^+,\Lambda^0,\Sigma^+, n,\pi^+),\nonumber\\
   S(A_8)_1[d,s,d,u]\,&=\, -\frac{1}{2}\Theta_{\beta^-,\beta^+}(\Xi^+_c,K^+,\Sigma^0,p, \pi^+, n)+\frac{1}{6}\Theta_{\beta^-,\beta^+}(\Xi^+_c,K^+,\Lambda^0,p, \pi^+, n)
,\nonumber\\
   S(A_8)_2[d,s,d,u]\,&=\, -\frac{1}{3}\Theta_{\beta^+,\beta^+}(\Xi^+_c,K^+,\Lambda^0,p, \pi^+, n).
\end{align}
Summing all long-distance contributions, we have
\begin{align}
  \mathcal{A}_L(\Xi_{c}^+\to n\pi^+)&=\frac{1}{2}\lambda_2\Delta_{\alpha^+,\gamma^-}(\Xi^+_c,K^+,\Sigma^0,K^{*0}, \pi^+,n)\nonumber\\&~~+\frac{1}{6}\lambda_2\Delta_{\alpha^+,(\gamma^--2\gamma^+)}(\Xi^+_c,K^+,\Lambda^0,K^{*0}, \pi^+,n)\nonumber\\&~~+\frac{1}{2}\lambda_2\Delta_{\beta^-,(\beta^+-\beta^-)}(\Xi^+_c,K^+,\Sigma^0,\Sigma^+, n,\pi^+)\nonumber\\&~~-\frac{1}{6}\lambda_2\Delta_{\beta^-,(\beta^++\beta^-)}(\Xi^+_c,K^+,\Lambda^0,\Sigma^+, n,\pi^+)\nonumber\\&~~+\frac{1}{2}\lambda_2\Theta_{\beta^-,\beta^+}(\Xi^+_c,K^+,\Sigma^0,p, \pi^+, n)\nonumber\\&~~+\frac{1}{6}\lambda_2\Theta_{(2\beta^+-\beta^-),\beta^+}(\Xi^+_c,K^+,\Lambda^0,p, \pi^+, n).
\end{align}
One can verify that the isospin sum rule
\begin{align}
\sqrt{2}\,\mathcal{A}( \Xi^0_c\to n\pi^0)+\mathcal{A}(\Xi^0_c\to p\pi^-)
-\sqrt{2}\,\mathcal{A}(\Xi^+_c\to p\pi^0)+\mathcal{A}(\Xi^+_c\to n\pi^+)=0
\end{align}
is satisfied.

\subsection{The $\Xi_c\to \Sigma\pi $ system}

The decay amplitude for the $\Xi_{c}^{0}\to \Sigma^-\pi^+$ mode is $\lambda_d(A_2+A_5+A_{12}+A_{14})+\lambda_s(A_{12}+A_{14})$.
The rescattering amplitudes contributing to the $\Xi_{c}^{0}\to \Sigma^-\pi^+$ decay include
\begin{align}
  U(A_2)_1[u,d,d,u]\,&=\, -\frac{1}{2}\Delta_{\alpha^+,\gamma^+}(\Xi^0_c,\pi^+,\Sigma^-,\rho^0, \pi^+,\Sigma^-)+\frac{1}{2}\Delta_{\alpha^+,\gamma^+}(\Xi^0_c,\pi^+,\Sigma^-,\omega, \pi^+,\Sigma^-),\nonumber\\
  U(A_2)_4[u,d,d,u]\,&=\, \frac{1}{2}\Delta_{\alpha^+,\gamma^-}(\Xi^0_c,\pi^+,\Sigma^-,\rho^0, \pi^+,\Sigma^-)-\frac{1}{2}\Delta_{\alpha^+,\gamma^-}(\Xi^0_c,\pi^+,\Sigma^-,\omega, \pi^+,\Sigma^-),\nonumber\\
  U(A_2)_5[u,d,d,u]\,&=\, \Delta_{\alpha^+,\gamma^0}(\Xi^0_c,\pi^+,\Sigma^-,\omega, \pi^+,\Sigma^-),\nonumber\\
  U(A_2)_6[u,d,d,u]\,&=\, -\Delta_{\alpha^+,\gamma^0}(\Xi^0_c,\pi^+,\Sigma^-,\omega, \pi^+,\Sigma^-),\nonumber\\
  U(A_5)[u,d,d,u]\,&=\, \frac{1}{2}\Delta_{\alpha^+,\gamma^-}(\Xi^0_c,\pi^+,\Sigma^-,\rho^0, \pi^+,\Sigma^-)+\frac{1}{2}\Delta_{\alpha^+,\gamma^-}(\Xi^0_c,\pi^+,\Sigma^-,\omega, \pi^+,\Sigma^-),\nonumber\\
  S(A_5)[u,d,d,u]\,&=\, \frac{1}{2}\Theta_{\beta^-,\beta^-}(\Xi^0_c,\pi^+,\Sigma^-,\Sigma^0, \pi^+,\Sigma^-)+\frac{1}{6}\Theta_{\beta^-,\beta^-}(\Xi^0_c,\pi^+,\Sigma^-,\Lambda^0, \pi^+,\Sigma^-),\nonumber\\
   U(A_{12})[u,d,d,u]\,&=\, -\frac{1}{2}\Delta_{\alpha^+,\gamma^+}(\Xi^0_c,\pi^+,\Sigma^-,\rho^0, \pi^+,\Sigma^-)-\frac{1}{2}\Delta_{\alpha^+,\gamma^+}(\Xi^0_c,\pi^+,\Sigma^-,\omega, \pi^+,\Sigma^-),\nonumber\\
  S(A_{12})[u,d,d,u]\,&=\, \frac{1}{2}\Theta_{\beta^+,\beta^+}(\Xi^0_c,\pi^+,\Sigma^-,\Sigma^0, \pi^+,\Sigma^-)+\frac{1}{6}\Theta_{\beta^+,\beta^+}(\Xi^0_c,\pi^+,\Sigma^-,\Lambda^0, \pi^+,\Sigma^-),\nonumber\\
  U(A_{12})[u,s,d,u]\,&=\, -\Delta_{\alpha^+,\gamma^+}(\Xi^0_c,K^+,\Xi^-,K^{*0}, \pi^+,\Sigma^-),\nonumber\\
  S(A_{12})[u,s,d,u]\,&=\, \frac{1}{2}\Theta_{\beta^+,\beta^+}(\Xi^0_c,K^+,\Xi^-,\Sigma^0, \pi^+,\Sigma^-)+\frac{1}{6}\Theta_{\beta^+,\beta^+}(\Xi^0_c,K^+,\Xi^-,\Lambda^0, \pi^+,\Sigma^-),\nonumber\\
  S(A_{14})_1[u,d,d,u]\,&=\, -\frac{1}{2}\Theta_{\beta^-,\beta^+}(\Xi^0_c,\pi^+,\Sigma^-,\Sigma^0, \pi^+,\Sigma^-)+\frac{1}{6}\Theta_{\beta^-,\beta^+}(\Xi^0_c,\pi^+,\Sigma^-,\Lambda^0, \pi^+,\Sigma^-),\nonumber\\
  S(A_{14})_4[u,d,d,u]\,&=\, -\frac{1}{2}\Theta_{\beta^+,\beta^-}(\Xi^0_c,\pi^+,\Sigma^-,\Sigma^0, \pi^+,\Sigma^-)+\frac{1}{6}\Theta_{\beta^+,\beta^-}(\Xi^0_c,\pi^+,\Sigma^-,\Lambda^0, \pi^+,\Sigma^-),\nonumber\\
  S(A_{14})_1[u,s,d,u]\,&=\, -\frac{1}{3}\Theta_{\beta^-,\beta^+}(\Xi^0_c,K^+,\Xi^-,\Lambda^0, \pi^+,\Sigma^-),\nonumber\\
  S(A_{14})_2[u,s,d,u]\,&=\, -\frac{1}{3}\Theta_{\beta^-,\beta^-}(\Xi^0_c,K^+,\Xi^-,\Lambda^0, \pi^+,\Sigma^-),\nonumber\\
  S(A_{14})_4[u,s,d,u]\,&=\, -\frac{1}{2}\Theta_{\beta^+,\beta^-}(\Xi^0_c,K^+,\Xi^-,\Sigma^0, \pi^+,\Sigma^-)+\frac{1}{6}\Theta_{\beta^+,\beta^-}(\Xi^0_c,K^+,\Xi^-,\Lambda^0, \pi^+,\Sigma^-).
\end{align}
Summing all long-distance contributions, we have
\begin{align}
  \mathcal{A}_L(\Xi_{c}^{0}\to \Sigma^-\pi^+)&=-\lambda_d\Delta_{\alpha^+,(\gamma^+-\gamma^-)}(\Xi^0_c,\pi^+,\Sigma^-,\rho^0, \pi^+,\Sigma^-)\nonumber\\&~~-\lambda_s\Delta_{\alpha^+,\gamma^+}(\Xi^0_c,K^+,\Xi^-,K^{*0}, \pi^+,\Sigma^-)\nonumber\\&~~+\frac{1}{2}\lambda_d\Theta_{(\beta^+-\beta^-),(\beta^+-\beta^-)}(\Xi^0_c,\pi^+,\Sigma^-,\Sigma^0, \pi^+,\Sigma^-)\nonumber\\&~~+\frac{1}{6}\lambda_d\Theta_{(\beta^++\beta^-),(\beta^++\beta^-)}(\Xi^0_c,\pi^+,\Sigma^-,\Lambda^0, \pi^+,\Sigma^-)\nonumber\\&~~+\frac{1}{2}\lambda_s\Theta_{\beta^+,(\beta^+-\beta^-)}(\Xi^0_c,K^+,\Xi^-,\Sigma^0, \pi^+,\Sigma^-)\nonumber\\&~~+\frac{1}{6}\lambda_s\Theta_{(\beta^+-2\beta^-),(\beta^++\beta^-)}(\Xi^0_c,K^+,\Xi^-,\Lambda^0, \pi^+,\Sigma^-).
\end{align}
The decay amplitude for the $\Xi_{c}^{0}\to \Sigma^0\pi^0$ mode is $\frac{1}{2}\lambda_d(-A_3-A_4+A_5+A_6+A_{11}+A_{12}+2A_{14})
+\frac{1}{2}\lambda_s(A_{11}+A_{12}+2A_{14})$.
The rescattering amplitudes contributing to the $\Xi_{c}^{0}\to \Sigma^0\pi^0$ decay include
\begin{align}
  U(A_3)[u,d,d,u]\,&=\, -\Delta_{\alpha^+,\gamma^-}(\Xi^0_c,\pi^+,\Sigma^-,\rho^+, \pi^0,\Sigma^0),\nonumber\\
  T(A_3)[u,d,d,u]\,&=\, \Delta_{\beta^-,\beta^-}(\Xi^0_c,\pi^+,\Sigma^-,\Sigma^-, \Sigma^0,\pi^0),\nonumber\\
  U(A_4)[u,d,u,d]\,&=\, \Delta_{\alpha^+,\gamma^+}(\Xi^0_c,\pi^+,\Sigma^-,\rho^+, \pi^0,\Sigma^0),\nonumber\\
  T(A_4)[u,d,u,d]\,&=\, \Delta_{\beta^+,\beta^+}(\Xi^0_c,\pi^+,\Sigma^-,\Sigma^-, \Sigma^0,\pi^0),\nonumber\\
  U(A_5)[u,d,d,d]\,&=\, \Delta_{\alpha^+,\gamma^-}(\Xi^0_c,\pi^+,\Sigma^-,\rho^+, \pi^0,\Sigma^0),\nonumber\\
  S(A_5)[u,d,d,d]\,&=\, \frac{1}{2}\Theta_{\beta^-,\beta^-}(\Xi^0_c,\pi^+,\Sigma^-,\Sigma^0, \pi^0,\Sigma^0)+ \frac{1}{6}\Theta_{\beta^-,\beta^-}(\Xi^0_c,\pi^+,\Sigma^-,\Lambda^0, \pi^0,\Sigma^0),\nonumber\\
  T(A_6)[u,d,d,d]\,&=\, \Delta_{\beta^-,\beta^+}(\Xi^0_c,\pi^+,\Sigma^-,\Sigma^-, \Sigma^0,\pi^0),\nonumber\\
  S(A_6)[u,d,d,d]\,&=\, \frac{1}{2}\Theta_{\beta^-,\beta^+}(\Xi^0_c,\pi^+,\Sigma^-,\Sigma^0, \pi^0,\Sigma^0)+ \frac{1}{6}\Theta_{\beta^-,\beta^+}(\Xi^0_c,\pi^+,\Sigma^-,\Lambda^0, \pi^0,\Sigma^0),\nonumber\\
  T(A_{11})[u,d,u,u]\,&=\, \Delta_{\beta^+,\beta^-}(\Xi^0_c,\pi^+,\Sigma^-,\Sigma^-, \Sigma^0,\pi^0),\nonumber\\
  S(A_{11})[u,d,u,u]\,&=\, \frac{1}{2}\Theta_{\beta^+,\beta^-}(\Xi^0_c,\pi^+,\Sigma^-,\Sigma^0, \pi^0,\Sigma^0)+ \frac{1}{6}\Theta_{\beta^+,\beta^-}(\Xi^0_c,\pi^+,\Sigma^-,\Lambda^0, \pi^0,\Sigma^0),\nonumber\\
  T(A_{11})[u,s,u,u]\,&=\, \Delta_{\beta^+,\beta^-}(\Xi^0_c,K^+,\Xi^-,\Xi^-, \Sigma^0,\pi^0),\nonumber\\
  S(A_{11})[u,s,u,u]\,&=\, \frac{1}{2}\Theta_{\beta^+,\beta^-}(\Xi^0_c,K^+,\Xi^-,\Sigma^0, \pi^0,\Sigma^0)+ \frac{1}{6}\Theta_{\beta^+,\beta^-}(\Xi^0_c,K^+,\Xi^-,,\Lambda^0, \pi^0,\Sigma^0),\nonumber\\
  U(A_{12})[u,d,u,u]\,&=\, -\Delta_{\alpha^+,\gamma^+}(\Xi^0_c,\pi^+,\Sigma^-,\rho^+, \pi^0,\Sigma^0),\nonumber\\
  S(A_{12})[u,d,u,u]\,&=\, \frac{1}{2}\Theta_{\beta^+,\beta^+}(\Xi^0_c,\pi^+,\Sigma^-,\Sigma^0, \pi^0,\Sigma^0)+ \frac{1}{6}\Theta_{\beta^+,\beta^+}(\Xi^0_c,\pi^+,\Sigma^-,\Lambda^0, \pi^0,\Sigma^0),\nonumber\\
  U(A_{12})[u,s,u,u]\,&=\, -\Delta_{\alpha^+,\gamma^+}(\Xi^0_c,K^+,\Xi^-,K^{*+}, \pi^0,\Sigma^0),\nonumber\\
  S(A_{12})[u,s,u,u]\,&=\, \frac{1}{2}\Theta_{\beta^+,\beta^+}(\Xi^0_c,K^+,\Xi^-,\Sigma^0, \pi^0,\Sigma^0)+ \frac{1}{6}\Theta_{\beta^+,\beta^+}(\Xi^0_c,K^+,\Xi^-,\Lambda^0, \pi^0,\Sigma^0),\nonumber\\
  S(A_{14})_1[u,d,u,u]\,&=\, -\frac{1}{2}\Theta_{\beta^-,\beta^+}(\Xi^0_c,\pi^+,\Sigma^-,\Sigma^0, \pi^0,\Sigma^0)+ \frac{1}{6}\Theta_{\beta^-,\beta^+}(\Xi^0_c,\pi^+,\Sigma^-,\Lambda^0, \pi^0,\Sigma^0),\nonumber\\
  S(A_{14})_2[u,d,u,u]\,&=\, -\frac{1}{2}\Theta_{\beta^-,\beta^-}(\Xi^0_c,\pi^+,\Sigma^-,\Sigma^0, \pi^0,\Sigma^0)+ \frac{1}{6}\Theta_{\beta^-,\beta^-}(\Xi^0_c,\pi^+,\Sigma^-,\Lambda^0, \pi^0,\Sigma^0),\nonumber\\
  S(A_{14})_3[u,d,d,d]\,&=\, -\frac{1}{2}\Theta_{\beta^+,\beta^+}(\Xi^0_c,\pi^+,\Sigma^-,\Sigma^0, \pi^0,\Sigma^0)+ \frac{1}{6}\Theta_{\beta^+,\beta^+}(\Xi^0_c,\pi^+,\Sigma^-,\Lambda^0, \pi^0,\Sigma^0),\nonumber\\
  S(A_{14})_4[u,d,d,d]\,&=\, -\frac{1}{2}\Theta_{\beta^+,\beta^-}(\Xi^0_c,\pi^+,\Sigma^-,\Sigma^0, \pi^0,\Sigma^0)+ \frac{1}{6}\Theta_{\beta^+,\beta^-}(\Xi^0_c,\pi^+,\Sigma^-,\Lambda^0, \pi^0,\Sigma^0),\nonumber\\
  S(A_{14})_1[u,s,u,u]\,&=\, -\frac{1}{3}\Theta_{\beta^-,\beta^+}(\Xi^0_c,K^+,\Xi^-,\Lambda^0, \pi^0,\Sigma^0),\nonumber\\
  S(A_{14})_2[u,s,u,u]\,&=\, -\frac{1}{3}\Theta_{\beta^-,\beta^-}(\Xi^0_c,K^+,\Xi^-,\Lambda^0, \pi^0,\Sigma^0),\nonumber\\
  S(A_{14})_1[u,s,d,d]\,&=\, -\frac{1}{3}\Theta_{\beta^-,\beta^+}(\Xi^0_c,K^+,\Xi^-,\Lambda^0, \pi^0,\Sigma^0),\nonumber\\
  S(A_{14})_2[u,s,d,d]\,&=\, -\frac{1}{3}\Theta_{\beta^-,\beta^-}(\Xi^0_c,K^+,\Xi^-,\Lambda^0, \pi^0,\Sigma^0),\nonumber\\
   S(A_{14})_3[u,s,d,d]\,&=\, -\frac{1}{2}\Theta_{\beta^+,\beta^+}(\Xi^0_c,K^+,\Xi^-,\Sigma^0, \pi^0,\Sigma^0)+\frac{1}{6}\Theta_{\beta^+,\beta^+}(\Xi^0_c,K^+,\Xi^-,\Lambda^0, \pi^0,\Sigma^0),\nonumber\\
   S(A_{14})_4[u,s,d,d]\,&=\, -\frac{1}{2}\Theta_{\beta^+,\beta^-}(\Xi^0_c,K^+,\Xi^-,\Sigma^0, \pi^0,\Sigma^0)+\frac{1}{6}\Theta_{\beta^+,\beta^-}(\Xi^0_c,K^+,\Xi^-,\Lambda^0, \pi^0,\Sigma^0).
\end{align}
Summing all long-distance contributions, we have
\begin{align}
  \mathcal{A}_L(\Xi_{c}^{0}\to \Sigma^0\pi^0)&=-\lambda_d\Delta_{\alpha^+,(\gamma^+-\gamma^-)}(\Xi^0_c,\pi^+,\Sigma^-,\rho^+, \pi^0,\Sigma^0)\nonumber\\&~~-\frac{1}{2}\lambda_d\Delta_{(\beta^+-\beta^-),(\beta^+-\beta^-)}(\Xi^0_c,\pi^+,\Sigma^-,\Sigma^-, \Sigma^0,\pi^0)\nonumber\\&~~-\frac{1}{2}\lambda_s\Delta_{\alpha^+,\gamma^+}(\Xi^0_c,K^+,\Xi^-,K^{*+}, \pi^0,\Sigma^0)\nonumber\\&~~+\frac{1}{2}\lambda_s\Delta_{\beta^+,\beta^-}(\Xi^0_c,K^+,\Xi^-,\Xi^-, \Sigma^0,\pi^0)\nonumber\\&~~+\frac{1}{6}\lambda_d\Theta_{(\beta^++\beta^-),
  (\beta^++\beta^-)}(\Xi^0_c,\pi^+,\Sigma^-,\Lambda^0, \pi^0,\Sigma^0)\nonumber\\&~~+\frac{1}{6}\lambda_s\Theta_{(\beta^+-2\beta^-),(\beta^++\beta^-)}(\Xi^0_c,K^+,\Xi^-,\Lambda^0, \pi^0,\Sigma^0).
\end{align}
The decay amplitude for the $\Xi_{c}^{0}\to \Sigma^+\pi^-$ mode is $\lambda_d(A_1+A_6+A_{11}+A_{14})+\lambda_s(A_{11}+A_{14})$.
The rescattering amplitudes contributing to the $\Xi_{c}^{0}\to \Sigma^+\pi^-$ decay include
\begin{align}
  T(A_1)_1[u,d,u,d]\,&=\, -\frac{1}{2}\Delta_{\beta^-,\beta^-}(\Xi^0_c,\pi^+,\Sigma^-,\Sigma^0, \Sigma^+,\pi^-)+\frac{1}{6}\Delta_{\beta^-,\beta^-}(\Xi^0_c,\pi^+,\Sigma^-,\Lambda^0, \Sigma^+,\pi^-),\nonumber\\
  T(A_1)_4[u,d,u,d]\,&=\, -\frac{1}{2}\Delta_{\beta^+,\beta^+}(\Xi^0_c,\pi^+,\Sigma^-,\Sigma^0, \Sigma^+,\pi^-)+\frac{1}{6}\Delta_{\beta^+,\beta^+}(\Xi^0_c,\pi^+,\Sigma^-,\Lambda^0, \Sigma^+,\pi^-),\nonumber\\
  T(A_6)[u,d,u,d]\,&=\, \frac{1}{2}\Delta_{\beta^-,\beta^+}(\Xi^0_c,\pi^+,\Sigma^-,\Sigma^0, \Sigma^+,\pi^-)+\frac{1}{6}\Delta_{\beta^-,\beta^+}(\Xi^0_c,\pi^+,\Sigma^-,\Lambda^0, \Sigma^+,\pi^-),\nonumber\\
  S(A_6)[u,d,u,d]\,&=\, \frac{1}{2}\Theta_{\beta^-,\beta^+}(\Xi^0_c,\pi^+,\Sigma^-,\Sigma^0, \pi^-, \Sigma^+)+\frac{1}{6}\Theta_{\beta^-,\beta^+}(\Xi^0_c,\pi^+,\Sigma^-,\Lambda^0,\pi^-, \Sigma^+),\nonumber\\
  T(A_{11})[u,d,u,d]\,&=\, \frac{1}{2}\Delta_{\beta^+,\beta^-}(\Xi^0_c,\pi^+,\Sigma^-,\Sigma^0, \Sigma^+,\pi^-)+\frac{1}{6}\Delta_{\beta^+,\beta^-}(\Xi^0_c,\pi^+,\Sigma^-,\Lambda^0, \Sigma^+,\pi^-),\nonumber\\
  S(A_{11})[u,d,u,d]\,&=\, \frac{1}{2}\Theta_{\beta^+,\beta^-}(\Xi^0_c,\pi^+,\Sigma^-,\Sigma^0, \pi^-, \Sigma^+)+\frac{1}{6}\Theta_{\beta^+,\beta^-}(\Xi^0_c,\pi^+,\Sigma^-,\Lambda^0,\pi^-, \Sigma^+),\nonumber\\
  T(A_{11})[u,s,u,d]\,&=\, \Delta_{\beta^+,\beta^-}(\Xi^0_c,K^+,\Xi^-,\Xi^0, \Sigma^+,\pi^-),\nonumber\\
   S(A_{11})[u,s,u,d]\,&=\, \frac{1}{2}\Theta_{\beta^+,\beta^-}(\Xi^0_c,K^+,\Xi^-,\Sigma^0, \pi^-, \Sigma^+)+\frac{1}{6}\Theta_{\beta^+,\beta^-}(\Xi^0_c,K^+,\Xi^-,\Lambda^0,\pi^-, \Sigma^+),\nonumber\\
  S(A_{14})_2[u,d,u,d]\,&=\, -\frac{1}{2}\Theta_{\beta^-,\beta^-}(\Xi^0_c,\pi^+,\Sigma^-,\Sigma^0, \pi^-, \Sigma^+)+\frac{1}{6}\Theta_{\beta^-,\beta^-}(\Xi^0_c,\pi^+,\Sigma^-,\Lambda^0,\pi^-, \Sigma^+),\nonumber\\
  S(A_{14})_3[u,d,u,d]\,&=\, -\frac{1}{2}\Theta_{\beta^+,\beta^+}(\Xi^0_c,\pi^+,\Sigma^-,\Sigma^0, \pi^-, \Sigma^+)+\frac{1}{6}\Theta_{\beta^+,\beta^+}(\Xi^0_c,\pi^+,\Sigma^-,\Lambda^0,\pi^-, \Sigma^+),\nonumber\\
   S(A_{14})_1[u,s,u,d]\,&=\, -\frac{1}{3}\Theta_{\beta^-,\beta^+}(\Xi^0_c,K^+,\Xi^-,\Lambda^0,\pi^-, \Sigma^+),\nonumber\\
   S(A_{14})_2[u,s,u,d]\,&=\, -\frac{1}{3}\Theta_{\beta^-,\beta^-}(\Xi^0_c,K^+,\Xi^-,\Lambda^0,\pi^-, \Sigma^+),\nonumber\\
   S(A_{14})_3[u,s,u,d]\,&=\, -\frac{1}{2}\Theta_{\beta^+,\beta^+}(\Xi^0_c,K^+,\Xi^-,\Sigma^0, \pi^-, \Sigma^+)+\frac{1}{6}\Theta_{\beta^+,\beta^+}(\Xi^0_c,K^+,\Xi^-,\Lambda^0,\pi^-, \Sigma^+).
\end{align}
Summing all long-distance contributions, we have
\begin{align}
  \mathcal{A}_L(\Xi_{c}^{0}\to \Sigma^+\pi^-)&=-\frac{1}{2}\lambda_d\Delta_{(\beta^+-\beta^-),(\beta^+-\beta^-)}(\Xi^0_c,\pi^+,\Sigma^-,\Sigma^0, \Sigma^+,\pi^-)\nonumber\\&~~+\frac{1}{6}\lambda_d\Delta_{(\beta^++\beta^-),(\beta^++\beta^-)}(\Xi^0_c,\pi^+,\Sigma^-,\Lambda^0, \Sigma^+,\pi^-)\nonumber\\&~~+\lambda_s\Delta_{\beta^+,\beta^-}(\Xi^0_c,K^+,\Xi^-,\Xi^0, \Sigma^+,\pi^-)\nonumber\\&~~-\frac{1}{2}\lambda_d\Theta_{(\beta^+-\beta^-),(\beta^+-\beta^-)}(\Xi^0_c,\pi^+,\Sigma^-,\Sigma^0, \pi^-, \Sigma^+)\nonumber\\&~~+\frac{1}{6}\lambda_d\Theta_{(\beta^++\beta^-),(\beta^++\beta^-)}(\Xi^0_c,\pi^+,\Sigma^-,\Lambda^0,\pi^-, \Sigma^+)\nonumber\\&~~-\frac{1}{2}\lambda_s\Theta_{\beta^+,(\beta^+-\beta^-)}(\Xi^0_c,K^+,\Xi^-,\Sigma^0, \pi^-, \Sigma^+)\nonumber\\&~~+\frac{1}{6}\lambda_s\Theta_{(\beta^+-2\beta^-),(\beta^++\beta^-)}(\Xi^0_c,K^+,\Xi^-,\Lambda^0,\pi^-, \Sigma^+).
\end{align}
The decay amplitude for the $\Xi_{c}^{+}\to \Sigma^0\pi^+$ mode is $\frac{1}{\sqrt{2}}\lambda_d(A_2+A_3-A_7+A_8-A_{11}+A_{12})
+\frac{1}{\sqrt{2}}\lambda_s(-A_{11}+A_{12})$.
The rescattering amplitudes contributing to the $\Xi_{c}^{+}\to \Sigma^0\pi^+$ decay include
\begin{align*}
  T(A_2)_1[d,d,d,u]\,&=\, -\frac{1}{4}\Delta_{\alpha^+,\gamma^+}(\Xi^+_c,\pi^+,\Sigma^0,\rho^0, \pi^+,\Sigma^0)+\frac{1}{4}\Delta_{\alpha^+,\gamma^+}(\Xi^+_c,\pi^+,\Sigma^0,\omega, \pi^+,\Sigma^0)\nonumber\\&~~~~-\frac{1}{12}\Delta_{\alpha^+,\gamma^+}(\Xi^+_c,\pi^+,\Lambda^0,\rho^0, \pi^+,\Sigma^0)+\frac{1}{12}\Delta_{\alpha^+,\gamma^+}(\Xi^+_c,\pi^+,\Lambda^0,\omega, \pi^+,\Sigma^0),\nonumber\\
  T(A_2)_2[d,d,d,u]\,&=\, -\frac{1}{4}\Delta_{\alpha^+,\gamma^-}(\Xi^+_c,\pi^+,\Sigma^0,\rho^0, \pi^+,\Sigma^0)+\frac{1}{4}\Delta_{\alpha^+,\gamma^-}(\Xi^+_c,\pi^+,\Sigma^0,\omega, \pi^+,\Sigma^0)\nonumber\\&~~~~-\frac{1}{12}\Delta_{\alpha^+,\gamma^-}(\Xi^+_c,\pi^+,\Lambda^0,\rho^0, \pi^+,\Sigma^0)+\frac{1}{12}\Delta_{\alpha^+,\gamma^-}(\Xi^+_c,\pi^+,\Lambda^0,\omega, \pi^+,\Sigma^0),\nonumber\\
  T(A_2)_5[d,d,d,u]\,&=\, \frac{1}{2}\Delta_{\alpha^+,\gamma^0}(\Xi^+_c,\pi^+,\Sigma^0,\omega, \pi^+,\Sigma^0)+\frac{1}{6}\Delta_{\alpha^+,\gamma^0}(\Xi^+_c,\pi^+,\Lambda^0,\omega, \pi^+,\Sigma^0),\nonumber\\
  T(A_2)_6[d,d,d,u]\,&=\, -\frac{1}{2}\Delta_{\alpha^+,\gamma^0}(\Xi^+_c,\pi^+,\Sigma^0,\omega, \pi^+,\Sigma^0)-\frac{1}{6}\Delta_{\alpha^+,\gamma^0}(\Xi^+_c,\pi^+,\Lambda^0,\omega, \pi^+,\Sigma^0),\nonumber\\
  T(A_3)_1[d,d,d,u]\,&=\, -\frac{1}{4}\Delta_{\alpha^+,\gamma^-}(\Xi^+_c,\pi^+,\Sigma^0,\rho^0, \pi^+,\Sigma^0)-\frac{1}{4}\Delta_{\alpha^+,\gamma^-}(\Xi^+_c,\pi^+,\Sigma^0,\omega, \pi^+,\Sigma^0)\nonumber\\&~~~~-\frac{1}{12}\Delta_{\alpha^+,\gamma^-}(\Xi^+_c,\pi^+,\Lambda^0,\rho^0, \pi^+,\Sigma^0)-\frac{1}{12}\Delta_{\alpha^+,\gamma^-}(\Xi^+_c,\pi^+,\Lambda^0,\omega, \pi^+,\Sigma^0),\nonumber\\
  T(A_3)_2[d,d,d,u]\,&=\, \frac{1}{2}\Delta_{\beta^-,\beta^-}(\Xi^+_c,\pi^+,\Sigma^0,\Sigma^-,\Sigma^0, \pi^+)+\frac{1}{6}\Delta_{\beta^-,\beta^-}(\Xi^+_c,\pi^+,\Lambda^0,\Sigma^-,\Sigma^0, \pi^+),\nonumber\\
  T(A_7)_1[d,d,u,u]\,&=\, -\frac{1}{4}\Delta_{\alpha^+,\gamma^-}(\Xi^+_c,\pi^+,\Sigma^0,\rho^0, \pi^+,\Sigma^0)-\frac{1}{4}\Delta_{\alpha^+,\gamma^-}(\Xi^+_c,\pi^+,\Sigma^0,\omega, \pi^+,\Sigma^0)\nonumber\\&~~~~+\frac{1}{12}\Delta_{\alpha^+,\gamma^-}(\Xi^+_c,\pi^+,\Lambda^0,\rho^0, \pi^+,\Sigma^0)+\frac{1}{12}\Delta_{\alpha^+,\gamma^-}(\Xi^+_c,\pi^+,\Lambda^0,\omega, \pi^+,\Sigma^0),\nonumber\\
  T(A_7)_2[d,d,u,u]\,&=\, -\frac{1}{4}\Delta_{\alpha^+,\gamma^+}(\Xi^+_c,\pi^+,\Sigma^0,\rho^0, \pi^+,\Sigma^0)-\frac{1}{4}\Delta_{\alpha^+,\gamma^+}(\Xi^+_c,\pi^+,\Sigma^0,\omega, \pi^+,\Sigma^0)\nonumber\\&~~~~+\frac{1}{12}\Delta_{\alpha^+,\gamma^+}(\Xi^+_c,\pi^+,\Lambda^0,\rho^0, \pi^+,\Sigma^0)+\frac{1}{12}\Delta_{\alpha^+,\gamma^+}(\Xi^+_c,\pi^+,\Lambda^0,\omega, \pi^+,\Sigma^0),\nonumber\\
  T(A_7)_3[d,d,u,u]\,&=\, -\frac{1}{2}\Delta_{\beta^+,\beta^+}(\Xi^+_c,\pi^+,\Sigma^0,\Sigma^-,\Sigma^0, \pi^+)+\frac{1}{6}\Delta_{\beta^+,\beta^+}(\Xi^+_c,\pi^+,\Lambda^0,\Sigma^-,\Sigma^0, \pi^+),\nonumber\\
 S(A_7)_{1}[d,d,u,u]\,&=\, -\frac{1}{2}\Theta_{\beta^-,\beta^-}(\Xi^+_c,\pi^+,\Sigma^0,\Sigma^+, \pi^+,\Sigma^0)+\frac{1}{6}\Theta_{\beta^-,\beta^-}(\Xi^+_c,\pi^+,\Lambda^0,\Sigma^+, \pi^+,\Sigma^0),\nonumber\\
  T(A_8)_3[d,d,u,u]\,&=\, -\frac{1}{2}\Delta_{\beta^-,\beta^+}(\Xi^+_c,\pi^+,\Sigma^0,\Sigma^-,\Sigma^0, \pi^+)+\frac{1}{6}\Delta_{\beta^-,\beta^+}(\Xi^+_c,\pi^+,\Lambda^0,\Sigma^-,\Sigma^0, \pi^+),\nonumber\\
 S(A_8)_{1}[d,d,u,u]\,&=\, -\frac{1}{2}\Theta_{\beta^-,\beta^+}(\Xi^+_c,\pi^+,\Sigma^0,\Sigma^+, \pi^+,\Sigma^0)+\frac{1}{6}\Theta_{\beta^-,\beta^+}(\Xi^+_c,\pi^+,\Lambda^0,\Sigma^+, \pi^+,\Sigma^0), \nonumber\\ T(A_{11})[d,d,u,u]\,&=\, \frac{1}{2}\Delta_{\beta^+,\beta^-}(\Xi^+_c,\pi^+,\Sigma^0,\Sigma^-,\Sigma^0, \pi^+)+\frac{1}{6}\Delta_{\beta^+,\beta^-}(\Xi^+_c,\pi^+,\Lambda^0,\Sigma^-,\Sigma^0, \pi^+),\nonumber\\
  S(A_{11})[d,d,u,u]\,&=\, \frac{1}{2}\Theta_{\beta^+,\beta^-}(\Xi^+_c,\pi^+,\Sigma^0,\Sigma^+,\pi^+,\Sigma^0 )+\frac{1}{6}\Theta_{\beta^+,\beta^-}(\Xi^+_c,\pi^+,\Lambda^0,\Sigma^+,\pi^+,\Sigma^0),\nonumber\\
  T(A_{11})[d,s,u,u]\,&=\, \Delta_{\beta^+,\beta^-}(\Xi^+_c,K^+,\Xi^0,\Xi^-,\Sigma^0, \pi^+),\nonumber\\
  S(A_{11})[d,s,u,u]\,&=\, \Theta_{\beta^+,\beta^-}(\Xi^+_c,K^+,\Xi^0,\Sigma^+,\pi^+,\Sigma^0 ),\nonumber\\
  T(A_{12})[d,d,d,u]\,&=
  -\frac{1}{4}\Delta_{\alpha^+,\gamma^+}(\Xi^+_c,\pi^+,\Sigma^0,\rho^0, \pi^+,\Sigma^0)-\frac{1}{4}\Delta_{\alpha^+,\gamma^+}(\Xi^+_c,\pi^+,\Sigma^0,\omega, \pi^+,\Sigma^0)\nonumber\\&~~~~-\frac{1}{12}\Delta_{\alpha^+,\gamma^+}(\Xi^+_c,\pi^+,\Lambda^0,\rho^0, \pi^+,\Sigma^0)-\frac{1}{12}\Delta_{\alpha^+,\gamma^+}(\Xi^+_c,\pi^+,\Lambda^0,\omega, \pi^+,\Sigma^0),\nonumber\\
  S(A_{12})[d,d,d,u]\,&=\, \frac{1}{2}\Theta_{\beta^+,\beta^+}(\Xi^+_c,\pi^+,\Sigma^0,\Sigma^+,\pi^+,\Sigma^0 )+\frac{1}{6}\Theta_{\beta^+,\beta^+}(\Xi^+_c,\pi^+,\Lambda^0,\Sigma^+,\pi^+,\Sigma^0),
  \nonumber\\
  T(A_{12})[d,s,d,u]\,&=\, -\Delta_{\alpha^+,\gamma^+}(\Xi^+_c,K^+,\Xi^0,K^{*0}, \pi^+,\Sigma^0),\nonumber\\
  S(A_{12})[d,s,u,u]\,&=\, \Theta_{\beta^+,\beta^+}(\Xi^+_c,K^+,\Xi^0,\Sigma^+,\pi^+,\Sigma^0 ),
\end{align*}
\begin{align}
  T(A_{19})_3[d,d,u,u]\,&=\, -\frac{1}{4}\Delta_{\alpha^+,\gamma^-}(\Xi^+_c,\pi^+,\Sigma^0,\rho^0, \pi^+,\Sigma^0)+\frac{1}{4}\Delta_{\alpha^+,\gamma^-}(\Xi^+_c,\pi^+,\Sigma^0,\omega, \pi^+,\Sigma^0)\nonumber\\&~~~~+\frac{1}{12}\Delta_{\alpha^+,\gamma^-}(\Xi^+_c,\pi^+,\Lambda^0,\rho^0, \pi^+,\Sigma^0)-\frac{1}{12}\Delta_{\alpha^+,\gamma^-}(\Xi^+_c,\pi^+,\Lambda^0,\omega, \pi^+,\Sigma^0),\nonumber\\
  T(A_{19})_4[d,d,u,u]\,&=\, -\frac{1}{4}\Delta_{\alpha^+,\gamma^+}(\Xi^+_c,\pi^+,\Sigma^0,\rho^0, \pi^+,\Sigma^0)+\frac{1}{4}\Delta_{\alpha^+,\gamma^+}(\Xi^+_c,\pi^+,\Sigma^0,\omega, \pi^+,\Sigma^0)\nonumber\\&~~~~+\frac{1}{12}\Delta_{\alpha^+,\gamma^+}(\Xi^+_c,\pi^+,\Lambda^0,\rho^0, \pi^+,\Sigma^0)-\frac{1}{12}\Delta_{\alpha^+,\gamma^+}(\Xi^+_c,\pi^+,\Lambda^0,\omega, \pi^+,\Sigma^0),\nonumber\\
   T(A_{19})_5[d,d,u,u]\,&=\, -\frac{1}{4}\Delta_{\alpha^+,\gamma^0}(\Xi^+_c,\pi^+,\Sigma^0,\omega, \pi^+,\Sigma^0)+\frac{1}{12}\Delta_{\alpha^+,\gamma^0}(\Xi^+_c,\pi^+,\Lambda^0,\omega, \pi^+,\Sigma^0),\nonumber\\
 T(A_{19})_{6}[d,d,u,u]\,&=\, \frac{1}{4}\Delta_{\alpha^+,\gamma^0}(\Xi^+_c,\pi^+,\Sigma^0,\omega, \pi^+,\Sigma^0)-\frac{1}{12}\Delta_{\alpha^+,\gamma^0}(\Xi^+_c,\pi^+,\Lambda^0,\omega, \pi^+,\Sigma^0).
\end{align}
Summing all long-distance contributions, we have
\begin{align}
  \mathcal{A}_L(\Xi_{c}^{+}\to \Sigma^0\pi^+)&=-\frac{1}{3\sqrt{2}}\lambda_d\Delta_{\alpha^+,(\gamma^++\gamma^-)}(\Xi^+_c,\pi^+,\Lambda^0,\rho^0, \pi^+,\Sigma^0)\nonumber\\&~~+\frac{1}{2\sqrt{2}}\lambda_d\Delta_{(\beta^+-\beta^-),(\beta^+-\beta^-)}(\Xi^+_c,\pi^+,\Sigma^0,\Sigma^-,\Sigma^0, \pi^+)\nonumber\\&~~-\frac{1}{6\sqrt{2}}\lambda_d\Delta_{(\beta^+-\beta^-),(\beta^++\beta^-)}(\Xi^+_c,\pi^+,\Lambda^0,\Sigma^-,\Sigma^0, \pi^+)\nonumber\\&~~-\frac{1}{\sqrt{2}}\lambda_s\Delta_{\beta^+,\beta^-}(\Xi^+_c,K^+,\Xi^0,\Xi^-,\Sigma^0, \pi^+)\nonumber\\&~~-\frac{1}{\sqrt{2}}\lambda_s\Delta_{\alpha^+,\gamma^+}(\Xi^+_c,K^+,\Xi^0,K^{*0}, \pi^+,\Sigma^0)\nonumber\\&~~+\frac{1}{2\sqrt{2}}\lambda_d\Theta_{(\beta^+-\beta^-),(\beta^+-\beta^-)}(\Xi^+_c,\pi^+,\Sigma^0,\Sigma^+,\pi^+,\Sigma^0 )\nonumber\\&~~+\frac{1}{6\sqrt{2}}\lambda_d\Theta_{(\beta^++\beta^-),
  (\beta^+-\beta^-)}(\Xi^+_c,\pi^+,\Lambda^0,\Sigma^+,\pi^+,\Sigma^0)
  \nonumber\\&~~+\frac{1}{\sqrt{2}}\lambda_s\Theta_{\beta^+,(\beta^+-\beta^-)}(\Xi^+_c,K^+,\Xi^0,\Sigma^+,\pi^+,\Sigma^0 ).
\end{align}
The decay amplitude for the $\Xi_{c}^{+}\to \Sigma^+\pi^0$ mode is $\frac{1}{\sqrt{2}}\lambda_d(A_1+A_4+A_7-A_8+A_{11}-A_{12})
+\frac{1}{\sqrt{2}}\lambda_s(A_{11}-A_{12})$.
The rescattering amplitudes contributing to the $\Xi_{c}^{+}\to \Sigma^+\pi^0$ decay include
\begin{align}
  T(A_1)_1[d,d,u,d]\,&=\, -\frac{1}{4}\Delta_{\beta^-,\beta^-}(\Xi^+_c,\pi^+,\Sigma^0,\Sigma^0, \Sigma^+,\pi^0)+\frac{1}{12}\Delta_{\beta^-,\beta^-}(\Xi^+_c,\pi^+,\Sigma^0,\Lambda^0, \Sigma^+,\pi^0)\nonumber\\&~~~~-\frac{1}{12}\Delta_{\beta^-,\beta^-}(\Xi^+_c,\pi^+,\Lambda^0,\Sigma^0, \Sigma^+,\pi^0)+\frac{1}{36}\Delta_{\beta^-,\beta^-}(\Xi^+_c,\pi^+,\Lambda^0,\Lambda^0, \Sigma^+,\pi^0),\nonumber\\
  T(A_1)_2[d,d,u,d]\,&=\, -\frac{1}{4}\Delta_{\beta^-,\beta^+}(\Xi^+_c,\pi^+,\Sigma^0,\Sigma^0, \Sigma^+,\pi^0)+\frac{1}{12}\Delta_{\beta^-,\beta^+}(\Xi^+_c,\pi^+,\Sigma^0,\Lambda^0, \Sigma^+,\pi^0)\nonumber\\&~~~~-\frac{1}{12}\Delta_{\beta^-,\beta^+}(\Xi^+_c,\pi^+,\Lambda^0,\Sigma^0, \Sigma^+,\pi^0)+\frac{1}{36}\Delta_{\beta^-,\beta^+}(\Xi^+_c,\pi^+,\Lambda^0,\Lambda^0, \Sigma^+,\pi^0),\nonumber\\
  U(A_4)[d,d,u,d]\,&=\, \frac{1}{2}\Delta_{\alpha^+,\gamma^+}(\Xi^+_c,\pi^+,\Sigma^0,\rho^+,\pi^0, \Sigma^+)+\frac{1}{6}\Delta_{\alpha^+,\gamma^+}(\Xi^+_c,\pi^+,\Lambda^0,\rho^+,\pi^0, \Sigma^+),\nonumber\\
  T(A_4)[d,d,u,d]\,&=\, \frac{1}{4}\Delta_{\beta^+,\beta^+}(\Xi^+_c,\pi^+,\Sigma^0,\Sigma^0, \Sigma^+,\pi^0)+\frac{1}{12}\Delta_{\beta^+,\beta^+}(\Xi^+_c,\pi^+,\Sigma^0,\Lambda^0, \Sigma^+,\pi^0)\nonumber\\&~~~~+\frac{1}{12}\Delta_{\beta^+,\beta^+}(\Xi^+_c,\pi^+,\Lambda^0,\Sigma^0, \Sigma^+,\pi^0)+\frac{1}{36}\Delta_{\beta^+,\beta^+}(\Xi^+_c,\pi^+,\Lambda^0,\Lambda^0, \Sigma^+,\pi^0),\nonumber\\
   U(A_7)_1[d,d,u,d]\,&=\, -\frac{1}{2}\Delta_{\alpha^+,\gamma^-}(\Xi^+_c,\pi^+,\Sigma^0,\rho^+,\pi^0, \Sigma^+)+\frac{1}{6}\Delta_{\alpha^+,\gamma^-}(\Xi^+_c,\pi^+,\Lambda^0,\rho^+,\pi^0, \Sigma^+),\nonumber\\
 S(A_{7})_1[d,d,u,d]\,&=\, -\frac{1}{2}\Theta_{\beta^-,\beta^-}(\Xi^+_c,\pi^+,\Sigma^0,\Sigma^+,\pi^0, \Sigma^+)+\frac{1}{6}\Theta_{\beta^-,\beta^-}(\Xi^+_c,\pi^+,\Lambda^0,\Sigma^+,\pi^0, \Sigma^+),\nonumber\\
    U(A_8)_1[d,d,u,u]\,&=\, \frac{1}{2}\Delta_{\alpha^+,\gamma^-}(\Xi^+_c,\pi^+,\Sigma^0,\rho^+,\pi^0, \Sigma^+)-\frac{1}{6}\Delta_{\alpha^+,\gamma^-}(\Xi^+_c,\pi^+,\Lambda^0,\rho^+,\pi^0, \Sigma^+),\nonumber\\
  T(A_8)_1[d,d,u,u]\,&=\, -\frac{1}{4}\Delta_{\beta^-,\beta^+}(\Xi^+_c,\pi^+,\Sigma^0,\Sigma^0, \Sigma^+,\pi^0)-\frac{1}{12}\Delta_{\beta^-,\beta^+}(\Xi^+_c,\pi^+,\Sigma^0,\Lambda^0, \Sigma^+,\pi^0)\nonumber\\&~~~~+\frac{1}{12}\Delta_{\beta^-,\beta^+}(\Xi^+_c,\pi^+,\Lambda^0,\Sigma^0, \Sigma^+,\pi^0)+\frac{1}{36}\Delta_{\beta^-,\beta^+}(\Xi^+_c,\pi^+,\Lambda^0,\Lambda^0, \Sigma^+,\pi^0),\nonumber\\
  T(A_8)_2[d,d,u,u]\,&=\, -\frac{1}{4}\Delta_{\beta^-,\beta^-}(\Xi^+_c,\pi^+,\Sigma^0,\Sigma^0, \Sigma^+,\pi^0)-\frac{1}{12}\Delta_{\beta^-,\beta^-}(\Xi^+_c,\pi^+,\Sigma^0,\Lambda^0, \Sigma^+,\pi^0)\nonumber\\&~~~~+\frac{1}{12}\Delta_{\beta^-,\beta^-}(\Xi^+_c,\pi^+,\Lambda^0,\Sigma^0, \Sigma^+,\pi^0)+\frac{1}{36}\Delta_{\beta^-,\beta^-}(\Xi^+_c,\pi^+,\Lambda^0,\Lambda^0, \Sigma^+,\pi^0),\nonumber\\
 S(A_{8})_1[d,d,u,u]\,&=\, -\frac{1}{2}\Theta_{\beta^-,\beta^+}(\Xi^+_c,\pi^+,\Sigma^0,\Sigma^+,\pi^0, \Sigma^+)+\frac{1}{6}\Theta_{\beta^-,\beta^+}(\Xi^+_c,\pi^+,\Lambda^0,\Sigma^+,\pi^0, \Sigma^+),\nonumber\\
  T(A_{10})_3[d,d,u,u]\,&=\, \frac{1}{4}\Delta_{\beta^+,\beta^+}(\Xi^+_c,\pi^+,\Sigma^0,\Sigma^0, \Sigma^+,\pi^0)-\frac{1}{12}\Delta_{\beta^+,\beta^+}(\Xi^+_c,\pi^+,\Sigma^0,\Lambda^0, \Sigma^+,\pi^0)\nonumber\\&~~~~-\frac{1}{12}\Delta_{\beta^+,\beta^+}(\Xi^+_c,\pi^+,\Lambda^0,\Sigma^0, \Sigma^+,\pi^0)+\frac{1}{36}\Delta_{\beta^+,\beta^+}(\Xi^+_c,\pi^+,\Lambda^0,\Lambda^0, \Sigma^+,\pi^0),\nonumber\\
  T(A_{10})_4[d,d,u,u]\,&=\, \frac{1}{4}\Delta_{\beta^+,\beta^-}(\Xi^+_c,\pi^+,\Sigma^0,\Sigma^0, \Sigma^+,\pi^0)-\frac{1}{12}\Delta_{\beta^+,\beta^-}(\Xi^+_c,\pi^+,\Sigma^0,\Lambda^0, \Sigma^+,\pi^0)\nonumber\\&~~~~-\frac{1}{12}\Delta_{\beta^+,\beta^-}(\Xi^+_c,\pi^+,\Lambda^0,\Sigma^0, \Sigma^+,\pi^0)+\frac{1}{36}\Delta_{\beta^+,\beta^-}(\Xi^+_c,\pi^+,\Lambda^0,\Lambda^0, \Sigma^+,\pi^0),\nonumber\\
  T(A_{11})[d,d,u,d]\,&=\, \frac{1}{4}\Delta_{\beta^+,\beta^-}(\Xi^+_c,\pi^+,\Sigma^0,\Sigma^0, \Sigma^+,\pi^0)+\frac{1}{12}\Delta_{\beta^+,\beta^-}(\Xi^+_c,\pi^+,\Sigma^0,\Lambda^0, \Sigma^+,\pi^0)\nonumber\\&~~~~+\frac{1}{12}\Delta_{\beta^+,\beta^-}(\Xi^+_c,\pi^+,\Lambda^0,\Sigma^0, \Sigma^+,\pi^0)+\frac{1}{36}\Delta_{\beta^+,\beta^-}(\Xi^+_c,\pi^+,\Lambda^0,\Lambda^0, \Sigma^+,\pi^0),\nonumber\\
  S(A_{11})[d,d,u,d]\,&=\, \frac{1}{2}\Theta_{\beta^+,\beta^-}(\Xi^+_c,\pi^+,\Sigma^0,\Sigma^+,\pi^0, \Sigma^+)+\frac{1}{6}\Theta_{\beta^+,\beta^-}(\Xi^+_c,\pi^+,\Lambda^0,\Sigma^+,\pi^0, \Sigma^+)\nonumber\\
  T(A_{11})[d,s,u,d]\,&=\, \Delta_{\beta^+,\beta^-}(\Xi^+_c,K^+,\Xi^0,\Xi^0, \Sigma^+,\pi^0),\nonumber\\
  S(A_{11})[d,s,u,d]\,&=\, \Theta_{\beta^+,\beta^-}(\Xi^+_c,K^+,\Xi^0,\Sigma^+,\pi^0, \Sigma^+),\\\nonumber
   U(A_{12})[d,d,u,u]\,&=\, -\frac{1}{2}\Delta_{\alpha^+,\gamma^+}(\Xi^+_c,\pi^+,\Sigma^0,\rho^+,\pi^0, \Sigma^+)-\frac{1}{6}\Delta_{\alpha^+,\gamma^+}(\Xi^+_c,\pi^+,\Lambda^0,\rho^+,\pi^0, \Sigma^+),\nonumber\\
  S(A_{12})[d,d,u,u]\,&=\, \frac{1}{2}\Theta_{\beta^+,\beta^+}(\Xi^+_c,\pi^+,\Sigma^0,\Sigma^+,\pi^0, \Sigma^+)+\frac{1}{6}\Theta_{\beta^+,\beta^+}(\Xi^+_c,\pi^+,\Lambda^0,\Sigma^+,\pi^0, \Sigma^+),\nonumber\\
   U(A_{12})[d,s,u,u]\,&=\, -\Delta_{\alpha^+,\gamma^+}(\Xi^+_c,K^+,\Xi^0,K^{*+},\pi^0, \Sigma^+),\nonumber\\
   S(A_{12})[d,s,u,u]\,&=\, \Theta_{\beta^+,\beta^+}(\Xi^+_c,K^+,\Xi^0,\Sigma^+,\pi^0, \Sigma^+).
\end{align}
Summing all long-distance contributions, we have
\begin{align}
  \mathcal{A}_L(\Xi_{c}^{+}\to \Sigma^+\pi^0)&=\frac{1}{\sqrt{2}}\lambda_d\Delta_{\alpha^+,(\gamma^+-\gamma^-)}(\Xi^+_c,\pi^+,\Sigma^0,\rho^+,\pi^0, \Sigma^+)\nonumber\\&~~+\frac{1}{3\sqrt{2}}\lambda_d\Delta_{\alpha^+,(\gamma^++\gamma^-)}(\Xi^+_c,\pi^+,\Lambda^0,\rho^+,\pi^0, \Sigma^+)\nonumber\\&~~+\frac{1}{6\sqrt{2}}\lambda_d\Delta_{(\beta^++\beta^-),(\beta^++\beta^-)}(\Xi^+_c,\pi^+,\Sigma^0,\Lambda^0, \Sigma^+,\pi^0)\nonumber\\&~~+\frac{1}{6\sqrt{2}}\lambda_d\Delta_{(\beta^+-\beta^-),(\beta^++\beta^-)}(\Xi^+_c,\pi^+,\Lambda^0,\Sigma^0, \Sigma^+,\pi^0)\nonumber\\&~~+\frac{1}{\sqrt{2}}\lambda_s\Delta_{\alpha^+,\gamma^+}(\Xi^+_c,K^+,\Xi^0,K^{*+},\pi^0, \Sigma^+)\nonumber\\&~~+\frac{1}{\sqrt{2}}\lambda_s\Delta_{\beta^+,\beta^-}(\Xi^+_c,K^+,\Xi^0,\Xi^0, \Sigma^+,\pi^0)\nonumber\\&~~-\frac{1}{2\sqrt{2}}\lambda_d\Theta_{(\beta^+-\beta^-),(\beta^+-\beta^-)}(\Xi^+_c,\pi^+,\Sigma^0,\Sigma^+,\pi^0, \Sigma^+)\nonumber\\&~~-\frac{1}{6\sqrt{2}}\lambda_d\Theta_{(\beta^++\beta^-),(\beta^+-\beta^-)}(\Xi^+_c,\pi^+,\Lambda^0,\Sigma^+,\pi^0, \Sigma^+)\nonumber\\&~~-\frac{1}{\sqrt{2}}\lambda_s \Theta_{\beta^+,(\beta^+-\beta^-)}(\Xi^+_c,K^+,\Xi^0,\Sigma^+,\pi^0, \Sigma^+).
\end{align}
One can verify that the isospin sum rule
\begin{align}
 &2\,\mathcal{A}( \Xi^0_c\to\Sigma^0\pi^0)-\mathcal{A}(\Xi^0_c\to\Sigma^+\pi^-)
-\mathcal{A}(\Xi^0_c\to\Sigma^-\pi^+)
+\sqrt{2}\mathcal{A}(\Xi^+_c\to\Sigma^+\pi^0)+
\sqrt{2}\mathcal{A}(\Xi^+_c\to\Sigma^0\pi^+)=0
\end{align}
is satisfied.

\section{Fitting results for all decay modes}\label{fit}

\begin{table*}[htp!]
\centering
\caption{Theoretical predictions for Cabibbo-favored and doubly Cabibbo-suppressed decays in the $SU(3)_F$-symmetric fit.}
\label{tab:su3-cf}
\small
\renewcommand{\arraystretch}{1.08}
\begin{tabular}{lcccc}
\hline\hline
Decay mode~~~~~ & ~~~~~~~~~~~~~$\mathcal B$~~~~~~~~~~~~~~~ & ~~~~~~~~~~~~~~~$\alpha$~~~~~~~~~~~ & ~~~~~~~~~~~$\beta$~~~~~~~~~~~ & ~~~~~~~~~~~$\gamma$~~~~~~~~~~ \\
\hline
$\Lambda_c^+\to \Sigma^0\pi^+$ & $(1.235\pm0.043)\times10^{-2}$ & $-0.473\pm0.015$ & $0.553\pm0.100$ & $-0.686\pm0.081$ \\
$\Lambda_c^+\to \Sigma^+\pi^0$ & $(1.245\pm0.043)\times10^{-2}$ & $-0.471\pm0.015$ & $0.552\pm0.100$ & $-0.688\pm0.080$ \\
$\Lambda_c^+\to \Lambda^0\pi^+$ & $(1.263\pm0.046)\times10^{-2}$ & $-0.779\pm0.009$ & $0.383\pm0.014$ & $0.496\pm0.011$ \\
$\Lambda_c^+\to \Sigma^+\eta$ & $(3.515\pm0.393)\times10^{-3}$ & $-0.979\pm0.059$ & $0.010\pm0.691$ & $0.206\pm0.283$ \\
$\Lambda_c^+\to \Sigma^+\eta'$ & $(4.028\pm0.689)\times10^{-3}$ & $-0.462\pm0.070$ & $-0.277\pm0.552$ & $0.842\pm0.184$ \\
$\Lambda_c^+\to \Xi^0K^+$ & $(4.247\pm0.316)\times10^{-3}$ & $-0.189\pm0.038$ & $-0.965\pm0.014$ & $0.180\pm0.067$ \\
$\Lambda_c^+\to pK_S^0$ & $(1.662\pm0.051)\times10^{-2}$ & $-0.753\pm0.010$ & $0.443\pm0.412$ & $-0.486\pm0.376$ \\
$\Lambda_c^+\to pK_L^0$ & $(1.585\pm0.048)\times10^{-2}$ & $-0.751\pm0.022$ & $0.440\pm0.375$ & $-0.492\pm0.350$ \\
$\Xi_c^0\to \Xi^-\pi^+$ & $(1.116\pm0.192)\times10^{-2}$ & $-0.644\pm0.050$ & $0.763\pm0.098$ & $-0.051\pm1.413$ \\
$\Xi_c^0\to \Sigma^+K^-$ & $(1.062\pm0.325)\times10^{-3}$ & $0.230\pm0.138$ & $0.138\pm5.974$ & $0.963\pm0.872$ \\
$\Xi_c^0\to \Xi^0\pi^0$ & $(6.465\pm1.255)\times10^{-3}$ & $-0.667\pm0.191$ & $0.695\pm0.615$ & $-0.269\pm1.173$ \\
$\Xi_c^0\to \Sigma^0K_S^0$ & $(5.604\pm1.386)\times10^{-4}$ & $-0.719\pm1.095$ & $-0.287\pm3.510$ & $0.633\pm2.584$ \\
$\Xi_c^0\to \Sigma^0K_L^0$ & $(4.291\pm2.190)\times10^{-4}$ & $-0.696\pm1.395$ & $-0.465\pm3.029$ & $0.548\pm3.939$ \\
$\Xi_c^0\to \Lambda^0 K_S^0$ & $(3.621\pm0.541)\times10^{-3}$ & $-0.988\pm0.224$ & $0.153\pm1.519$ & $-0.035\pm1.477$ \\
$\Xi_c^0\to \Lambda^0 K_L^0$ & $(3.946\pm0.689)\times10^{-3}$ & $-0.989\pm0.229$ & $0.139\pm1.676$ & $-0.059\pm1.471$ \\
$\Xi_c^0\to \Xi^0\eta$ & $(1.571\pm0.398)\times10^{-3}$ & $-0.897\pm1.530$ & $-0.105\pm2.580$ & $0.429\pm2.785$ \\
$\Xi_c^0\to \Xi^0\eta'$ & $(1.284\pm0.383)\times10^{-3}$ & $-0.682\pm2.640$ & $-0.460\pm2.334$ & $-0.568\pm1.774$ \\
$\Xi_c^+\to \Xi^0\pi^+$ & $(7.474\pm1.081)\times10^{-3}$ & $-0.933\pm0.034$ & $0.349\pm0.119$ & $-0.086\pm0.249$ \\
$\Xi_c^+\to \Sigma^+K_S^0$ & $(2.502\pm0.831)\times10^{-3}$ & $-0.843\pm0.909$ & $0.535\pm1.556$ & $-0.057\pm1.544$ \\
$\Xi_c^+\to \Sigma^+K_L^0$ & $(1.004\pm0.610)\times10^{-3}$ & $-0.830\pm1.563$ & $0.547\pm2.058$ & $0.108\pm2.175$ \\
\hline
$\Lambda_c^+\to nK^+$ & $(1.078\pm0.202)\times10^{-5}$ & $-0.883\pm0.057$ & $0.330\pm0.128$ & $-0.334\pm0.223$ \\
$\Xi_c^0\to n\pi^0$ & $(0.412\pm2.274)\times10^{-5}$ & $0.894\pm0.549$ & $0.447\pm1.169$ & $-0.019\pm2.242$ \\
$\Xi_c^0\to p\pi^-$ & $(3.283\pm0.705)\times10^{-6}$ & $0.311\pm0.156$ & $0.186\pm8.008$ & $0.932\pm1.594$ \\
$\Xi_c^0\to \Sigma^-K^+$ & $(3.113\pm0.706)\times10^{-5}$ & $-0.640\pm0.085$ & $0.759\pm0.169$ & $-0.121\pm1.396$ \\
$\Xi_c^0\to n\eta$ & $(2.624\pm1.430)\times10^{-5}$ & $-0.884\pm0.495$ & $0.199\pm1.551$ & $-0.423\pm1.055$ \\
$\Xi_c^0\to n\eta'$ & $(1.863\pm0.989)\times10^{-5}$ & $-0.415\pm1.400$ & $0.042\pm1.088$ & $-0.909\pm0.651$ \\
$\Xi_c^+\to n\pi^+$ & $(7.060\pm0.461)\times10^{-5}$ & $-0.174\pm0.034$ & $-0.885\pm0.027$ & $-0.432\pm0.056$ \\
$\Xi_c^+\to p\pi^0$ & $(2.959\pm0.946)\times10^{-5}$ & $-0.115\pm0.504$ & $-0.134\pm0.368$ & $-0.984\pm0.043$ \\
$\Xi_c^+\to \Sigma^0K^+$ & $(1.222\pm0.041)\times10^{-4}$ & $-0.702\pm0.039$ & $0.701\pm0.031$ & $0.127\pm0.058$ \\
$\Xi_c^+\to \Lambda^0 K^+$ & $(4.321\pm0.407)\times10^{-5}$ & $-0.252\pm0.142$ & $-0.329\pm0.129$ & $-0.910\pm0.036$ \\
$\Xi_c^+\to p\eta$ & $(1.314\pm0.298)\times10^{-4}$ & $-0.514\pm0.172$ & $0.189\pm0.107$ & $-0.837\pm0.111$ \\
$\Xi_c^+\to p\eta'$ & $(6.606\pm1.255)\times10^{-5}$ & $-0.829\pm0.101$ & $-0.058\pm0.956$ & $0.556\pm0.178$ \\
\hline
\end{tabular}
\end{table*}

\begin{table*}[htp!]
\centering
\caption{Theoretical predictions for singly Cabibbo-suppressed decays in the $SU(3)_F$-symmetric fit.}
\label{tab:su3-scs}
\small
\renewcommand{\arraystretch}{1.08}
\begin{tabular}{lcccc}
\hline\hline
Decay mode~~~~~ & ~~~~~~~~~~~~~$\mathcal B$~~~~~~~~~~~~~~~ & ~~~~~~~~~~~~~~~$\alpha$~~~~~~~~~~~ & ~~~~~~~~~~~$\beta$~~~~~~~~~~~ & ~~~~~~~~~~~$\gamma$~~~~~~~~~~\\
\hline
$\Lambda_c^+\to \Lambda^0 K^+$ & $(6.645\pm0.289)\times10^{-4}$ & $-0.535\pm0.032$ & $0.304\pm0.060$ & $-0.788\pm0.025$ \\
$\Lambda_c^+\to \Sigma^0K^+$ & $(4.348\pm0.257)\times10^{-4}$ & $-0.530\pm0.142$ & $0.782\pm0.064$ & $0.328\pm0.124$ \\
$\Lambda_c^+\to \Sigma^+K^0$ & $(6.419\pm1.034)\times10^{-4}$ & $-0.826\pm0.239$ & $0.469\pm0.315$ & $-0.312\pm0.260$ \\
$\Lambda_c^+\to p\pi^0$ & $(1.960\pm0.391)\times10^{-4}$ & $-0.466\pm0.600$ & $-0.576\pm0.371$ & $-0.672\pm0.710$ \\
$\Lambda_c^+\to p\eta$ & $(1.518\pm0.078)\times10^{-3}$ & $-0.633\pm0.245$ & $0.363\pm0.628$ & $-0.684\pm0.249$ \\
$\Lambda_c^+\to p\eta'$ & $(4.841\pm0.752)\times10^{-4}$ & $-0.741\pm0.196$ & $0.027\pm0.838$ & $0.671\pm0.227$ \\
$\Lambda_c^+\to n\pi^+$ & $(6.542\pm0.724)\times10^{-4}$ & $-0.661\pm0.141$ & $-0.687\pm0.071$ & $0.301\pm0.159$ \\
$\Xi_c^0\to \Sigma^-\pi^+$ & $(6.975\pm2.088)\times10^{-4}$ & $-0.633\pm0.137$ & $0.751\pm0.235$ & $-0.186\pm1.367$ \\
$\Xi_c^0\to \Lambda^0\pi^0$ & $(1.781\pm2.483)\times10^{-4}$ & $0.158\pm0.999$ & $0.680\pm0.578$ & $-0.716\pm0.409$ \\
$\Xi_c^0\to \Lambda^0\eta$ & $(2.801\pm0.641)\times10^{-4}$ & $-0.927\pm0.355$ & $0.363\pm0.923$ & $0.096\pm1.067$ \\
$\Xi_c^0\to \Lambda^0\eta'$ & $(2.596\pm1.279)\times10^{-4}$ & $-0.428\pm2.163$ & $-0.021\pm1.494$ & $-0.903\pm1.009$ \\
$\Xi_c^0\to \Sigma^0\pi^0$ & $(2.710\pm2.293)\times10^{-4}$ & $-0.657\pm0.895$ & $0.750\pm0.853$ & $0.072\pm1.369$ \\
$\Xi_c^0\to \Sigma^0\eta$ & $(7.576\pm8.423)\times10^{-5}$ & $-0.727\pm1.140$ & $0.007\pm1.586$ & $0.687\pm1.215$ \\
$\Xi_c^0\to \Sigma^0\eta'$ & $(8.753\pm3.086)\times10^{-5}$ & $-0.736\pm1.349$ & $-0.414\pm1.396$ & $-0.536\pm1.270$ \\
$\Xi_c^0\to n\overline K^0$ & $(6.941\pm5.425)\times10^{-4}$ & $-0.932\pm0.571$ & $0.024\pm2.358$ & $-0.362\pm1.429$ \\
$\Xi_c^0\to \Xi^-K^+$ & $(4.846\pm0.837)\times10^{-4}$ & $-0.644\pm0.088$ & $0.764\pm0.042$ & $0.036\pm1.415$ \\
$\Xi_c^0\to \Xi^0K^0$ & $(4.164\pm1.271)\times10^{-4}$ & $-0.998\pm0.098$ & $0.025\pm2.530$ & $0.054\pm1.639$ \\
$\Xi_c^0\to \Sigma^+\pi^-$ & $(6.276\pm1.620)\times10^{-5}$ & $0.245\pm0.139$ & $0.147\pm6.356$ & $0.958\pm0.989$ \\
$\Xi_c^0\to pK^-$ & $(5.665\pm1.045)\times10^{-5}$ & $0.299\pm0.150$ & $0.179\pm7.719$ & $0.937\pm1.477$ \\
$\Xi_c^+\to \Lambda^0\pi^+$ & $(3.522\pm0.626)\times10^{-4}$ & $-0.079\pm0.545$ & $-0.134\pm0.176$ & $0.988\pm0.055$ \\
$\Xi_c^+\to \Sigma^0\pi^+$ & $(3.198\pm0.093)\times10^{-3}$ & $-0.569\pm0.018$ & $0.563\pm0.048$ & $-0.600\pm0.042$ \\
$\Xi_c^+\to \Sigma^+\pi^0$ & $(2.808\pm0.213)\times10^{-3}$ & $-0.520\pm0.279$ & $0.578\pm0.429$ & $-0.629\pm0.241$ \\
$\Xi_c^+\to \Sigma^+\eta$ & $(1.320\pm0.310)\times10^{-3}$ & $-0.962\pm0.209$ & $0.230\pm0.332$ & $0.144\pm0.986$ \\
$\Xi_c^+\to \Sigma^+\eta'$ & $(7.355\pm1.747)\times10^{-4}$ & $-0.663\pm0.160$ & $-0.444\pm0.687$ & $0.603\pm0.504$ \\
$\Xi_c^+\to p\overline K^0$ & $(2.923\pm0.410)\times10^{-3}$ & $-0.679\pm0.245$ & $0.385\pm0.245$ & $-0.625\pm0.175$ \\
$\Xi_c^+\to \Xi^0K^+$ & $(1.152\pm0.101)\times10^{-3}$ & $-0.549\pm0.145$ & $-0.570\pm0.037$ & $0.612\pm0.109$ \\
\hline
\end{tabular}
\end{table*}

\begin{table*}[htp!]
\centering
\caption{Theoretical predictions for Cabibbo-favored and doubly Cabibbo-suppressed decays in the fit with $SU(3)_F$ breaking.}
\label{tab:br-cf}
\small
\renewcommand{\arraystretch}{1.08}
\begin{tabular}{lcccc}
\hline\hline
Decay mode~~~~~ & ~~~~~~~~~~~~~$\mathcal B$~~~~~~~~~~~~~~~ & ~~~~~~~~~~~~~~~$\alpha$~~~~~~~~~~~ & ~~~~~~~~~~~$\beta$~~~~~~~~~~~ & ~~~~~~~~~~~$\gamma$~~~~~~~~~~\\
\hline
$\Lambda_c^+\to \Sigma^0\pi^+$ & $(1.292\pm0.044)\times10^{-2}$ & $-0.473\pm0.015$ & $0.643\pm0.103$ & $-0.602\pm0.111$ \\
$\Lambda_c^+\to \Sigma^+\pi^0$ & $(1.302\pm0.045)\times10^{-2}$ & $-0.472\pm0.015$ & $0.642\pm0.103$ & $-0.605\pm0.110$ \\
$\Lambda_c^+\to \Lambda^0\pi^+$ & $(1.313\pm0.048)\times10^{-2}$ & $-0.779\pm0.009$ & $0.382\pm0.014$ & $0.497\pm0.011$ \\
$\Lambda_c^+\to \Sigma^+\eta$ & $(3.566\pm0.395)\times10^{-3}$ & $-0.965\pm0.058$ & $-0.260\pm0.293$ & $0.032\pm1.769$ \\
$\Lambda_c^+\to \Sigma^+\eta'$ & $(4.122\pm0.835)\times10^{-3}$ & $-0.470\pm0.069$ & $-0.474\pm0.416$ & $0.744\pm0.266$ \\
$\Lambda_c^+\to \Xi^0K^+$ & $(4.599\pm0.395)\times10^{-3}$ & $-0.190\pm0.038$ & $-0.963\pm0.021$ & $0.190\pm0.103$ \\
$\Lambda_c^+\to pK_S^0$ & $(1.694\pm0.052)\times10^{-2}$ & $-0.753\pm0.010$ & $0.449\pm0.656$ & $-0.482\pm0.611$ \\
$\Lambda_c^+\to pK_L^0$ & $(1.601\pm0.049)\times10^{-2}$ & $-0.754\pm0.016$ & $0.442\pm0.659$ & $-0.486\pm0.607$ \\
$\Xi_c^0\to \Xi^-\pi^+$ & $(1.176\pm0.200)\times10^{-2}$ & $-0.647\pm0.050$ & $0.746\pm0.254$ & $-0.157\pm1.217$ \\
$\Xi_c^0\to \Sigma^+K^-$ & $(1.158\pm0.291)\times10^{-3}$ & $0.232\pm0.130$ & $0.103\pm6.374$ & $0.967\pm0.692$ \\
$\Xi_c^0\to \Xi^0\pi^0$ & $(6.472\pm1.124)\times10^{-3}$ & $-0.608\pm0.189$ & $0.700\pm0.651$ & $-0.374\pm0.967$ \\
$\Xi_c^0\to \Sigma^0K_S^0$ & $(6.138\pm1.383)\times10^{-4}$ & $-0.896\pm0.474$ & $-0.319\pm4.384$ & $0.310\pm4.215$ \\
$\Xi_c^0\to \Sigma^0K_L^0$ & $(4.282\pm2.675)\times10^{-4}$ & $-0.864\pm0.663$ & $-0.444\pm4.038$ & $0.240\pm6.185$ \\
$\Xi_c^0\to \Lambda^0 K_S^0$ & $(3.770\pm0.553)\times10^{-3}$ & $-0.987\pm0.320$ & $0.127\pm2.191$ & $-0.102\pm1.553$ \\
$\Xi_c^0\to \Lambda^0 K_L^0$ & $(4.049\pm0.657)\times10^{-3}$ & $-0.984\pm0.373$ & $0.127\pm2.436$ & $-0.123\pm1.578$ \\
$\Xi_c^0\to \Xi^0\eta$ & $(1.481\pm0.394)\times10^{-3}$ & $-0.883\pm1.544$ & $0.249\pm3.997$ & $0.397\pm4.737$ \\
$\Xi_c^0\to \Xi^0\eta'$ & $(1.437\pm0.375)\times10^{-3}$ & $-0.426\pm4.269$ & $-0.150\pm2.501$ & $-0.892\pm1.768$ \\
$\Xi_c^+\to \Xi^0\pi^+$ & $(7.138\pm1.805)\times10^{-3}$ & $-0.902\pm0.078$ & $0.324\pm0.168$ & $-0.284\pm0.278$ \\
$\Xi_c^+\to \Sigma^+K_S^0$ & $(3.326\pm0.799)\times10^{-3}$ & $-0.714\pm0.693$ & $0.669\pm1.010$ & $-0.205\pm1.208$ \\
$\Xi_c^+\to \Sigma^+K_L^0$ & $(1.558\pm0.709)\times10^{-3}$ & $-0.640\pm1.191$ & $0.750\pm1.243$ & $-0.168\pm1.503$ \\
\hline
$\Lambda_c^+\to nK^+$ & $(1.080\pm0.318)\times10^{-5}$ & $-0.814\pm0.116$ & $0.293\pm0.162$ & $-0.502\pm0.226$ \\
$\Xi_c^0\to n\pi^0$ & $(0.418\pm3.204)\times10^{-5}$ & $0.805\pm1.765$ & $0.588\pm2.550$ & $-0.074\pm1.552$ \\
$\Xi_c^0\to p\pi^-$ & $(3.573\pm0.788)\times10^{-6}$ & $0.314\pm0.156$ & $0.139\pm8.592$ & $0.939\pm1.269$ \\
$\Xi_c^0\to \Sigma^-K^+$ & $(3.305\pm0.693)\times10^{-5}$ & $-0.639\pm0.078$ & $0.736\pm0.312$ & $-0.225\pm1.185$ \\
$\Xi_c^0\to n\eta$ & $(2.714\pm1.401)\times10^{-5}$ & $-0.805\pm0.604$ & $0.185\pm2.148$ & $-0.564\pm0.876$ \\
$\Xi_c^0\to n\eta'$ & $(2.346\pm1.186)\times10^{-5}$ & $-0.223\pm1.758$ & $0.220\pm0.923$ & $-0.950\pm0.430$ \\
$\Xi_c^+\to n\pi^+$ & $(7.596\pm0.810)\times10^{-5}$ & $-0.175\pm0.036$ & $-0.889\pm0.040$ & $-0.423\pm0.087$ \\
$\Xi_c^+\to p\pi^0$ & $(3.967\pm3.276)\times10^{-5}$ & $-0.035\pm0.670$ & $0.067\pm0.809$ & $-0.997\pm0.037$ \\
$\Xi_c^+\to \Sigma^0K^+$ & $(1.297\pm0.053)\times10^{-4}$ & $-0.691\pm0.037$ & $0.703\pm0.026$ & $0.170\pm0.064$ \\
$\Xi_c^+\to \Lambda^0 K^+$ & $(4.224\pm0.506)\times10^{-5}$ & $-0.266\pm0.148$ & $-0.203\pm0.159$ & $-0.942\pm0.041$ \\
$\Xi_c^+\to p\eta$ & $(1.337\pm0.271)\times10^{-4}$ & $-0.484\pm0.206$ & $0.236\pm0.287$ & $-0.843\pm0.159$ \\
$\Xi_c^+\to p\eta'$ & $(7.462\pm2.092)\times10^{-5}$ & $-0.821\pm0.255$ & $-0.418\pm0.561$ & $0.390\pm0.616$ \\
\hline
\end{tabular}
\end{table*}

\begin{table*}[htp!]
\centering
\caption{Theoretical predictions for singly Cabibbo-suppressed decays in the fit with $SU(3)_F$ breaking.}
\label{tab:br-scs}
\small
\renewcommand{\arraystretch}{1.08}
\begin{tabular}{lcccc}
\hline\hline
Decay mode~~~~~ & ~~~~~~~~~~~~~$\mathcal B$~~~~~~~~~~~~~~~ & ~~~~~~~~~~~~~~~$\alpha$~~~~~~~~~~~ & ~~~~~~~~~~~$\beta$~~~~~~~~~~~ & ~~~~~~~~~~~$\gamma$~~~~~~~~~~\\
\hline
$\Lambda_c^+\to \Lambda^0 K^+$ & $(6.606\pm0.297)\times10^{-4}$ & $-0.534\pm0.031$ & $0.284\pm0.061$ & $-0.797\pm0.024$ \\
$\Lambda_c^+\to \Sigma^0K^+$ & $(3.992\pm0.279)\times10^{-4}$ & $-0.493\pm0.142$ & $0.700\pm0.099$ & $0.517\pm0.141$ \\
$\Lambda_c^+\to \Sigma^+K^0$ & $(4.961\pm0.983)\times10^{-4}$ & $-0.854\pm0.308$ & $0.513\pm0.639$ & $-0.084\pm0.804$ \\
$\Lambda_c^+\to p\pi^0$ & $(1.983\pm0.387)\times10^{-4}$ & $-0.385\pm0.795$ & $-0.443\pm0.800$ & $-0.810\pm0.798$ \\
$\Lambda_c^+\to p\eta$ & $(1.396\pm0.075)\times10^{-3}$ & $-0.562\pm0.118$ & $0.472\pm0.479$ & $-0.679\pm0.296$ \\
$\Lambda_c^+\to p\eta'$ & $(4.324\pm0.804)\times10^{-4}$ & $-0.771\pm0.362$ & $-0.195\pm0.614$ & $0.606\pm0.495$ \\
$\Lambda_c^+\to n\pi^+$ & $(6.019\pm0.994)\times10^{-4}$ & $-0.611\pm0.175$ & $-0.792\pm0.135$ & $-0.012\pm0.179$ \\
$\Xi_c^0\to \Sigma^-\pi^+$ & $(4.408\pm1.574)\times10^{-4}$ & $-0.628\pm0.119$ & $0.723\pm0.363$ & $-0.288\pm1.145$ \\
$\Xi_c^0\to \Lambda^0\pi^0$ & $(1.279\pm2.287)\times10^{-4}$ & $0.192\pm1.610$ & $0.735\pm0.555$ & $-0.650\pm0.431$ \\
$\Xi_c^0\to \Lambda^0\eta$ & $(2.427\pm0.917)\times10^{-4}$ & $-0.896\pm0.578$ & $0.404\pm1.317$ & $-0.186\pm0.809$ \\
$\Xi_c^0\to \Lambda^0\eta'$ & $(2.500\pm1.117)\times10^{-4}$ & $-0.185\pm2.885$ & $0.271\pm1.279$ & $-0.945\pm0.733$ \\
$\Xi_c^0\to \Sigma^0\pi^0$ & $(1.761\pm1.609)\times10^{-4}$ & $-0.753\pm1.256$ & $0.650\pm1.445$ & $-0.098\pm0.890$ \\
$\Xi_c^0\to \Sigma^0\eta$ & $(8.147\pm9.557)\times10^{-5}$ & $-0.809\pm0.960$ & $-0.309\pm1.562$ & $0.501\pm1.173$ \\
$\Xi_c^0\to \Sigma^0\eta'$ & $(8.235\pm3.290)\times10^{-5}$ & $-0.563\pm1.856$ & $-0.318\pm1.431$ & $-0.763\pm1.063$ \\
$\Xi_c^0\to n\overline K^0$ & $(4.411\pm3.311)\times10^{-4}$ & $-0.913\pm0.651$ & $-0.071\pm4.152$ & $-0.402\pm1.638$ \\
$\Xi_c^0\to \Xi^-K^+$ & $(4.475\pm0.808)\times10^{-4}$ & $-0.654\pm0.080$ & $0.753\pm0.178$ & $-0.072\pm1.242$ \\
$\Xi_c^0\to \Xi^0K^0$ & $(3.085\pm1.853)\times10^{-4}$ & $-0.963\pm0.981$ & $0.263\pm3.410$ & $-0.060\pm1.976$ \\
$\Xi_c^0\to \Sigma^+\pi^-$ & $(4.044\pm1.171)\times10^{-5}$ & $0.247\pm0.133$ & $0.110\pm6.788$ & $0.963\pm0.785$ \\
$\Xi_c^0\to pK^-$ & $(5.455\pm1.019)\times10^{-5}$ & $0.302\pm0.150$ & $0.134\pm8.275$ & $0.944\pm1.175$ \\
$\Xi_c^+\to \Lambda^0\pi^+$ & $(3.046\pm0.804)\times10^{-4}$ & $0.015\pm0.707$ & $0.227\pm0.470$ & $0.974\pm0.108$ \\
$\Xi_c^+\to \Sigma^0\pi^+$ & $(1.964\pm0.421)\times10^{-3}$ & $-0.566\pm0.035$ & $0.589\pm0.047$ & $-0.578\pm0.049$ \\
$\Xi_c^+\to \Sigma^+\pi^0$ & $(1.895\pm0.334)\times10^{-3}$ & $-0.498\pm0.166$ & $0.639\pm0.301$ & $-0.586\pm0.243$ \\
$\Xi_c^+\to \Sigma^+\eta$ & $(8.400\pm3.230)\times10^{-4}$ & $-0.975\pm0.145$ & $0.223\pm0.685$ & $-0.009\pm1.341$ \\
$\Xi_c^+\to \Sigma^+\eta'$ & $(6.160\pm1.815)\times10^{-4}$ & $-0.644\pm0.238$ & $-0.709\pm0.333$ & $0.288\pm0.588$ \\
$\Xi_c^+\to p\overline K^0$ & $(1.897\pm0.592)\times10^{-3}$ & $-0.672\pm0.363$ & $0.199\pm0.860$ & $-0.714\pm0.531$ \\
$\Xi_c^+\to \Xi^0K^+$ & $(7.489\pm1.495)\times10^{-4}$ & $-0.632\pm0.135$ & $-0.303\pm0.193$ & $0.713\pm0.133$ \\
\hline
\end{tabular}
\end{table*}

\end{appendix}

\end{document}